\documentclass[sigconf, authorversion]{acmart}

\AtBeginDocument{%
  }

\copyrightyear{2026}
\acmYear{2026}
\setcopyright{cc}
\setcctype{by-nc-nd}
\acmConference[CHI '26]{Proceedings of the 2026 CHI Conference on Human Factors in Computing Systems}{April 13--17, 2026}{Barcelona, Spain}
\acmBooktitle{Proceedings of the 2026 CHI Conference on Human Factors in Computing Systems (CHI '26), April 13--17, 2026, Barcelona, Spain}
\acmPrice{}
\acmDOI{10.1145/3772318.3790712}
\acmISBN{979-8-4007-2278-3/2026/04}

\usepackage{subcaption}
\usepackage{makecell}
\usepackage{colortbl}
\usepackage{color}

\usepackage{framed}
\usepackage{fancyhdr}
\usepackage{setspace}

\definecolor{highlight}{rgb}{0.9,0.9,0.9}
\definecolor{Hypercolor}{HTML}{18a1e7} 
\definecolor{TFFrameColor}{rgb}{0,0,0}
\newcommand{\hypcol}[1]{\textcolor{Hypercolor}{#1}}

\usepackage{float}
\usepackage{multirow}

\newcommand{\PromptBox}[2]{%
  \par\bigskip\noindent
  \textbf{#1}\par\smallskip
  \begingroup
  \setlength{\fboxsep}{6pt}% padding inside the box 
  \noindent\fbox{%
    \begin{minipage}{\dimexpr\linewidth-2\fboxsep-2\fboxrule\relax}\small
      \setlength{\parindent}{0pt}
      \setlength{\parskip}{3pt}
      #2
    \end{minipage}
  }
  \endgroup
  \par\medskip
}

\begin{document}

\title[Agent-Supported Foresight for AI Systemic Risk]{Agent-Supported Foresight for AI Systemic Risks:\\AI Agents for Breadth, Experts for Judgment}


\author{Leon Fröhling}
\orcid{0000-0002-5339-7019}
\affiliation{
  \institution{GESIS - Leibniz Institute for the Social Sciences}
  \city{Cologne}
  \country{Germany}}
\email{leon.froehling@gesis.org}

\author{Alessandro Giaconia}
\orcid{0009-0005-3268-8027}
\affiliation{
  \institution{ETH Zurich}
  \city{Zurich}
  \country{Switzerland}}
\email{agiaconia@ethz.ch}

\author{Edyta Paulina Bogucka}
\orcid{0000-0002-8774-2386}
\affiliation{
  \institution{Nokia Bell Labs}
  \city{Cambridge}
  \country{United Kingdom}}
\affiliation{
  \institution{University of Cambridge}
  \city{Cambridge}
  \country{United Kingdom}}
\email{edyta.bogucka@nokia-bell-labs.com}

\author{Daniele Quercia}
\orcid{0000-0001-9461-5804}
\affiliation{
  \institution{Nokia Bell Labs}
  \city{Cambridge}
  \country{United Kingdom}}
\affiliation{
  \institution{Politecnico di Torino}
  \city{Turin}
  \country{Italy}}
\email{quercia@cantab.net}

\renewcommand{\shortauthors}{Fröhling et al.}

\begin{abstract}
AI impact assessments often stress near-term risks because human judgment degrades over longer horizons, exemplifying the Collingridge dilemma: foresight is most needed when knowledge is scarcest. To address long-term systemic risks, we introduce a scalable approach that simulates in-silico agents using the strategic foresight method of the Futures Wheel. We applied it to four AI uses spanning Technology Readiness Levels (TRLs): Chatbot Companion (TRL 9, mature), AI Toy (TRL 7, medium), Griefbot (TRL 5, low), and Death App (TRL 2, conceptual). Across 30 agent runs per use, agents produced 86–110 consequences, condensed into 27–47 unique risks. To benchmark the agent outputs against human perspectives, we collected evaluations from 290 domain experts and 7 leaders, and conducted Futures Wheel sessions with 42 experts and 42 laypeople. Agents generated many systemic consequences across runs. Compared with these outputs, experts identified fewer risks, typically less systemic but judged more likely, whereas laypeople surfaced more emotionally salient concerns that were generally less systemic. We propose a hybrid foresight workflow, wherein agents broaden systemic coverage, and humans provide contextual grounding. Our dataset is available at: \href{https://social-dynamics.net/ai-risks/foresight}{\hypcol{https://social-dynamics.net/ai-risks/foresight}.}
\end{abstract}



%\keywords{responsible AI, ethical AI, AI governance, empirical ethics, value alignment, AI fears, AI hopes, AI influencers, participatory AI ethics, crowdsourcing}

%\keywords{Artifact or System, Quantitative Methods, Crowdsourced, Policy/Politics/Legal Issues }


\begin{teaserfigure}
  \centering
  \includegraphics[width=0.75\textwidth]{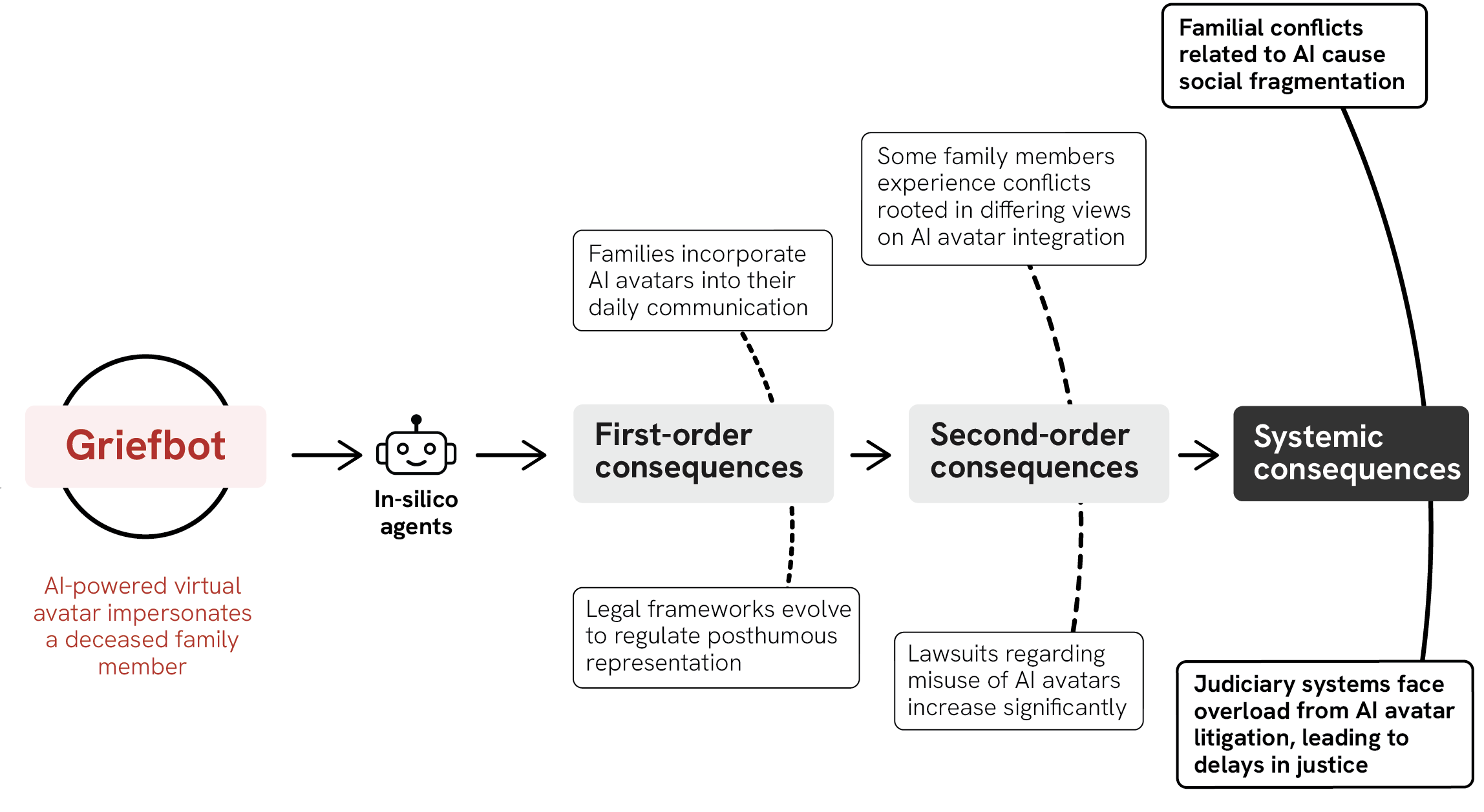}
  \caption{\textbf{Our scalable approach to agent-supported foresight supporting the ``Science Fiction Science'' method \cite{rahwan2025science}.} Starting from an AI use case (e.g., Griefbot), in-silico agents simulate the Futures Wheel process to map cascading consequences. This unfolds across first-order, second-order, and systemic rounds, highlighting potential social, legal, and institutional impacts.}
  \Description{Flowchart illustrating a causal chain of systemic risks originating from the Griefbot AI use case, which features an AI-powered virtual avatar impersonating a deceased family member. The diagram shows the progression from first-order consequences (such as families incorporating AI avatars into daily communication and legal frameworks evolving to regulate posthumous representation), to second-order consequences (familial conflicts and lawsuits over avatar misuse), and finally to systemic consequences (social fragmentation and judicial system overload).}
  \label{fig:teaser}
\end{teaserfigure}

\maketitle

\section{Introduction}
\label{sec:introduction}

Foresight is hardest when it is most needed. The Collingridge Control Dilemma \cite{collingridge1982social} captures this paradox: early in a technology's development, there is the greatest opportunity to shape its trajectory, yet also the least knowledge about its consequences. To address this difficulty, \citet{rahwan2025science} propose ``Science Fiction Science'', which uses simulated future environments to study how people might interact with emerging technologies. Their framework draws on the Technology Readiness Level (TRL) scale, which ranges from 1 (basic principles of a technology observed) to 9 (systems using the technology in full operation), and suggests that such simulations become informative only once a technology reaches around level 4.

Uncertainty at early stages also complicates governance. As \citet{pearson2024science} notes in her review of \citet{mulgan2023science}'s ``When Science Meets Power'', modern technologies increasingly exceed the capacity of existing institutions to oversee them. The EU AI Act \cite{EUAIAct2024} reflects one attempt to anticipate such challenges, requiring providers of certain high-risk AI systems to identify and mitigate systemic risks, defined as wide-reaching and potentially cascading social effects \cite{uuk2024taxonomy}. This raises a fundamental question: how can such risks be foreseen when technologies are still in formative stages?

Human–computer interaction (HCI) offers several approaches for thinking ahead about emerging technologies \cite{sanchez2025letstalkfutures}. Speculative methods such as design fiction \cite{designWorkbook2017,DesignFictionAssistans2018,AILiteracyBooklet2023} help articulate early visions of AI systems and surface value tensions. Participatory and expert-driven methods such as workshops with laypeople \cite{GlobalAIDialogues2025} and expert panels \cite{marinkovic2022corporate,glenn2003futures} help ground foresight in lived experience and domain expertise.

Despite their strengths, these methods face limits when applied to early-stage or rapidly evolving AI. They are typically anchored in concrete use contexts and near-term horizons, which restricts their ability to reveal long-range or system-level consequences \cite{sanchez2025letstalkfutures}. These methodological limits are compounded by cognitive ones: people undervalue long-term outcomes \cite{kahneman1979prospect,laibson1997golden}, misread complex systems \cite{sterman1989misperceptions}, overlook cascading failures \cite{perrow1999normal}, and downplay rare but high-impact events \cite{slovic2007mass}. These tendencies make systemic risks hard to anticipate.

Recent advances in large language models (LLMs) may offer a complementary way to support foresight. LLMs can generate many ideas, represent diverse perspectives, and be integrated into structured brainstorming methods under human oversight \cite{chen2025coexploreds,heyman2024supermind,lin2025seeking,liu2025personaflow,qin2024charactermeet}. While these capabilities do not replace human judgment, they may help broaden the considerations available during early-stage reflection. This has led to interest in using LLMs as ``in-silico agents'', simulated participants that propose possible consequences of a technology \cite{ashkinaze2025plurals}, and in combining them with foresight tools such as scenario planning \cite{perez2024prediction}. We asked two research questions:
\begin{enumerate}
  \item[(RQ1)] Can in-silico agents generate systemic risks of sufficient quality to support foresight?  
  \item[(RQ2)] How do agent-generated risks compare with human-ideated ones?  
\end{enumerate}

To address these questions, we made two main contributions: 
\begin{enumerate}
    \item \textbf{We built a pipeline for systemic risk generation with in-silico agents (\S\ref{sec:methodology}).} 
    By embedding agents in the process used for the Futures Wheel (Figure \ref{fig:methodology}), we elicited cascading consequences (Figure \ref{fig:pipeline}) for four AI use cases (Table \ref{tab:selected_uses}) of decreasing levels of TRL: a chatbot companion, an AI toy, a griefbot (Figure \ref{fig:teaser}), and a death-ordering app. These consequences were then classified, filtered, and de-duplicated into 103 systemic risks, adapting the Plurals agentic framework \cite{ashkinaze2025plurals} for pluralistic deliberation.
    
    \item \textbf{We evaluated the pipeline across three empirical studies (\S\ref{sec:results}).} 
    We first built a custom Futures Wheel interface that mirrored the pipeline, and ran it with 42 experts and 42 laypeople in two conditions: a human–plus-AI version in which participants could request AI-generated consequences, and a human-only version, producing the human-identified risks used for comparison. We then ran three studies: \emph{Study 1} involved 170 domain experts who rated 103 agent-generated risks using a shared evaluation rubric assessing each risk's systemic scope, likelihood, severity, specificity, novelty, usability, and applicability; \emph{Study 2} involved 120 domain experts who rated 89 human-identified risks using the same rubric; \emph{Study 3} involved seven domain leaders in semi-structured interviews, where they prioritized 85 agent-generated risks for the three most speculative use cases, added 19 new risks, and rated both agent- and human-generated risks with the shared rubric. Even with AI assistance, participants matched the pipeline in volume (they mentioned 13–33 risks per case), but produced narrower sets that were far less systemic (as low as 24\% of human-generated risks).
\end{enumerate}

Based on these findings, we sketched a hybrid governance workflow: in-silico agents expand the search space and help stakeholders avoid starting from scratch in risk brainstorming (Figure \ref{fig:rubric_results}), while experts and laypeople add contextual understanding and lived experience. We mapped where the pipeline performed well, identified where expert judgment was essential, and outlined how this division of work can be applied in future systemic risk identification workflows (\S\ref{sec:discussion}). To support researchers in advancing this research direction, we have publicly released the pipeline, along with agent- and human-generated systemic risks, at \textbf{\url{https://social-dynamics.net/ai-risks/foresight}}.
\section{Related Work}\label{sec:related}

We review three strands of literature that inform our work:
\emph{(1)} the limits of human foresight, which make systemic risks difficult to anticipate (\S\ref{subsec:human_limits}); 
\emph{(2)} foresight methods for AI, covering speculative design and emerging uses of AI in this domain (\S\ref{subsec:ai_foresight}); 
and \emph{(3)} approaches to identifying AI risks (\S\ref{subsec:ai_risk_identification}). 
We end by highlighting how these strands point to a common research gap that our work sets out to address.

\subsection{Limits of Human Foresight}
\label{subsec:human_limits}

Trope and Liberman's Construal Level Theory (CLT) offers insight into why the systemic risks of new technologies are difficult to foresee. CLT suggests that psychological distance across time, space, social relevance, or certainty, leads people to think in more abstract ways \citep{trope2012construal}. Systemic risks exhibit all of these four types of distance. They develop over \emph{long time horizons}, rely on \emph{complex and uncertain causal pathways}, have \emph{widespread societal consequences}, and often involve \emph{low likelihood but extreme impact}. This leads people to construct mental models that are too abstract to reveal the interactions and cascading effects that make risks systemic.

This insight connects with Garbuio and Lin's accounts of abductive reasoning in innovation. They emphasizes that mental models are essential for simplifying reality and enabling the generation of novel hypotheses. Richer, more complex mental models improve the ability to anticipate outcomes in uncertain environments \citep{garbuio2021innovative}. However, because systemic risks of future AI uses are psychologically distant, the mental models people construct about them are too abstract to capture their structural complexity, making them especially difficult for humans to foresee and reason about.

Empirical research further supports this theoretical account by showing that humans struggle with each of the defining characteristics of systemic risk. With respect to \emph{long time horizons}, studies on hyperbolic discounting and the planning fallacy demonstrate that people systematically undervalue distant outcomes and underestimate the resources required for long-term projects \citep{kahneman1979prospect,ainslie1975specious,laibson1997golden,buehler1994exploring}. 
When risks involve \emph{complex and uncertain causal pathways}, laboratory experiments in system dynamics and ``microworld'' simulations reveal persistent misperceptions of feedback, accumulation, and nonlinearity, leading to policy resistance and unintended consequences \citep{sterman1989misperceptions,dorner1996logic,forrester1971counterintuitive}. 
Regarding \emph{widespread societal consequences}, research on tightly coupled socio-technical systems shows that cascading failures emerge as ``normal accidents'' that defy intuitive hazard analysis \citep{perrow1999normal,reason1991human}. 
Finally, when risks are of \emph{low likelihood but potentially extreme impact}, people often neglect rare but catastrophic events, displaying scope insensitivity, probability neglect, and ``psychic numbing'', consistent with findings on the distorted weighting of small probabilities \citep{kahneman1979prospect,tversky1992advances,tversky1974judgment,fetherstonhaugh1997insensitivity,slovic2007mass}. 

\subsection{Foresight Methods for AI}
\label{subsec:ai_foresight}

HCI has long relied on foresight-oriented methods such as design fiction \cite{designWorkbook2017, JudgmentCall2019} and value-sensitive design \cite{systemicEffets2008, VSD2024} to interrogate emerging technologies. \citet{nathan2007value} proposed four criteria for doing so: attending not only to direct users but also to indirect stakeholders; considering how technologies may support or undermine key human values; examining longer-term consequences rather than only immediate effects; and recognizing that technologies often become pervasive across different social contexts.

However, recent analyses reveal a disconnect between these criteria and contemporary HCI practice. \citet{sanchez2025letstalkfutures} find that HCI futures research is constrained by what appears technologically plausible today, with limited attention to early-stage or speculative technologies. Long-term horizons (10–50 years) are rare. Cascading effects across political, economic, societal, technological, environmental, and legal domains (commonly referred to as ``PESTEL'' domains) \cite{glenn2003futures, systemicEffets2008} are seldom examined.

%%%
As a result, foresight practices applied to AI have often been scoped to individual applications and specific users, or to broader domains of use. For example, work in this space has examined home-monitoring technologies for households \cite{designWorkbook2017}, conversational agents for intimate interactions \cite{DesignFictionAssistans2018}, and AI learning tools for children \cite{AILiteracyBooklet2023}. \citet{GlobalAIDialogues2025} added a participatory perspective by inviting laypeople to envision potential consequences of generative AI in education, public service, and arts and culture.

More recently, emerging research has begun to test whether AI systems themselves can augment foresight. \citet{perez2024prediction} propose ``responsible computational foresight'', arguing that AI can expand scenario spaces and accelerate ideation under human judgment. \citet{ferrer2025time} show that integrating generative AI into scenario generation can diversify perspectives and stimulate debate. \citet{jung2023toward} show that LLMs can write short stories about how technology might affect vulnerable groups, helping designers think about what matters to them. Along similar lines, \citet{davidson2024exploring} compares traditional and AI-simulated Delphi processes, finding that AI can mirror and extend expert brainstorming.

\subsection{AI Risk Research}
\label{subsec:ai_risk_identification}
Research on AI risks spans four main approaches. First, taxonomies and high-level syntheses map the current landscape by reviewing the literature \cite{weidinger2022taxonomy,uuk2024taxonomy}, analyzing principles and policy documents \cite{zeng2024ai,arda2024taxonomy}, compiling incidents \cite{zhang2025dark,bagehorn2025ai,abercrombie2024collaborative}, or synthesizing existing taxonomies \cite{slattery2024ai}. Expert-led reports similarly highlight broad, catastrophic, or systemic risks \cite{hendrycks2023overview,bengio2024managing}. These works provide valuable structure, but they are largely retrospective and descriptive.  

Second, system-level and quantitative analyses evaluate risks through system features or component properties. Examples include benchmarks and taxonomies for LLM risks \cite{cui2024risk,harandizadeh2024risk}, models of risk sources \cite{steimers2022sources}, and quantitative metrics for system safety \cite{schmitz2025global}. Complementary to these are policy-to-practice translations, which operationalize regulatory frameworks into concrete risk assessment methods \cite{novelli2024ai,novelli2024taking,barrett2022actionable,bogucka2024co,broughton2023elements}. These approaches are more actionable than descriptive taxonomies, yet remain bounded by existing systems and regulatory contexts.  

Third, participatory and experiential approaches seek to surface risks grounded in lived experience \cite{delgado2023participatory,lee2019webuildai,zytko2022participatory}. \citet{kieslich2025scenario} use participatory scenario writing to capture diverse societal perspectives on AI risks, while \citet{mun2024particip} propose democratic surveying frameworks to anticipate future uses. \citet{datey2024just} conducted interviews with women as potential victims of online dating harms to inform a risk detection model. Other methods extrapolate risks from past incidents. \citet{pang2024blip} facilitate exploration of undesirable consequences of digital technologies through interactive tools, and \citet{wang2024farsight} integrate foresight into prototyping processes to sensitize practitioners to possible harms. These approaches are contextually rich but resource-intensive.

Fourth, recent work has begun to explore generative LLM-based approaches. These methods assist practitioners by generating candidate risks in structured formats \cite{buccinca2023aha,constantinides2024good,herdel2024exploregen,wang2024farsight}. Despite their scalability and efficiency, such tools are often application-specific, and tend to prioritize immediate harms over systemic consequences.

\smallskip
\noindent\textbf{Research Gap}. These research strands reveal three limitations. First, humans face cognitive barriers that make systemic risks particularly hard to anticipate. Second, foresight research has begun to test AI's role in augmenting human scenario generation, but its integration into practice remains partial. Third, most AI risk research either systematizes what is already known, or narrows in on immediate harms, leaving systemic risks underexplored. These three limitations point to a clear gap: \emph{the need for scalable methods that combine the generative breadth of AI with the contextual judgment of humans to anticipate systemic risks of novel AI uses}. Our work aims at addressing this gap.
\section{Author Positionality Statement}
\label{sec:positionality}

Before outlining our methodology, we first position ourselves in relation to the approach we present in this work. The team consists of individuals with expertise in Computer Science, AI, and Data Visualization, three men and one woman, bringing together diverse experiences from both industrial research labs and academic institutions. We have cultural and professional backgrounds spanning all parts of Europe and North America. We also represent a range of religious affiliations. We acknowledge that our positionality may influence various aspects of our research, including, but not limited to, our design decisions for the choice of use cases, the implementation of the in-silico agents, and the topics emphasized in quantitative analyses. We recognize the importance of including a broader range of voices from academia and the public.
\section{Methodology}
\label{sec:methodology}

\begin{figure*}[t!]
    \centering
    \includegraphics[width=\linewidth]{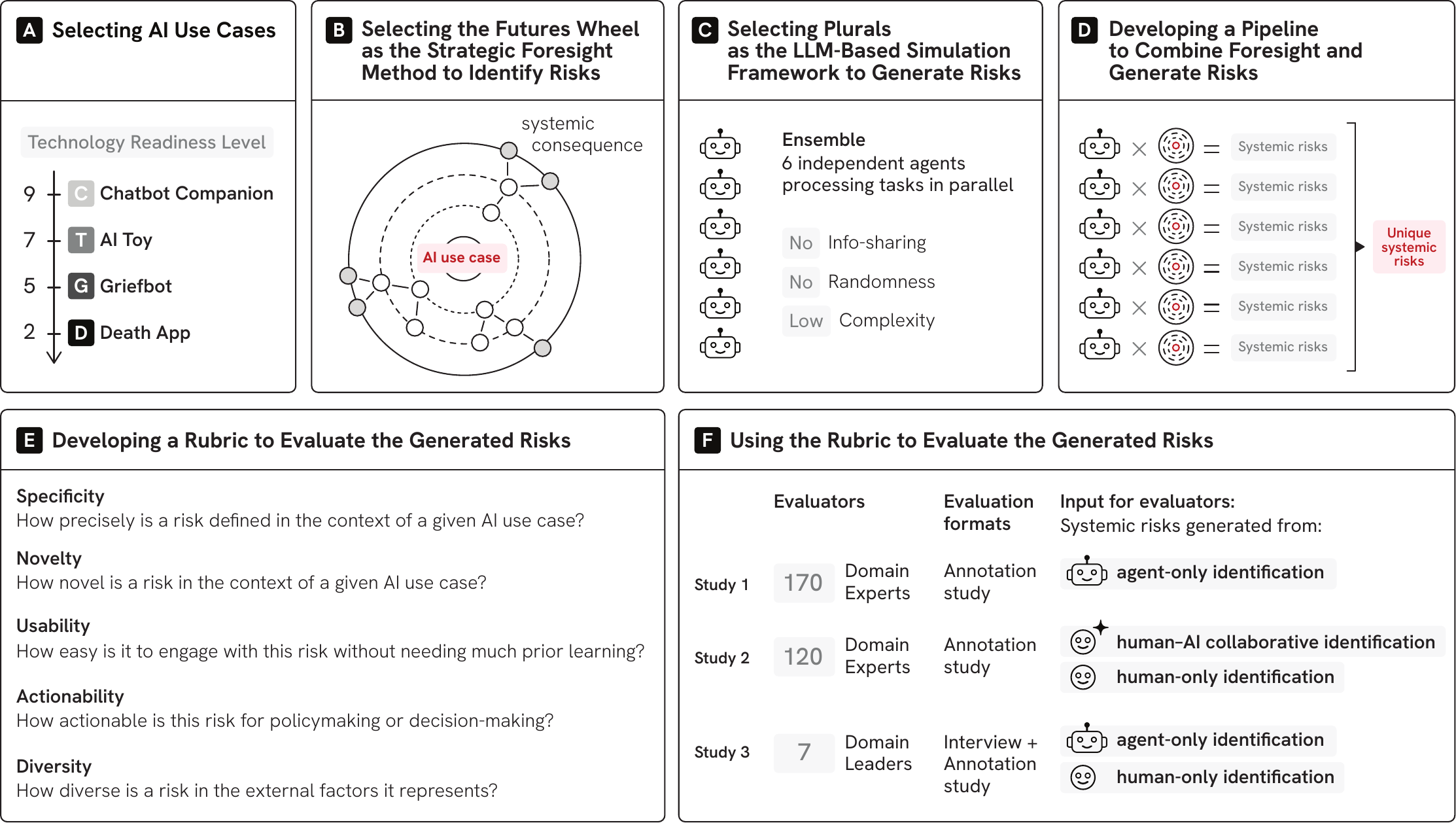}
    \caption{\textbf{Overview of our five-step approach for generating and evaluating systemic risks of novel AI uses.} The process combines strategic foresight (via the Futures Wheel), LLM-based agentic simulation (via Plurals), and a structured evaluation rubric. Together, these steps allow us to generate systemic risks across AI use cases of varying technological maturity, and to compare agent-generated risks against human-identified ones.}
    
    \Description{Diagram showing a five-step approach for generating and evaluating systemic risks of novel AI uses. The process begins with selecting the AI use cases Chatbot Companions, AI Toy, Griefbot, and Death App. Next, risks are generated by combining a strategic foresight method (the Futures Wheel) with LLM-based simulation (Plurals). An evaluation rubric with five criteria --- specificity, novelty, usability, applicability, and diversity --- assesses the generated risks. Finally, risks are compared against those identified by human evaluators, including AI domain experts and laypeople.}
    \label{fig:methodology}
\end{figure*}

To develop our approach for generating and evaluating systemic risks of novel AI uses, we followed a five-step method that combines strategic foresight with in-silico agent simulation (Figure \ref{fig:methodology}A–E). First, we selected four AI use cases with decreasing levels of technological maturity, from widely deployed applications to speculative concepts (\S\ref{subsec:use_cases}). Second, we adopted the Futures Wheel as our foresight method to structure the identification of potential systemic risks for each use case (\S\ref{subsec:futures_wheel}). Third, we instantiated in-silico agents using the Plurals framework \cite{ashkinaze2025plurals}, an ensemble-based approach for simulating pluralistic deliberation, as the mechanism for generating risks from those consequences (\S\ref{subsec:plurals}). Fourth, we combined foresight and simulation into a pipeline that systematically produces systemic risks across use cases (\S\ref{subsec:pipeline}). Finally, we developed a rubric to evaluate the generated risks along five dimensions (specificity, novelty, usability, applicability, and diversity). We then used it with domain experts and domain leaders to assess the quality of agent-generated risks, comparing them against two conditions: risks identified by humans alone, and those identified through human–AI collaborations (\S\ref{subsec:rubric}).

\subsection{Selecting AI Use Cases}
\label{subsec:use_cases}

To identify our AI use cases, we followed best practices in case-study research \cite{Greenhalgh2025}. First, all authors independently collected candidate AI use cases, and combined them into a longlist of 13, spanning domains such as education, healthcare, and family, reflecting the principle of broad initial exploration before focused selection. Second, through two structured discussions, this longlist was narrowed using four criteria: (1) the use case could plausibly produce systemic risks; (2) it occupied a clearly identifiable position on the TRL scale assessing product maturity \cite{hicks2009methodology}; (3) it was conceptually distinct from other candidates; and (4) it showed sufficient grounding in public discourse, appearing in existing societal or expert debates. Two use cases were discarded because they were unlikely to lead to a broad and diverse enough set of systemic implications (e.g., using AI-powered exoskeletons to support workers), three because of their unclear maturity (e.g., AI-powered co-workers), two because they were conceptually too similar to others that ended up being included (e.g., celebrities as AI companion dolls), and two because they lack sufficient grounding in current public discourse (e.g., digital clones of human minds used as characters in video games). This resulted in selecting four cases spanning distinct domains of AI application and maturity levels (Table \ref{tab:selected_uses}): Chatbot Companion (TRL 9, mature), AI Toy (TRL 7, medium), Griefbot (TRL 5, low), and Death App (TRL 2, conceptual). Despite their differences, all four use cases involve AI mediating forms of personal support. These range from everyday companionship \cite{malfacini2025impacts} to childhood learning \cite{AIChildren2025} to grief-related connection \cite{Hollanek2024, Kozlovski_Makhortykh_2025} to end-of-life guidance \cite{ForbesDeath2025}, placing them on a shared spectrum of human–AI interaction. By covering the full spectrum of TRLs, from widely deployed systems to early-stage concepts, we test our approach across the full ``cone of uncertainty''~\cite{rahwan2025science}: from contexts where risks are clearly defined (lowest uncertainty) \cite{gall2022visualise} to contexts where risks remain speculative and distant (highest uncertainty).

Prior work on ``Science Fiction Science'' recommends focusing only on technologies with TRL 4 or higher, arguing that humans struggle to imagine long-term or disruptive effects when technologies are too speculative \cite{rahwan2025science}. Our aim is precisely to test this boundary: can in-silico agents surface systemic risks even for early-stage, disruptive concepts that humans find hardest to envision?

\begin{table*}
    \centering
    \small
    \setlength{\tabcolsep}{7pt}
    \renewcommand{\arraystretch}{1.1}
    \caption{\textbf{Overview of the four novel AI use cases selected for evaluation, alongside their  TRLs.} The set was chosen to span the full range of readiness, from early-stage speculative ideas (Death App, TRL 2) to widely deployed applications (Chatbot Companion, TRL 9), in order to test how well our approach captures risks across different levels of maturity.}
    
    \Description{A table describing four AI use cases across a spectrum of technological maturity. The table has three columns ('AI Use', 'Description', 'TRL') and four data rows. The use cases are presented in descending order of readiness, from most mature to least: Chatbot Companion is TRL 9 (proven through real-world operation); AI Toy is TRL 7 (beta prototype); Griefbot is TRL 5 (validated in a relevant environment); and Death App is TRL 2 (conceptual). Each row provides a detailed description of the use case and a justification for its TRL rating.}

    \label{tab:selected_uses}
    \begin{tabular}{p{1.2cm}p{6.55cm}p{8.4cm}}
    \toprule
    AI Use & Description & Technology Readiness Level (TRL)  \\
    \toprule
    Chatbot \newline Companion & 
    An AI chatbot that provides conversation and emotional support at any time of day. It is designed to help people who feel lonely or who want a steady source of dialogue and reassurance. The intended users are people who seek companionship outside normal social or family circles. & 
    \textbf{TRL 9}~\textemdash~\textit{Proven through real-world operation.} Conversational agents providing companionship are already deployed at scale on mobile and web platforms. OpenAI has released ChatGPT as a conversational chatbot in November 2022~\footnote{\url{https://openai.com/index/chatgpt/}}.\\
    \addlinespace[0.3em]
    
    AI Toy & 
    A soft toy with built-in AI that answers children’s questions about science in clear, spoken language. It is meant to spark curiosity and give comfort while children explore ideas. The main users are children aged 5–12 and their parents or carers, who want both play and learning. & 
    \textbf{TRL 7}~\textemdash~\textit{Beta prototype demonstrated in operational environment.} Voice-activated educational toys are currently in beta release, operating in households but not yet certified against regulatory standards for child products. Curio's beta toys have been available for pre-order since December 2023~\footnote{\url{https://x.com/CurioBeta/status/1735372169939652785}}.\\
    \addlinespace[0.3em]
    
    Griefbot & 
    A digital avatar that imitates a deceased family member. It produces text or voice responses based on past records, such as old messages, in order to give users the sense of ongoing contact. The intended users are people who wish to maintain a form of connection with the dead. & 
    \textbf{TRL 5}~\textemdash~\textit{Validation of basic elements in relevant environment.} Systems that simulate interaction with deceased individuals are undergoing small-scale trials. Core elements are validated in relevant environments, though fidelity, ethical safeguards, and regulation remain unresolved. In 2025, companies like HereAfterAI~\footnote{\url{https://www.hereafter.ai/}} are experimenting with features that allow for the imitation of the dead through interactive avatars.\\
    \addlinespace[0.3em]
    
    Death App & 
    An AI-powered platform that matches individuals seeking to end their lives with service providers and shows transition plans similar to those provided by end-of-life doulas. The intended users are people considering assisted dying.
    & 
    \textbf{TRL 2}~\textemdash~\textit{Invention of concept or application.} Applications connecting individuals with assisted dying services remain conceptual, with pilot exploration under restrictive legal conditions. The idea of an ``AirBnB for death'' has been presented during art exhibition ``The Future is Present'' in 2024~\footnote{\url{https://designmuseum.dk/udstilling/the-future-is-present}}.\\
    \bottomrule
    \end{tabular}
\end{table*}

\subsection{Selecting the Futures Wheel as the Strategic Foresight Method to Identify Risks}
\label{subsec:futures_wheel}

The Futures Wheel is a simple method for identifying primary, secondary, and tertiary consequences of the subject \cite{glenn1972futurizing}. As described by \citet{glenn2003futures}, the method requires no more than blank paper and a pen, and can be run individually or in groups. The process always starts with a central subject such as a trend, novel idea, or recent event (as shown in Step 1 of Figure~\ref{fig:pipeline}). Participants are then asked the simple question: ``If this occurs, then what happens next?'', generating first-order consequences. After a number of such first-order consequences have been identified, the next round begins: for each first-order consequence, the same question is asked to elicit second-order consequences. In principle, this cycle can continue for as many rounds as desired, although foresight practice shows that three rounds are typically sufficient to surface the main cascading effects. In the final round, the question is often adjusted to reconnect emerging consequences back to the central subject \cite{hines2006thinking}. For example, participants may be asked, ``If this occurs, what new risks and benefits emerge around the subject?''.

Futurists widely use the Futures Wheel because it supports systematic exploration of complex and interconnected outcomes. \citet{glenn2003futures} argue that the method helps move from ``linear, hierarchical, and simplistic'' reasoning toward a more ``network-oriented, and organic'' view of future developments. The visual structure of the wheel makes these interactions explicit, providing a clear map of how consequences interrelate and evolve over time.

We select the Futures Wheel as the most suitable method for LLM-based foresight for three reasons:
\begin{enumerate}
    \item \emph{It encourages thinking about complex interactions and unintended consequences.} The Futures Wheel is designed to surface cascading and multi-order effects, revealing consequences of a trend or change that may otherwise remain unconsidered. This aligns directly with the type of reasoning needed to identify systemic risks.
    \item \emph{It does not require advanced expertise.} The method can be used equally well by laypeople and experts, individually or in groups. This makes it well suited for instantiating diverse LLM-based agents as brainstorming participants.
    \item \emph{It does not require complex inputs.} The Futures Wheel is designed to operate without detailed technical specifications or extensive preparatory materials. This enables rapid application to new use cases without needing specialized system knowledge or large input datasets.
\end{enumerate}

\subsection{Selecting Plurals as the LLM-Based Simulation Framework to Generate Risks}
\label{subsec:plurals}

\citet{ashkinaze2025plurals} introduce Plurals as a framework for simulating pluralistic AI deliberations. At its core, Plurals comprises three abstractions: Agents, Structures, and Moderators. \emph{Agents} are LLMs initialized with roles and profile descriptions, tasked with completing deliberative steps and optionally given instructions for processing other Agents' outputs. \emph{Structures} are interaction topologies determining the flow of information between Agents, including Chains, Debates, Directed Acyclic Graphs, and Ensembles, which vary in information sharing, randomness, and repetition cycles. \emph{Moderators} are LLMs initialized with role descriptions and responsible for combining and summarizing Agents' outputs at the end of each deliberation phase. 

These three abstractions can be customized and arranged to simulate different forms of deliberation. \citet{ashkinaze2025plurals} demonstrate Plurals through six such forms, including LLM debates, role-based discussions, and audience-targeted argument generation. Their studies show that assigning different roles to LLMs and placing them in different interaction structures reliably produces contributions that differ from, and are often stronger than, those produced by simpler prompting baselines.

\bigskip

We select Plurals as the most suitable framework for simulating brainstorming participants for three reasons:
\begin{enumerate}
    \item \emph{It is scalable across multiple use cases.} Plurals allows to run many independent simulations across different use cases with minimal overhead. Its LiteLLM backend \cite{liteLLM} supports fast access to state-of-the-art LLMs, enabling large-scale experimentation without additional infrastructure.
    \item \emph{It is adaptable to diverse methodologies.} Plurals provides several pre-implemented deliberation structures and task prompts for Agents and Moderators, while also allowing their easy modification and integration of alternative approaches.
    \item \emph{It enables the introduction of different in-silico participants.} Plurals offers built-in functionality to sample a diverse set of in-silico participants from clearly defined, representative populations, while also allowing full customization of agent personas by specifying their roles, goals, or traits.
\end{enumerate}

In our implementation, we use the term \emph{in-silico agent} to refer to an individual LLM instance configured with a distinct role, background, or attitude, and instructed to participate in a structured, multi-step deliberation. These agents lack autonomy or persistent internal state; their reasoning is governed entirely by prompts specifying what information they may access, and how they should respond. This definition builds on \citet{ashkinaze2025plurals}; in our setup, each agent plays the role of a standardized, controllable stand-in for a human participant, enabling comparisons across runs.

\medskip

\subsection{Developing a Pipeline to Combine Foresight and Generate Risks}
\label{subsec:pipeline}

\begin{figure*}[t!]
    \centering
    \includegraphics[width=\linewidth]{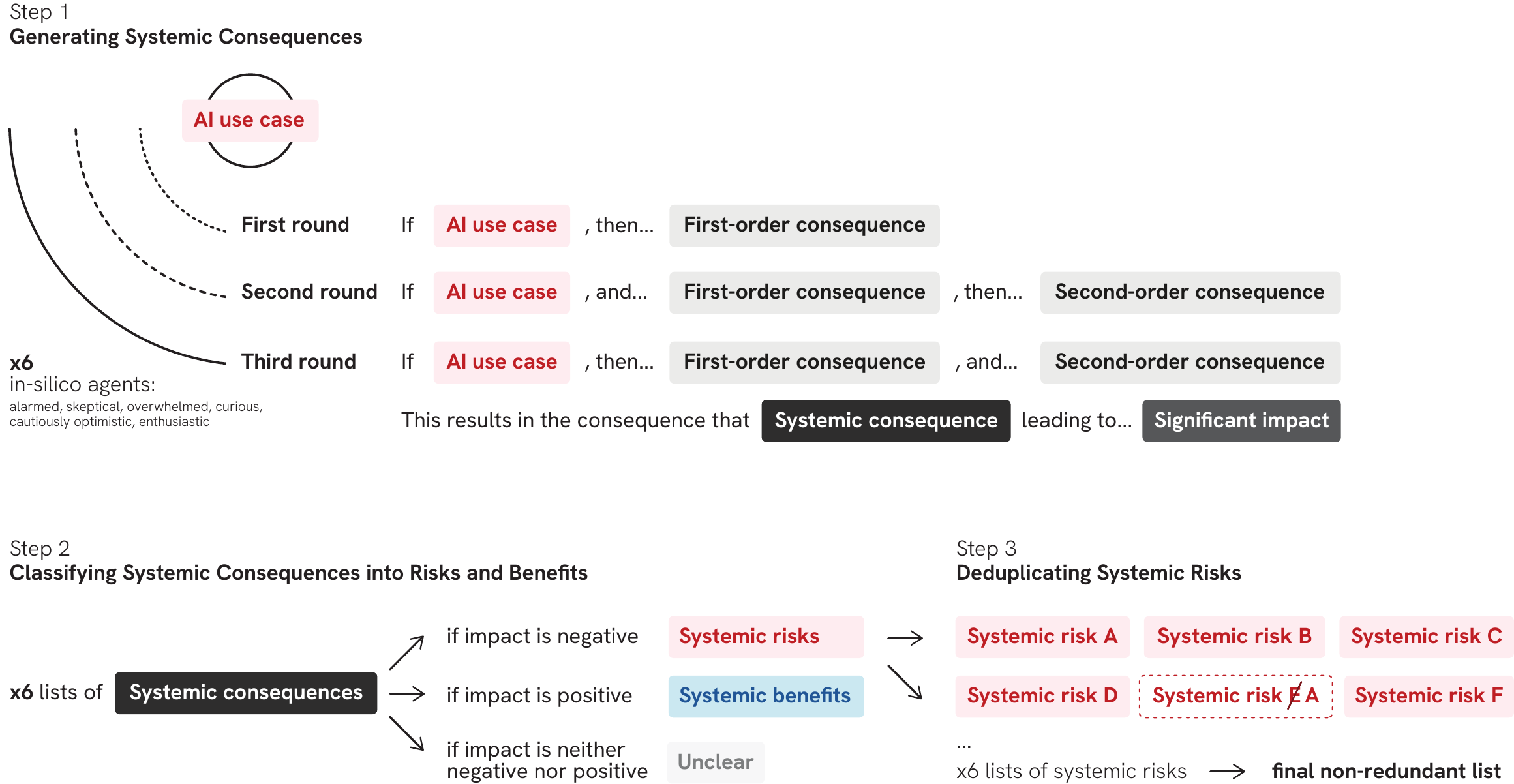}
    \caption{\textbf{Simulation of the Futures Wheel within our systemic risk pipeline.} We adapt the Futures Wheel for use with six in-silico agents to move from initial AI use cases to consolidated lists of systemic risks. Agents first generate multi-order systemic consequences (Step 1), which are then classified into risks or benefits (Step 2), and finally deduplicated into non-redundant sets (Step 3). This process mirrors structured human foresight while ensuring diversity of outputs across agents and producing consistent, machine-generated inputs for later evaluation.}
    \Description{Diagram illustrating the simulation of the Futures Wheel within the systemic risk pipeline. Six in-silico agents, each with different dispositions (alarmed, skeptical, overwhelmed, curious, cautiously optimistic, enthusiastic), generate consequences of increasing order following the Future Wheel process, starting from an AI use case. Consequences are first produced in parallel (Step 1), then classified as risks, benefits, or unclear (Step 2), and finally deduplicated across the six in-silico agents into a consolidated non-redundant list of systemic risks (Step 3).}
    \label{fig:pipeline}
\end{figure*}

We designed a three-step pipeline that takes a short description of an AI use and produces a list of its potential systemic risks. In the Futures Wheel, the AI use sits at the center of the wheel (Figure \ref{fig:pipeline}, Step 1); systemic risks emerge after three rounds of consequence generation. The pipeline follows the Futures Wheel structure, implemented with the Plurals framework. For robustness, we ran the full pipeline 30 times per AI use case, following standard practice in computational studies \cite{Song2025, Ouyang2025}. To avoid information leakage between steps, each step was executed in a separate API session with no shared memory or state.

\mbox{ }\\
\noindent
\textbf{Step 1. Generating Systemic Consequences.}  
We implemented the Futures Wheel using six individual agents, run in parallel through Plural’s Ensemble structure without interaction. Each agent received a short task description and a persona: a simple one-word profile reflecting its attitude toward AI. Appendix~\ref{app:generation_prompts} provides the full prompts used in this step.

The task unfolded in three rounds (Figure~\ref{fig:pipeline}, Step 1). First, agents listed immediate consequences of the AI use case. Second, they combined these with the use to suggest second-order effects. Third, they linked the use with first- and second-order consequences to describe the resulting systemic consequence and significant impact, guided by definitions adapted from the EU AI Act's Code of Practice\footnote{A systemic consequence has a significant impact on international markets due to its reach, or through actual or foreseeable effects on public health, safety, security, fundamental rights, or society at large, capable of spreading across the value chain.}. Agents always saw the full chain of their own earlier inputs but never the outputs of other agents.

Guided by foresight best practices \cite{hines2006thinking, Maertins2016} and two pilot studies, we used six agents: a group size large enough to capture diverse perspectives, while keeping outcomes manageable for human analysis \cite{hines2006thinking, Maertins2016}. We followed standard Futures Wheel guidance by adopting three rounds \cite{hines2006thinking}; a four-round pilot proved redundant because systemic consequences were consistently reached by round three. To avoid overly risk-heavy outputs and support a balanced range of thinking, we assigned each agent a distinct attitude (alarmed, skeptical, overwhelmed, curious, cautiously optimistic, and enthusiastic). These attitudes were selected to span the positive--negative spectrum of public perspectives on AI and were informed by labels used in existing attitude surveys \cite{genAI_Attitudes_2025} and prior research on role specialization and cognitive diversity in multi-agent systems \cite{liu2025personaflow, agentPersonas_2025}. In our first pilot, without persona variation, the model's third-round outputs skewed heavily negative: over 90\% were risks rather than benefits. To support consistent yet diverse outputs, agents operated under a bounded creativity setup. Each round was guided by structured prompts that defined the type of consequence to generate, provided operational definitions, and enforced a consistent output format, while still allowing agents to develop their own reasoning and ideas within those constraints. 

To implement this setup at scale, we evaluated three language models for the generation step: the open-source \textsc{Llama3.3-70B} from Meta AI and the proprietary \textsc{GPT-4.1 mini} and \textsc{GPT-5} from OpenAI. For each model, we conducted 10 independent generation runs under identical prompts. We compared models based on the number and consistency of generated consequences, benchmark performance, and cost. Using \textsc{Llama3.3-70B} yielded $\mu=83.6$ consequences per run ($\sigma=11.4$), \textsc{GPT-4.1 mini} produced $\mu=81.4$ ($\sigma=9.0$), and \textsc{GPT-5} produced $\mu=121.6$ ($\sigma=16.1$). While \textsc{GPT-5} was the most prolific, \textsc{GPT-4.1 mini} and \textsc{Llama3.3-70B} produced similar numbers. Across all runs and models, social consequences dominated the outputs, while environmental impacts were almost entirely absent. We report exemplary consequences from each model in Appendix~\ref{app:supplementary_pipeline_analysis}, Table~\ref{tab:generated_consequences} and make the complete set available on the project website.

We opted for \textsc{GPT-4.1 mini} as our model of choice for the generation step. It shows strong performance on instruction following benchmarks (MultiChallenge and the OpenAI API instruction-following benchmark) and very strong performance on long-context detail extraction (BrowseComp Long Context 128k benchmark)\footnote{\url{https://openai.com/index/gpt-4-1}}. Compared to \textsc{GPT-5}, \textsc{GPT-4.1 mini} achieves the same accuracy on long-input extraction task, and only slightly lower performance on instruction following, while running approximately 4 times faster and costing roughly 75 times less per run\footnote{\url{https://openai.com/index/introducing-gpt-5-for-developers}}. These properties were essential for repeatedly running foresight simulations without compromising output quality, and for making the pipeline accessible for replication by other practitioners.

We implemented the pipeline using \textsc{GPT-4.1 mini} via LiteLLM \cite{liteLLM} at a temperature of 1 (Appendix~\ref{app:supplementary_pipeline_analysis}, Figure  \ref{fig:agents_wheel}). To test its stability, we repeated the process 30 times, following standard practice in computational studies to account for stochastic variation in LLM outputs \cite{Song2025, Ouyang2025}. We assessed variation in the number, type, and thematic coverage of consequences across runs and agents. The number of consequences per run was stable $(\mu = 81.4,\ \sigma = 9.0$, see Appendix~\ref{app:supplementary_pipeline_analysis}, Figures~\ref{fig:overview_pipeline_analysis}a and ~\ref{fig:rq1_boxplots_generations}), while the number of unique risks and benefits increased with additional runs, but plateaued after approximately 20 (Appendix~\ref{app:supplementary_pipeline_analysis}, Figure~\ref{fig:overview_pipeline_analysis}b), suggesting a reliable balance between diversity and saturation. Varying agent attitudes increased the diversity of consequences: pessimistic agents generated mostly risks, while optimistic ones contributed more benefits (Appendix~\ref{app:supplementary_pipeline_analysis}, Figure~\ref{fig:overview_pipeline_analysis}c). Thematic coverage was highly consistent across runs, with over 0.95 average similarity in the distribution of risk types across PESTEL categories (Political, Economic, Societal, Technological, Environmental, and Legal) \cite{glenn2003futures} (Appendix~\ref{app:supplementary_pipeline_analysis}, Figure~\ref{fig:overview_pipeline_analysis}d).

\begin{figure*}[t!]
    \centering
    \includegraphics[width=0.98\linewidth]{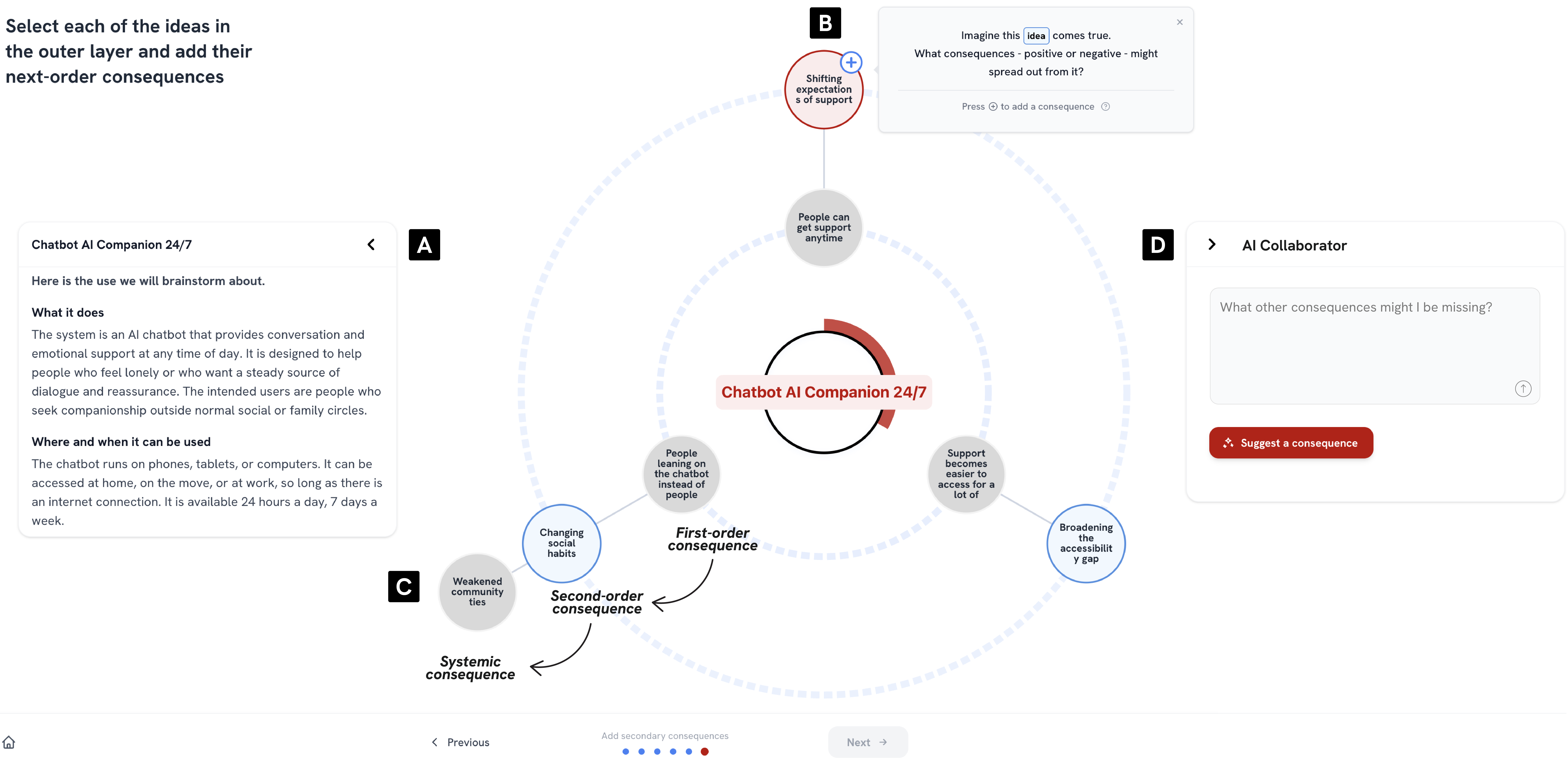}
    \caption{\textbf{Futures Wheel interface used to collect \emph{human}-ideated risks in both the human-only and human–AI collaboration conditions.} The left panel (A) provides a structured description of the studied AI use case, following ISO 42005 guidelines by outlining its intended function and users, context of use, known limitations, and deployment environment. At the center, the focal use case is displayed together with round-specific prompts shown in pop-ups (B). Participants brainstorm first-order consequences, which then branch into second- and third-order consequences (C), allowing cascading impacts to be visualized across multiple Futures Wheel rounds. The right panel (D) provides optional support through a chat window for exploring potentially missing consequences, and a button to generate some of them automatically with AI. Interface elements A–C were used in both conditions, while D appeared only in the human–AI collaboration condition.}
    
    \Description{Screenshot of the Futures Wheel interface used to elicit human-ideated risks in both the human-only and human–AI collaboration conditions. The interface presents structured information about an AI use case (Panel A, to the left), prompts participants to brainstorm cascading consequences (Panel B, center, and visualizes first-, second-, and third-order impacts (Panel C, branching visualization next to the ideated consequences of the different rounds of the Futures Wheel) he right panel (D) provides optional support through a chat window for exploring potentially missing consequences and a button to generate them automatically with AI. Interface elements A–C were used in both conditions, while D appeared only in the human–AI collaboration condition.}
    \label{fig:futures_wheel_viz}
\end{figure*}

\begin{figure*}
    \centering
    \includegraphics[width=\linewidth]{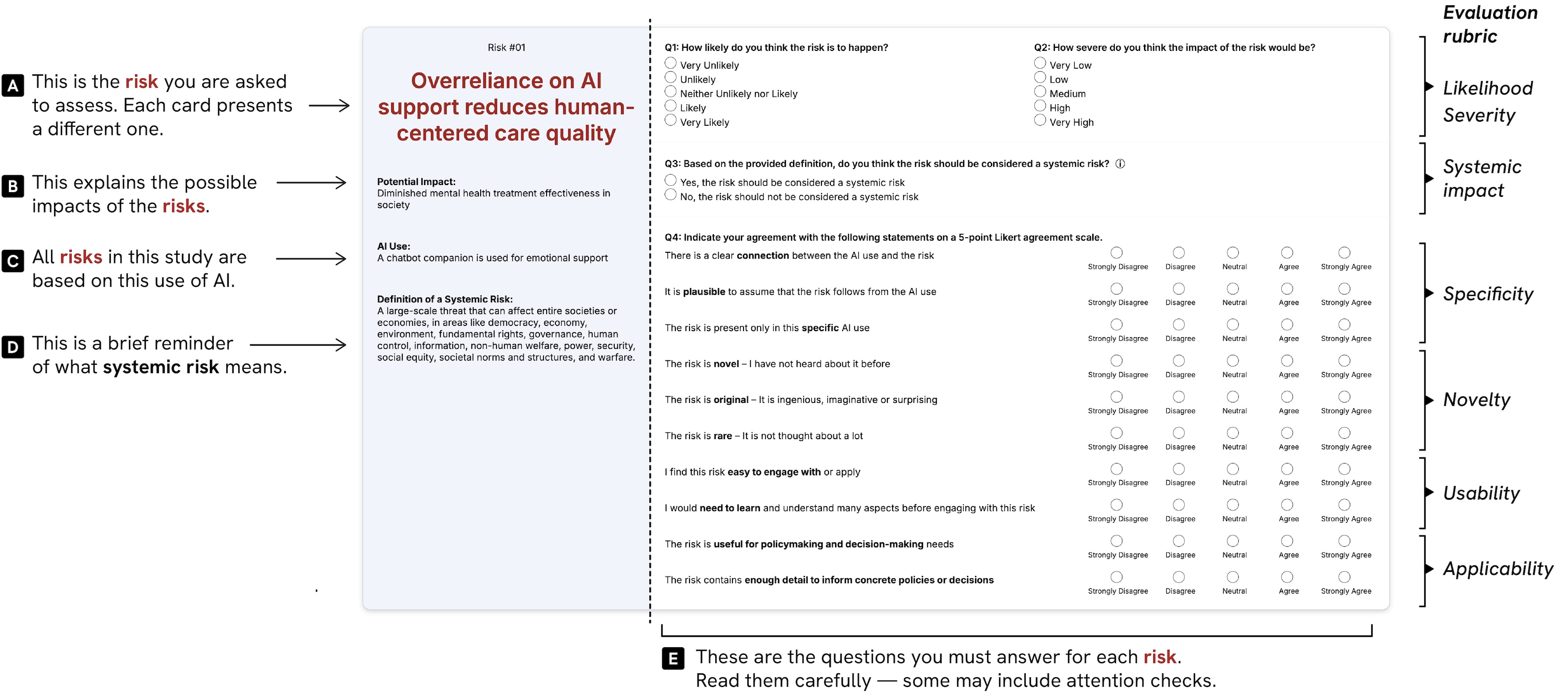}
    \caption{\textbf{Example annotation card shown to domain experts for evaluating the risks generated by in-silico agents.} The left panel illustrates a sample risk statement (i.e., ``Overreliance on AI support reduces human-centered care quality'', A), its potential impact (B), the AI use case it belongs to (i.e., ``Chatbot companion'', C), and a brief definition of systemic risk (D). The right panel (E) presents the evaluation metric from our rubric. It includes questions on likelihood, severity, and systemic classification, as well as ten risk characteristics organized across four dimensions: specificity, novelty, usability, and applicability.}
    
    \Description{Example annotation card used by domain experts to evaluate systemic risks generated by in-silico agents. The card presents a sample risk statement with potential impact, the AI use case, and the definition of systemic risks as supporting context on the left panel, and the evaluation rubric with structured questions on the right panel. The evaluation questions ask for a likelihood classification on a five-point scale from very unlikely to very likely, for a severity classification on a five-point scale from very low to very high, for a binary classification of whether the risk should be considered systemic, and for the level of agreement on five-point Likert scale with the ten statements for specificity, novelty, usability, and applicability}
    \label{fig:annotation_card}
\end{figure*}

\smallskip
\textbf{Step 2. Classifying Systemic Consequences into Risks and Benefits.} 
The Futures Wheel produces systemic consequences, but our focus is on systemic risks. We therefore added a step in which each agent classified its own systemic consequences as risks, as benefits, or, if neither applied, unclear (Figure \ref{fig:pipeline}, Step 2). 

We used OpenAI's \textsc{GPT-4.1 mini} via LiteLLM \cite{liteLLM}, initializing agents with the same personas as in Step~1. Appendix~\ref{app:generation_prompts} lists the classification prompt used in this step. To validate the approach, three co-authors independently annotated all 75 consequences from a randomly chosen generation run. Their annotations were aggregated by majority vote, forming a human ground truth against which we compared the agents' classifications. The weighted $F1$-score was 0.86, with 0.96 for the risk class. All 25 risks identified by the human annotators were correctly classified; only two items were labeled ``unclear'' by humans but marked as ``risks'' by the agents. These results indicate that the method is sufficiently reliable for classifying generated consequences into risks and benefits. 

To assess whether this performance was specific to \textsc{GPT-4.1 mini} or stable across models, we repeated the classification step with the open-source \textsc{Llama3.3-70B}, and with the proprietary \textsc{GPT-5}. Their agreement with the human ground truth was comparable ($F1$-scores of 0.88 and 0.91, respectively \emph{vs.} \ 0.86), and all three models showed very high mutual agreement (Krippendorff's $\alpha = 0.91$), exceeding the agreement among human annotators ($\alpha = 0.73$). These results indicate that the classification step was stable across models.

%\mbox{ }\\
%\noindent
\smallskip
\textbf{Step 3. Deduplicating Systemic Risks.}
Finally, we consolidated the systemic risks produced by independent agents into a single list. While duplication was rare within one agent's outputs, overlaps were common across agents. We used an LLM to deduplicate iteratively: we started with the risks from the first agent, then compared each subsequent agent's list against the growing set, adding only items not already present (Figure \ref{fig:pipeline}, Step 3). The result is a consolidated, non-redundant list. We used OpenAI's \textsc{o4-mini} via LiteLLM \cite{liteLLM} for the deduplication (prompt in Appendix \ref{app:generation_prompts}), and \textsc{text-embedding-3-small}\footnote{https://platform.openai.com/docs/models/text-embedding-3-small} to compute pairwise cosine similarities. On one run with 27 risks, the average pairwise similarity fell from 0.3418 before deduplication to 0.3355 after. In that run, two risks were flagged as duplicates, reducing the final set from 27 to 25. We verified that these removals corresponded to the highest pairwise similarities (Appendix \ref{app:supplementary_pipeline_analysis}, Table \ref{tab:deduplication_validation} report examples of risks removed as too similar, and those retained as sufficiently distinct).  

To assess whether deduplication outcomes depended on the specific model used, we repeated this step with the open-source \textsc{Llama3.3-70B} and with the proprietary \textsc{GPT-5} (Appendix~\ref{app:supplementary_pipeline_analysis}, Figure~\ref{fig:deduplication_models}). All models reduced similarity by a comparable magnitude, with \textsc{GPT-4.1 mini} producing the strongest decrease ($-0.0062$; \textsc{GPT-5}: $-0.0023$; \textsc{Llama3.3-70B}: $-0.0025$).

\subsection{Developing a Rubric to Evaluate the Generated Risks}
\label{subsec:rubric}

Because no established framework exists for assessing the quality of generated risks, we developed a new rubric through a four-step process (Appendix \ref{app:rubric_process}). First, three co-authors independently reviewed criteria used to evaluate ideas, scenarios, and socio-technical systems \cite{diakopoulos2021anticipating, dean2006identifying, kieslich2025scenario, doshi2024generative, brooke1996sus, glenn2003futures, Wang1996BeyondAccuracy, sanchez2025letstalkfutures}, and proposed eight candidate dimensions. Second, we consolidated these proposals, and removed dimensions that could not be applied consistently across both agent-generated and human-ideated risks. This resulted in five core dimensions: \emph{specificity}, \emph{novelty}, \emph{usability}, \emph{actionability}, and \emph{diversity}. Specificity draws on plausibility, connectivity, and uniqueness, building on scenario research \cite{diakopoulos2021anticipating, dean2006identifying, kieslich2025scenario}; novelty builds on creativity metrics in generative systems \cite{doshi2024generative}; usability adapts two items from the System Usability Scale \cite{brooke1996sus}; actionability adapts two items from the Data Quality Framework \cite{Wang1996BeyondAccuracy}; and diversity is operationalized through the PESTEL categories \cite{glenn2003futures,sanchez2025letstalkfutures}.

Third, we translated each dimension into one or two rating items on a 5-point Likert scale (1 = ``strongly disagree'', 5 = ``strongly agree''), and refined the wording for clarity and consistency. Finally, we piloted the rubric with 20 Prolific participants, who rated a mixed set of 25 agent-generated and human-ideated risks, and provided feedback that helped further refine the wording. Table~\ref{tab:evaluation_rubric} in Appendix~\ref{app:rubric_process} shows the final items for each dimension, along with the rationale for how they were adapted from prior work.

After the rubric was finalized and all risks had been collected, we used OpenAI's \textsc{o3} model to assign each risk to a single PESTEL category (see the full classification prompt in Appendix~\ref{app:rubric_process}). This step involves applying a fixed coding scheme rather than exercising substantive judgment, and could therefore be performed by humans; however, automating it allowed us to apply the same criteria consistently across all risks while keeping judgments in human hands. We validated the model's classifications through a two-step process: first, a co-author independently annotated a random sample of 25 risks from the Chatbot Companion use case, yielding a weighted $F1$-score of $0.86$; second, the full author team reviewed the model's rationales to ensure each label aligned with the meaning of the risk \cite{corbin2014basics}. We then computed the normalized Shannon Diversity Index \cite{shannon1948mathematical} over the PESTEL labels for each AI use case to assess the diversity of identified risks.

\newpage
\subsection{Using the Rubric to Evaluate the Generated Risks}

To assess the quality of risks produced by our pipeline, we compared them with risks identified by humans. We first collected human-ideated risks through a Futures Wheel interface under two conditions (human only, and human-plus-AI condition). We then examined both agent-generated and human-identified risks across three complementary studies.

To generate human comparison data, we built a custom Futures Wheel interface (Figure~\ref{fig:futures_wheel_viz}) that guided participants through the production of first-, second-, and third-order consequences. The interface was implemented in two experimental conditions that reflect common ways humans engage in foresight tasks today. In the human-only condition, participants generated consequences without any assistance. In the human-plus-AI condition, participants could optionally request AI input through a chat window that helped explore potentially missing directions or through a button that generated candidate consequences. We recruited through Prolific \cite{prolific} 24 domain experts, and 24 laypeople for human-only condition and 18 domain experts and 18 laypeople for human-plus-AI condition, resulting in 84 participants in total. Their demographics are detailed in Appendix~\ref{app:participants}, Tables~\ref{tab:participant_demographics_laypeople}--\ref{tab:participant_demographics_hybrid_domainexperts}.
All participants completed the task individually, and their cascading consequences were processed using the same classification and deduplication steps applied to agent-generated consequences. The resulting human-ideated risks formed the input for Studies 2 and 3.

%\mbox{ }\\
\smallskip
\noindent
\textbf{Study 1: Scoring \emph{Agent}-Generated Risks with Domain \emph{Experts} as Evaluators.} To evaluate risks generated by in-silico agents, we conducted a large-scale annotation study with 170 domain experts recruited through Prolific \cite{prolific}, and sampled across five stakeholder cohorts: decision makers ($N = 39$), designers ($N = 38$), developers ($N = 46$), legal experts ($N = 33$), and healthcare professionals ($N = 14$). Healthcare experts were included specifically for the two health-related use cases (\emph{Griefbot} and \emph{Death App}). All participants were screened for relevant expertise, prior experience with AI, and familiarity with risk assessment practices, with comprehension and attention checks ensuring data quality (Appendix~\ref{appendix:using_rubric} reports the recruitment details). Each evaluator was randomly assigned between 12 and 16 risks, ensuring multiple independent assessments of every risk. In total, this yielded 2331 annotations across 110 unique agent-generated risks. Risks were presented via structured annotation cards (Figure~\ref{fig:annotation_card}), which displayed the risk statement, its potential impact, the focal use case, and the definition of systemic risk. Evaluators rated likelihood and severity on 5-point scales, indicated whether the risk should be considered systemic, and scored agreement with ten rubric dimensions. Evaluators were blinded to whether risks were AI-generated or human-identified.

%\mbox{ }\\
\smallskip
\noindent
\textbf{Study 2: Scoring \emph{Human}-Ideated Risks with Domain \emph{Experts} as Evaluators.}
To evaluate risks generated through the Futures Wheel interface, we conducted a large-scale annotation study with an additional 120 domain experts recruited via Prolific \cite{prolific}. We used the same screening protocol and sampled across the same five stakeholder cohorts as in Study 1: decision makers ($N = 28$), designers ($N = 25$), developers ($N = 23$), legal experts ($N = 28$), and healthcare professionals ($N = 16$). Each evaluator was assigned between 14 and 18 human-ideated risks, ensuring multiple independent assessments per item. Annotations were collected using the annotation cards (Figure~\ref{fig:annotation_card}), with evaluators blinded to whether risks originated from human-only or human-plus-AI ideation. This yielded 1030 annotations across 89 unique risks. Because fewer risks were evaluated than in Study 1, fewer evaluators were required to achieve comparable coverage per risk.

\smallskip
\noindent
\textbf{Study 3: Scoring Agent-Generated Risks and Human-Ideated Risks with Domain \emph{Leaders} as Evaluators.} To evaluate how domain leaders interpret and assess both agent-generated and human-ideated risks, we conducted a semi-structured interview and annotation study with seven leaders recruited through purposive sampling (4 female, 3 male; ages 27–60). The first two participants were identified via an internal mailing list at our organization, and each referred to additional external leaders. Their disciplinary backgrounds and relevance to the three speculative use cases are listed in Appendix \ref{appendix:using_rubric}, Table~\ref{tab:domain_leaders}. Each leader completed a 45–90 minute recorded session following a standardized protocol (Appendix~\ref{subsec:scoring_leaders}) that included a warm-up, briefing, vignette immersion, risk prioritization, gap analysis, and debriefing. After discussing and prioritizing risks, they annotated each one and identified missing, unclear, or underdeveloped risks. Following the interview, leaders also completed a follow-up annotation survey using the same rubric (Section~\ref{subsec:rubric}). They were sent a link to the same structured annotation cards used in Studies 1 and 2, and rated both the agent-generated and human-ideated risks. As in the expert studies, leaders were blinded to the origin of each risk.

We then conducted a qualitative analysis of the interview recordings and transcripts. This included leaders' vignette responses, risk discussions, prioritization decisions, and gap analyses. Two authors thematically analyzed these materials following an inductive approach~\cite{saldana2015coding, miles1994qualitative, mcdonald2019reliability, braun2006thematic}. The authors used Figma~\cite{figma} to collaboratively create affinity diagrams based on leaders' responses. Over the course of 4 meetings, totaling 10 hours, they discussed and resolved any disagreements that arose during the coding and theme refinement process. The final themes describe how leaders contextualized, reframed, or extended the risks, revealing differences in emphasis and diversity across agent-generated and human-identified risks.
\section{Results}
\label{sec:results}

Next, we discuss our results in two parts. First, we examine the output of the in-silico risk generation pipeline across four AI use cases, assessing whether the generated risks are sufficiently systemic, plausible, specific, diverse, and actionable to support anticipatory decision-making (RQ1). Second, we compare these risks to those ideated by human participants across two conditions (human-only, and human–plus-AI), analyzing differences in diversity, tone, emotional depth, and systemic framing (RQ2).

\subsection{RQ1: Can in-silico agents generate systemic risks of sufficient quality to support foresight?}

\subsubsection{Quantitative Analysis} We found that agent-based Futures Wheels generate (Figure \ref{fig:rubric_results}, Tables~\ref{tab:pestel_results_llm}--\ref{tab:deathapp_hybrid} in Appendix~\ref{app:quantitative_risk_evaluation}):

\vspace{3pt}
\noindent\textbf{Numerous risks.} Across all 30 wheels, Step 1 (Figure \ref{fig:pipeline}) generated an average of 86 consequences for the AI Toy, and 110 for the Griefbot. After Step 2 (risk classification), these were reduced to 32 and 58 risks, which Step 3 (deduplication) further condensed into 27 and 47 unique systemic risks, respectively. Appendix~\ref{app:supplementary_pipeline_analysis}, Figure~\ref{fig:rq1_boxplots_generations}, shows the number of items at each step for each AI use case, and Figure \ref{fig:rq1_boxplots_resources} the resources needed to generate them. Tables \ref{tab:chatbot_agent_risks}, \ref{tab:aitoy_llm_risks}, \ref{tab:griefbot_llm_risks}, and \ref{tab:deathapp_llm_risks} list the representative, deduplicated systemic risks generated by the agents for four use cases: Chatbot Companion ($n{=}21$), AI Toy ($n{=}27$), Griefbot ($n{=}31$), and Death App ($n{=}24$).
Items are phrased at societal/institutional scale, and often expressed as cascades; the symbol ``$\rightarrow$'' denotes downstream effects.
These lists are illustrative rather than exhaustive, and provide a consolidated basis our qualitative analysis. We see that, across cases, the lists enumerate systemic, cascading risks that operate at societal and institutional scales.
For Chatbot Companion (Table~\ref{tab:chatbot_agent_risks}), items progress from health-system strain (e.g., ``diminished mental health treatment effectiveness'') to macro-social outcomes (e.g., ``widespread social unrest and instability'', ``pervasive disruptions in international digital commerce'').
For the AI Toy (Table~\ref{tab:aitoy_llm_risks}), risks center on child development and equity (e.g., ``critical thinking skills decline systemically'', ``large-scale data exploitation targeting minors'', ``lifelong profiling and discrimination begin in childhood''). The Griefbot list (Table~\ref{tab:griefbot_llm_risks}) is notable for legal-cultural rupture (e.g., ``legal and moral frameworks governing personal identity collapse'', ``grief is exploited as a marketable asset'', ``judiciary systems face overload from AI-avatar litigation'').
Finally, the Death App (Table~\ref{tab:deathapp_llm_risks}) shifts to public-health and governance shocks (e.g., ``mass public health crises from uncontrolled harmful interventions'', ``vulnerable populations face systemic coercion'').
Interestingly, even for mature systems, the agent-generated risks reach institutional instability, while more speculative cases add geopolitical failure modes.

\vspace{3pt}
\noindent\textbf{Risks on specific topics.} PESTEL classification showed Social as the most prominent category across all four AI use cases, ranging from 42\% of the generated risks being (Death App) to 78\% (AI Toy). Legal followed, with 11\% (AI Toy) to 34\% (Griefbot). Political risks were either absent or limited to one for the first three use cases: the Death App had indeed 5 (19\%). Appendix~\ref{app:quantitative_risk_evaluation}, Table~\ref{tab:pestel_results_llm}, shows the full PESTEL breakdown for agent-generated risks.

\vspace{3pt}
\noindent\textbf{Systemic risks for any TRL.} Across all four AI use cases, domain leaders judged about 75\% of the risks to be systemic, while domain experts judged 93\% as systemic. Among the leader ratings, the lowest share was for Chatbot Companion (72\%), and the highest for AI Toy (85\%). On a five-point scale, evaluators rated the risks as at least ``likely'' to occur (mean likelihood score of 3.57 or above), and ``serious'' in impact (mean severity score of 3.51 or above), as shown in Figure~\ref{fig:rubric_results}. Likelihood was highest for AI Toy (3.84), and lowest for Griefbot (3.57). Severity was highest for AI Toy and Death App (both 3.70), and lowest for Chatbot Companion (3.51).

\begin{figure*}[h!]
    \centering
    \includegraphics[width=0.9\linewidth]{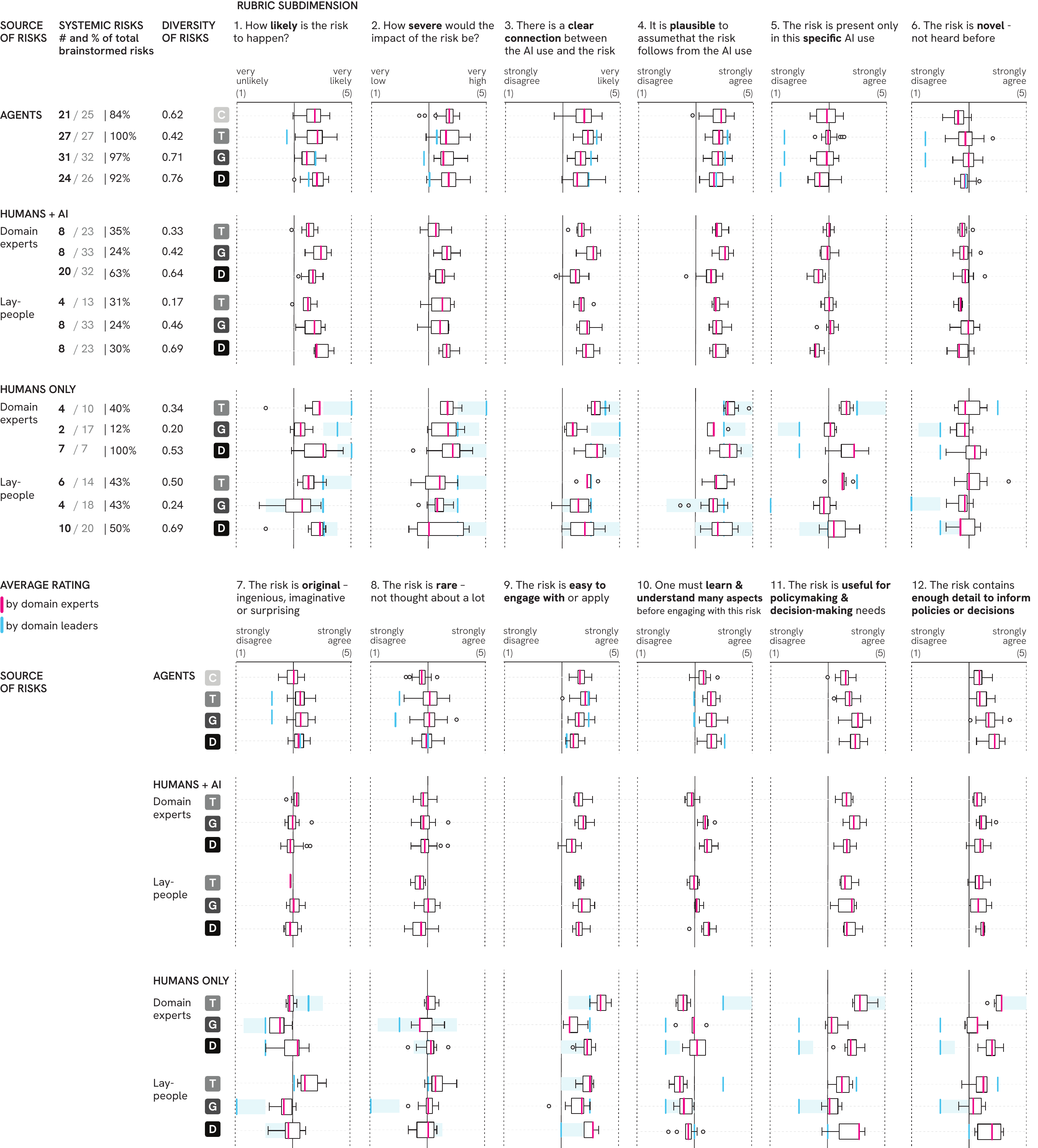}
    \caption{\textbf{Ratings of systemic risks generated by in-silico agents and humans across AI use cases and quality subdimensions.} In-silico agents produced substantially more systemic risks than humans, who generated comparatively few. Boxplots with mean scores are shown for risks associated with four AI use cases: Chatbot Companion (C, TRL~9), AI Toy (T, TRL~7), Griefbot (G, TRL~5), and Death App (D, TRL~2). Each colored bar reflects the average domain expert rating (pink bar) or domain leader rating (blue bar) on a 5-point Likert scale across ten evaluation subdimensions from our rubric developed in Section~\ref{subsec:rubric}. Domain experts generally rated the risks generated by in-silico agents as connected to the AI use, plausible, and moderately usable. Dimensions such as specificity, novelty, and originality received more varied ratings across use cases, especially for less mature concepts such as the Death App and Griefbot. Appendix~\ref{app:quantitative_risk_evaluation} (Tables~\ref{tab:llm_human_comparison}--\ref{tab:deathapp_hybrid}) reports the statistical tests and quantitative comparisons.}
    
    \Description{Boxplot chart showing ratings of systemic risks across the four AI use cases and ten evaluation subdimensions. The chart compares risks generated by in-silico agents versus humans, with separate ratings from domain experts (pink bars) and domain leaders (blue bars). Four AI use cases are shown: Chatbot Companion (C), AI Toy (T), Griefbot (G), and Death App (D). In-silico agents generated substantially more systemic risks than humans across all use cases, with counts ranging from 21--32 systemic risks for in-silico agents, versus 2--7 for domain experts and 4--10 for laypeople. Mean ratings on the 5-point Likert scale vary by subdimension and use case. The ratings for specificity, novelty, and originality showed more variation, particularly for less mature concepts like Death App and Griefbot. The percentage of total risks classified as systemic ranged from 12\% to 100\% across different AI use cases.}
    \label{fig:rubric_results}
\end{figure*}

\vspace{3pt}
\noindent\textbf{Novel risks, even for low TRL.} Griefbot risks scored highest on all three novelty subdimensions: 2.99 (novelty), 3.24 (originality), and 3.08 (rarity). By contrast, Chatbot Companion risks were rated least novel across the board (2.64, 2.94, and 2.79). Risks generated for the AI Toy scored highest on use case connectivity (3.92), plausibility (3.81), and uniqueness (3.02) (Figure~\ref{fig:rubric_results}). In contrast, Death App risks were rated least plausible (3.73), and least specific (2.78). AI Toy risks were considered easiest to engage with (3.81), while Chatbot Companion risks required the least additional information to be fully understood (3.32). Death App risks were seen as hardest to engage with (3.48), and required the most explanation (need to learn: 3.59).

\vspace{3pt}
\noindent\noindent\textbf{Hard-to-use risks at times, but policy-relevant.} We found that generated risks are harder to readily interpret as the AI use cases become more speculative, but, paradoxically, they become more relevant for policy discussions, especially in lower-TRL scenarios like the Death App.

\subsubsection{Qualitative Analysis} While the quantitative results show that the agents can generate systemic risks at scale, the qualitative analyses presented next complement this by revealing how domain leaders judged the systemic nature of the risks, how they assessed their comprehensiveness, and how they complemented the agents with risks grounded in lived experience. Expert quotes are referenced using $E_N$, corresponding to their anonymized ID.

\vspace{3pt}
\noindent\textbf{Domain leaders judged most agent-generated risks to be truly systemic, even for speculative use cases.} For the AI Toy (TRL 7), agreement was high: 70–78\% of risks were rated systemic, and 48–79\% were placed in the \emph{critical risk zone}, the upper-right quadrant where likelihood and impact are both high (as evidenced in expert E2's risk prioritization grid in Appendix \ref{appendix:using_rubric}, Figure \ref{fig:matrix_toy}). For the Griefbot (TRL 5), views diverged more (53\% \emph{vs.} 72\% \emph{vs.} 78\%), with 43–71\% still assigned to the critical risk zone though (as shown in expert E4's risk prioritization grid in Appendix \ref{appendix:using_rubric}, Figure \ref{fig:matrix_griefbot}). The most surprising case was the Death App (TRL 2): despite low technological readiness, domain leaders marked 58\% and 92\% of risks as systemic, with 33–73\% in the critical risk zone (as shown in expert E6's risk prioritization grid in Appendix \ref{appendix:using_rubric}, Figure \ref{fig:matrix_deathapp}). These findings indicate that the agents surfaced systemic risks across readiness levels (including low-TRL domains where uncertainty is greatest), and that many of these risks were judged both likely and consequential.

The leaders set aside some of the generated risks because they did not perceive them as systemic, and did so for five main reasons. First, some were too generic, applying to ``any AI rather than the specific use case''. As E3 put it, ``Data breaches happen all the time. Does it actually connect specifically to this case? No, it's anything with AI''. Second, some were judged as not plausibly caused by the technology itself. In the griefbot case, E6 asked whether ``a collapse of social cohesion'' could really be traced back to griefbots, concluding that ``the systemic risk has to be caused by the creation of griefbots as a widespread phenomenon''. E2 made a similar point about suicide rates, noting that while tragic, it would be ``a stretch to say that like this type of companies are causing these situations''. Third, some risks were considered too vague or poorly phrased such as ``changes in family values'', which E7 felt could mean almost anything. Fourth, some were seen as overly individual rather than systemic. E1, for example, dismissed social skill decline among children as ``not a systemic risk: life will find a way and there will be a contra movement'', while E7 argued that ``family-level conflicts over assisted dying did not automatically rise to systemic scale without evidence of wider disruption''. Fifth, some were excluded because they appeared more positive than negative. E6, for instance, set aside the expansion of international health services on the grounds that they ``don't see how that can be a negative''.

\vspace{3pt}
\noindent\textbf{Domain leaders judged the systemic risks generated by the agents as comprehensive.} They described the list as broad and often exceeding what they would have anticipated themselves (E5: ``There were a few risks that I never thought about in this way''). In terms of coverage, leader confirmed that the agents reliably surfaced the generalized systemic dynamics they expected to see across domains such as bias amplification in sensitive settings like education and healthcare, emotional harms and psychosocial spillovers in grief technologies and child-facing systems, and dependence or erosion of skills when AI substitutes for human judgment or learning. In-silico agents also articulated second-order effects (e.g., erosion of democratic processes, loss of human oversight) that resonated with leader's own systemic framing. Several noted that reviewing a pre-generated list reduced the ``blank-page problem'', allowing them to focus their energy on filtering, prioritizing, and contextualizing risks. These results suggest that the agents not only provided stimulus for reflection but also captured the backbone of systemic risks leaders already recognized.

\vspace{3pt}
\noindent\textbf{Domain leaders complemented the agents with risks grounded in lived experience.} Across the three use cases, leaders introduced 19 additional risks in total (6 for the AI Toy, 6 for the Griefbot, and 7 for the Death App). For the AI Toy, leaders emphasized developmental distinctions across educational systems, with E1 noting, ``Children from 3 to 12 are completely different people and represent completely different audiences from a design point of view''. They also cautioned against potential cultural backlash: ``Maybe there's a cultural risk. Will some cultures push back against girls becoming enthusiastic about STEM through the toy?''. E2 warned of systemic child protection gaps if disclosures of abuse were made only to the toy: ``What if the child tells the toy something serious, like someone is harming them? Will it ever reach the school or the police?''. Finally, both leaders added that mass attachment to commercial toys could create a new form of consumer-driven grief when products are discontinued, and that widespread adoption could contribute to global environmental burdens through e-waste.

In the Griefbot case, leaders pointed to institutional and cultural shifts overlooked by the agents. E5 stressed that griefbots might take over the logistics of dying: ``How do I arrange the funeral in this country? What are the legal steps?''. These tasks have historically been managed by families, religious communities, or professional services. This displacement could reconfigure how societies organize death and mourning. Others emphasized that digital companions may undermine human help-seeking: ``You need to rely on your network: people can’t heal for you, but they can help you'' (E5). Leaders also raised the risk of harmful new ``cultures of grief'', where AI-mediated rituals normalize shallow or exploitative practices, and E6 highlighted design-level unpredictability, describing griefbots as ``the curse of flexibility, it can be used in so many different ways that you can't really foresee the outcomes''.

In the Death App scenario, leaders added political-economic and cultural risks that the model had not surfaced. E7 highlighted the proliferation of unregulated provision: ``If it's not opening a black market, like a completely illegal one, it can open a grey market with pharmaceuticals''. They also warned of state misuse where local governments could weaponize platforms to pressure vulnerable groups, and of a possible normalization of eugenics, if such technologies were deployed unethically. Gendered harms were also raised, with leaders questioning how such platforms might reproduce or exacerbate existing inequalities. For instance, women could be ``disproportionately steered toward assisted dying in contexts where access to healthcare and social support is limited''. Finally, leaders suggested that assisted dying could become institutionalized as a distinct ``death industry'' separate from healthcare, creating new boundaries and weakening existing oversight mechanisms.

\subsection{RQ2: How do agent-generated risks compare with human-ideated ones?}

\subsubsection{Quantitative Analysis}

In comparison to those generated by agents, human-ideated risks were (Figure \ref{fig:rubric_results}, Tables~\ref{tab:pestel_results_llm}--\ref{tab:deathapp_hybrid} in Appendix~\ref{app:quantitative_risk_evaluation}):

\vspace{3pt}
\noindent\textbf{Not many.} Without AI assistance, humans understandably generated fewer unique risks than agents (Appendix \ref{app:results}, Tables~\ref{tab:chatbot_human_risks}, \ref{tab:aitoy_human_risks}, \ref{tab:griefbot_human_risks}, \ref{tab:deathapp_human_risks}, \ref{tab:aitoy_hybrid_risks}, \ref{tab:griefbot_hybrid_risks}, and~\ref{tab:deathapp_hybrid_risks}). Domain experts generated 7-17 risks per use case, and laypeople generated 14-20 across all cases. More interestingly, with AI assistance, participants matched or exceeded the pipeline in volume: domain experts generated 23–32 risks per case, and laypeople generated 13–23.

\vspace{3pt}
\noindent\textbf{Narrowly focused.} Human risks cluster narrowly around what feels directly impactful: overwhelmingly social in the case of Chatbot Companion and Griefbot (86\% and 89\%, respectively), and heavily legal for Death App (45\%), where moral and personal stakes are highest. Political and economic concerns are absent from human responses entirely. This suggests that humans filter systemic threats through personal salience \cite{Schweizer2021}, interpreting policy-level risk only when it intersects with perceived bodily or existential vulnerability. Appendix \ref{app:quantitative_risk_evaluation},  Tables~\ref{tab:pestel_results_llm}--\ref{tab:pestel_results_hybrid_experts} show the full table of PESTEL classifications for the risks generated by agents and ideated by humans, both with and without using AI .

\vspace{3pt}
\noindent\textbf{Partly systemic.} Only a fraction of risks were judged to be systemic. With AI assistance, this ranged from 43\% (laypeople and domain experts, griefbot) to 63\% (domain experts, death app). In the human-only condition, systemic coverage was lower, from just 12\% (domain experts, griefbot) to 43\% (laypeople, toy and griefbot), reaching 100\% only in one case (domain experts, death app).

\vspace{3pt}
\noindent\textbf{Neither novel nor original.} Human risks scored consistently low on novelty and originality, with average ratings in the lower range (1-2.3 out of 5) across most use cases, but were judged more likely to occur and more severe in their potential impact.

\subsubsection{Qualitative Analysis} Our analysis shows that agents, leaders, and laypeople frame systemic risks in different ways, diverging in emotional depth, interpretive lens, and uncertainty avoidance.  

\vspace{3pt}
\noindent\textbf{Emotional depth of systemic risks diverges: agents bureaucratize, leaders contextualize, laypeople dramatize.}
Agent risks adopt managerial language (e.g., ``regulatory fragmentation'', ``compliance costs'', ``market volatility'') that abstracts away from lived suffering. Expert contributions weave emotion into culture (``Dependency on fragile products could traumatize children'', ``Talking about death declines''). Laypeople contributions are blunt and affectively charged, capturing immediate fear or outrage (``Individuals may lose the ability to interact with reality effectively'', ``People could lose their jobs''). These differences in register shape how each human imagines the seriousness and felt impact of systemic risks.

\vspace{3pt}
\noindent\textbf{Systemic risks are made sense of through different lenses: agents scale up, leaders translate, laypeople scale down.}
Agents construct risk as structural malfunction (``Judiciary systems worldwide face overload'', ``Global AI technology adoption is severely hindered''). Leaders act as translators, showing how these failures ripple into norms and institutions (``Talking about death declines'', ``Generational miscommunication grows''). Laypeople invert the scale, interpreting systemic risks as personal threats (``The app could be misused against unwilling individuals'', ``Individuals are put on lists or lose their jobs'). For each group, the meaning of risk begins and ends at a different level of experience.

\vspace{3pt}
\noindent\textbf{Uncertainty is navigated through different strategies: agents contain it, leaders interrogate it, laypeople absorb it.}
Agents frame uncertainty as a variable to be managed, terms like ``slower deployment'', ``reduced market confidence'', or ``legal uncertainty'' suggest problems solvable through better design or governance. Leaders highlight uncertainty as a structural feature of emerging technologies (``Curse of flexibility spreads'', ``Systemic framings miss lived realities''), not something to eliminate, but something that complicates meaning and control. Laypeople do not manage or analyze uncertainty, but they live inside it: ``People could die'', ``I wouldn't know what's real''. For each group, uncertainty reflects not just what is unknown, but how the unknown is cognitively, emotionally, and socially processed.

\section{Discussion}
\label{sec:discussion}

We begin by consolidating our findings on agent-based foresight and its alignment with prior work on systemic risks (\S\ref{subsec:inline_results}). We then examine how our results extend beyond prior work (\S\ref{subsec:challenging_results}). Finally, we outline implications for hybrid workflows that integrate agents, domain leaders, and laypeople perspectives in systemic risk assessment (\S\ref{subsec:implications}).

\subsection{In-line with Previous Literature}
\label{subsec:inline_results}
We found that agent-based foresight consistently broadened the range of systemic risks identified. In our study, agents generated a higher volume of cascading consequences across all four use cases, confirming prior claims that AI support is most effective in expanding the search space of possible outcomes rather than deepening contextual analysis \cite{wang2024farsight, herdel2024exploregen, constantinides2024good}. Expert evaluation revealed that, while many of these agent-generated risks required refinement, a majority were judged systemic, plausible, and severe; findings that resonate with \citet{algorithmicImprint2022} argument that systemic harms, once surfaced, persist in decision-making even when the underlying technology is speculative. Moreover, our results resonate with \citet{algorithmicPluralism2024} proposal of algorithmic pluralism, which cautions against systemic bottlenecks and monocultures: our hybrid foresight workflow similarly emphasizes diversity of perspectives (through multiple agents) as a safeguard against narrowing risk imaginaries.

\subsection{Contributions to Previous Literature}
\label{subsec:challenging_results}
Our results also challenge and extend prior accounts of AI risk assessment. Whereas \citet{uuk2024taxonomy} emphasize the need of taxonomies of systemic risks, our findings demonstrate that generative exploration with agents can surface novel risks that have never been categorized before, even for technologies at very low TRLs. For instance, while chatbot companions (TRL 9) primarily yielded risks already well-documented in the literature such as overreliance on AI for social support \cite{chatbotDependenceHarms2024}, our foresight pipeline uncovered previously underexplored risks for speculative systems like griefbots (TRL 5), and death apps (TRL 2), including cascading impacts on healthcare ethics, and the normalization of assisted dying platforms. This contrast suggests that foresight does not have to wait for systems to reach maturity, countering the assumption in prior literature that systemic risks become tractable only once technologies are widely deployed. Finally, expert interviews highlighted that plausibility judgments, absent from  much prior taxonomic work, are critical in filtering speculative risks. By embedding plausibility as an explicit evaluation criterion, our approach adds a new step to systemic risk workflows, producing evidence that a hybrid division of labor (agents for breadth, experts for judgment) can extend existing risk ideation methods.

%%% Chatbot companion %%%
\begin{table*}[t]
    \setlength{\tabcolsep}{3pt}
    \renewcommand{\arraystretch}{1.04}
    \small
    \caption[\textbf{Chatbot Companion in-silico risks.}]%
    {\textbf{List of representative risks ($n=21$) generated by in-silico agents for the Chatbot Companion use case using our pipeline.} 
    \Description{A table listing 21 risks generated by in-silico agents following our pipeline for the Chatbot Companion use case.}}
    \label{tab:chatbot_agent_risks}
    \begin{tabular}{p{0.2cm} p{16.5cm}}
    \toprule
    \textbf{ID} & \textbf{Risk text} \\
    \midrule
    1  & Overreliance on AI support reduces human-centered care quality $\rightarrow$ diminished mental health treatment effectiveness in society \\
    2  & Social and interpersonal communication skills weaken across populations $\rightarrow$ reduced collaboration and increased social fragmentation \\
    3  & Availability of professional mental health expertise drops sharply $\rightarrow$ overwhelmed healthcare systems during crisis events \\
    4  & Loss of clinical experience hampers holistic mental health care $\rightarrow$ fragmented treatment approaches internationally \\
    5  & Confidential mental health information is frequently exposed $\rightarrow$ widespread loss of trust in digital mental health solutions \\
    %6  & Market consolidation and reduced innovation in mental health AI technologies \\
    6  & Market entry barriers increase substantially $\rightarrow$ slower deployment of innovative mental health AI solutions worldwide \\
    7  & Traditional social support networks weaken globally $\rightarrow$ higher vulnerability to health crises and greater reliance on technology-driven services \\
    8  & Population-wide physical and mental health deteriorates significantly $\rightarrow$ increased healthcare burdens and economic costs \\
    9  & Social fragmentation undermines collective action $\rightarrow$ weakened societal stability \\
    10 & Widespread mental health decline burdens healthcare systems $\rightarrow$ increased public health crises \\
    11 & Populations become less capable of managing stress independently $\rightarrow$ greater societal vulnerability during crises \\
    12 & Intergenerational divides intensify $\rightarrow$ reduced social cohesion across communities \\
    13 & Mental healthcare quality deteriorates widely $\rightarrow$ extended suffering and inefficiencies in health services \\
    14 & Trust in mental health systems collapses nationally $\rightarrow$ decreased care-seeking behavior \\
    15 & Consumer protections fail at scale $\rightarrow$ massive economic losses and reduced market confidence \\
    16 & Adoption of digital health innovations stalls globally $\rightarrow$ slowed progress in mental healthcare delivery \\
    17 & Cybersecurity threats multiply rapidly $\rightarrow$ pervasive disruptions in international digital commerce \\
    18 & Societal inequalities deepen significantly $\rightarrow$ widespread social unrest and instability \\
    19 & Future generations exhibit reduced social competencies $\rightarrow$ long-term societal and economic challenges worldwide \\
    20 & Reliance on AI emotional support systems becomes entrenched globally $\rightarrow$ heightened risks for social isolation and mental health vulnerability \\
    %22 & Increased compliance costs and market realignments \\
    21 & Mental health emergencies rise globally $\rightarrow$ increased burdens on healthcare systems and public safety sectors \\
    %24 & Varying international compliance challenges affecting market entry and innovation \\
    %25 & Slower market growth and calls for increased transparency worldwide \\
    \bottomrule
    \end{tabular}
\end{table*}

%%% AI Toy %%%
\begin{table*}[t]
    \small
    \setlength{\tabcolsep}{3pt}
    \renewcommand{\arraystretch}{1.04}
    \caption[\textbf{AI Toy in-silico risks.}]%
    {\textbf{List of representative risks ($n=27$) generated by in-silico agents for the AI Toy use case using our pipeline.} 
    \Description{A table listing 27 risks generated by in-silico agents following our pipeline for the AI Toy use case.}}
    \label{tab:aitoy_llm_risks}
    \begin{tabular}{p{0.2cm} p{16.5cm}}
    \toprule
    \textbf{ID} & \textbf{Risk text} \\
    \midrule
    1  & Unequal STEM skill distribution is reinforced across societies $\rightarrow$ reduced global workforce competitiveness and innovation equity \\
    2  & Marginalized populations face entrenched educational exclusion $\rightarrow$ increased socioeconomic divides and broad societal tensions \\
    3  & Unverified information is widely accepted unquestioningly $\rightarrow$ erosion of democratic discourse \\
    4  & Critical thinking skills decline systemically among youth $\rightarrow$ diminished innovation and weaker democratic participation \\
    5  & Social skill deficits become widespread among children $\rightarrow$ increased social isolation and long-term challenges for social cohesion \\
    6  & Emotional intelligence development is impaired broadly $\rightarrow$ rising societal conflict and decreased community trust \\
    7  & Population-wide relational dysfunctions become pervasive $\rightarrow$ increased burden on mental health and public health systems \\
    8  & Parental and family-based educational support systems weaken significantly $\rightarrow$ greater reliance on external institutions \\
    9  & Gaps in early educational interventions increase widely $\rightarrow$ higher youth dropout rates and workforce shortages \\
    10 & Child developmental monitoring becomes ineffective at scale $\rightarrow$ systemic inefficiencies in education policy and resource allocation \\
    11 & Shallow knowledge acquisition becomes widespread among learners $\rightarrow$ diminished innovation capacity in knowledge-intensive industries \\
    12 & Critical analytical skills are systematically underdeveloped $\rightarrow$ reduced national problem-solving capabilities \\
    13 & Misalignment of educational content with learner needs disrupts schooling effectiveness at scale $\rightarrow$ increased inequities \\
    14 & Education systems become fragmented globally $\rightarrow$ instability in qualification standards and cross-border employment challenges \\
    15 & False knowledge and misinformation become normalized among youth $\rightarrow$ poorer public health decisions and destabilized social trust \\
    16 & Bias propagation in AI content spreads widely $\rightarrow$ discriminatory practices embedded in workforce and institutions \\
    17 & Dependence on AI for cognitive tasks rises system-wide $\rightarrow$ reduced innovation and adaptability in global markets \\
    18 & Skill development deficiencies in critical reasoning become pervasive $\rightarrow$ lower productivity in knowledge economies \\
    19 & Social isolation among children is widespread $\rightarrow$ decreased societal cohesion and increased mental health issues \\
    20 & Developmental delays and emotional difficulties are prevalent $\rightarrow$ long-term societal costs in healthcare and education \\
    21 & Trust in educational institutions collapses widely $\rightarrow$ fragmented education markets and uneven skill development globally \\
    22 & Educational monopolies distort knowledge dissemination $\rightarrow$ international market imbalances and cultural homogenization \\
    23 & Widespread cultural myopia is entrenched early in life $\rightarrow$ international conflicts and weakened global cooperation \\
    24 & Psychological crises linked to AI withdrawal are widespread $\rightarrow$ increased demand on health services and societal instability \\
    25 & Large-scale data exploitation targeting minors is normalized $\rightarrow$ eroded privacy rights and reduced trust in technology \\
    26 & Lifelong profiling and discrimination of individuals begin in childhood $\rightarrow$ systemic inequality and restricted mobility \\
    27 & Traditional educational employment sectors contract $\rightarrow$ market consolidation and global shifts in labor markets \\
    \bottomrule
    \end{tabular}
\end{table*}

%%% Griefbot %%%
\begin{table*}[t]
    \setlength{\tabcolsep}{3pt}
    \renewcommand{\arraystretch}{0.92}
    \small
    \caption[\textbf{Griefbot in-silico risks.}]%
    {\textbf{List of representative risks ($n=31$) generated by in-silico agents for the Griefbot use case using our pipeline.} 
    \Description{A table listing 31 risks generated by in-silico agents following our pipeline for the Griefbot use case.}}
    \label{tab:griefbot_llm_risks}
    \begin{tabular}{p{0.2cm} p{16.5cm}}
    \toprule
    \textbf{ID} & \textbf{Risk text} \\
    \midrule
    1  & Mental health systems face overwhelming demand and widespread strain $\rightarrow$ reduced quality of care along with decreased public health resilience \\
    2  & Productivity in workplaces declines due to higher rates of psychological distress $\rightarrow$ significant economic losses in markets \\
    3  & Large-scale social fragmentation and weakened cultural cohesion occur $\rightarrow$ increased mental health crises and reduced societal stability \\
    4  & Family support networks diminish substantially $\rightarrow$ rise in social isolation affecting population mental health at large \\
    5  & Mental health systems face increased demand for new therapies and social services $\rightarrow$ greater pressure on healthcare infrastructure \\
    6  & Mental health care costs escalate internationally $\rightarrow$ resource allocation challenges in health policy and budgeting \\
    7  & Consumer trust and mass trust in digital services deteriorate widely $\rightarrow$ decreased engagement, economic contraction in AI sectors \\
    8  & Regulatory compliance costs surge sharply for international companies $\rightarrow$ constrained innovation and reduced AI competitiveness \\
    9  & Fragmented data jurisdiction laws and legal uncertainty disrupt international digital economy operations $\rightarrow$ friction in global trade and investment \\
    10 & Protracted legal conflicts impose heavy costs on tech companies $\rightarrow$ market volatility and investor wariness \\
    11 & Consumer protection crises emerge internationally $\rightarrow$ heightened regulatory intervention and market distrust \\
    12 & Specialized mental health services expand internationally $\rightarrow$ increased public health expenditures and resource reallocation \\
    13 & Mental health treatment efficacy declines broadly $\rightarrow$ longer recovery times and higher societal healthcare costs \\
    14 & International legal frameworks are destabilized by competing claims $\rightarrow$ prolonged litigation and uncertainty in data governance \\
    15 & Global AI technology adoption is severely hindered $\rightarrow$ reduced innovation and fractured international cooperation \\
    16 & Social isolation and entrenched loneliness increase among large segments of society $\rightarrow$ widespread public health challenges and increased mortality \\
    17 & Cybercrime and fraud involving AI avatars become endemic $\rightarrow$ escalated public safety risks and weakened trust in digital ecosystems \\
    18 & Legal and moral frameworks governing personal identity collapse $\rightarrow$ deep societal conflict and regulatory chaos \\
    19 & Grief is exploited as a marketable asset on a global scale $\rightarrow$ ethical erosion and significant emotional harm across societies \\
    20 & Mental health systems are overburdened by AI-induced dependency cases $\rightarrow$ widespread challenges in public health management \\
    21 & Social relationships increasingly rely on artificial interactions $\rightarrow$ degradation of interpersonal social skills at scale \\
    22 & Traditional mourning rituals lose legal and cultural recognition $\rightarrow$ fragmentation of societal cohesion across cultures \\
    23 & Cultural homogenization occurs due to techno-centric mourning practices $\rightarrow$ loss of cultural diversity in international societies \\
    24 & Legal services related to digital legacy surge worldwide $\rightarrow$ increased legal market specialization and cross-border disputes \\
    25 & Large-scale data breaches compromise personal information globally $\rightarrow$ significant threats to public security and trust in markets \\
    26 & Judiciary systems worldwide face overload from AI avatar litigation $\rightarrow$ delays in justice and increased public distrust in legal institutions \\
    %27 & Regulatory agencies enact stricter and expanded controls on AI avatar commercialization globally $\rightarrow$ increased compliance costs for technology companies and slowed innovation \\
    27 & Emotional dependence on AI systems becomes widespread $\rightarrow$ significant challenges in regulating psychological impacts \\
    28 & Conventional community-based mental health services face disruption globally $\rightarrow$ shifts in how social support networks operate across societies \\
    29 & Familial conflicts related to AI cause social fragmentation at scale $\rightarrow$ increased social tension and challenges to social cohesion \\
    30 & Data protection laws face widespread challenges $\rightarrow$ global legal conflicts and regulatory fragmentation \\
    31 & AI systems exhibit inconsistent behaviors across user bases $\rightarrow$ risks in trust and safety in international digital ecosystems \\
    \bottomrule
    \end{tabular}
\end{table*}

%%% Death App %%%
\begin{table*}[t]
    \setlength{\tabcolsep}{3pt}
    \renewcommand{\arraystretch}{0.92}
    \small
    \caption[\textbf{Death App in-silico risks.}]%
    {\textbf{List of representative risks ($n=24$) generated by in-silico agents for the Death App use case using our pipeline.} 
    \Description{A table listing 24 risks generated by in-silico agents following our pipeline for the Death App use case.}}
    \label{tab:deathapp_llm_risks}
    \begin{tabular}{p{0.2cm} p{16.5cm}}
    \toprule
    \textbf{ID} & \textbf{Risk text} \\
    \midrule
    1  & Public health systems face increased demand and strain $\rightarrow$ overwhelmed healthcare infrastructure \\
    2  & Healthcare systems experience resource reallocation pressures $\rightarrow$ reduced quality of care overall \\
    3  & Fragmented and inconsistent international regulations disrupt cooperation $\rightarrow$ increased illegal activities and market uncertainty \\
    4  & Social instability hampers economic activity $\rightarrow$ decreased investor confidence in affected markets \\
    5  & Market consolidation occurs $\rightarrow$ reduced competition and innovation in assisted death services \\
    6  & Therapeutic relationships weaken broadly $\rightarrow$ worsened mental health outcomes in affected populations \\
    %7  & User engagement across digital health platforms declines sharply $\rightarrow$ setbacks in digital health innovation \\
    7  & Ideological polarization deepens across societies $\rightarrow$ fragmented social cohesion and political instability \\
    8  & Widespread cybersecurity failures in sensitive health sectors occur $\rightarrow$ compromised public safety and urgency for regulation \\
    9  & Criminal exploitation undermines trust in AI health services $\rightarrow$ demands for stringent oversight and legal enforcement \\
    10 & Mass public health crises arise from uncontrolled harmful interventions $\rightarrow$ widespread loss of life and international outrage \\
    11 & Global health service market faces regulatory crackdowns and legal turmoil $\rightarrow$ destabilized market confidence and fragmented standards \\
    12 & Trust in AI-enabled health platforms collapses globally $\rightarrow$ widescale rejection of AI interventions in critical services \\
    13 & Societal mental health deteriorates broadly at scale $\rightarrow$ increasing suicide rates and overwhelming public health systems \\
    14 & Vulnerable populations face systemic coercion pressures $\rightarrow$ erosion of fundamental human rights protections worldwide \\
    15 & Widespread privacy violations cause international data security crises $\rightarrow$ reduced trust and investment in digital health technologies \\
    16 & Stagnation of digital health innovation occurs internationally $\rightarrow$ slowed progress in healthcare improvements \\
    17 & Coordinated exploitation campaigns cause widespread social harm $\rightarrow$ increased global insecurity and destabilization \\
    18 & Rising extremist influence fractures international relations $\rightarrow$ increased geopolitical conflicts and market volatility \\
    19 & Global mental health crises worsen profoundly $\rightarrow$ substantial economic losses and reduced workforce productivity \\
    20 & Ethical decay spreads widely through societies $\rightarrow$ deteriorated social cohesion and increased risk of unrest \\
    21 & National healthcare priorities are reorganized $\rightarrow$ international tensions over care standards \\
    22 & Public opinion drives legislative changes internationally $\rightarrow$ polarized markets and policy instability \\
    23 & International legal frameworks around AI and assisted dying harmonize unevenly $\rightarrow$ regulatory fragmentation in cross-border services \\
    %25 & AI companies face significant legal accountability in multiple jurisdictions $\rightarrow$ changes in liability insurance and compliance costs \\
    24 & Accelerated legislative reforms create uneven legal landscapes $\rightarrow$ challenges in international healthcare service provision \\
    \bottomrule
    \end{tabular}
\end{table*}

\subsection{Implications for Hybrid Workflows in Systemic Risk Assessment}
\label{subsec:implications}

Our findings have three main implications: (1) for how agent and human contributions complement each other in systemic risk generation, (2) for how hybrid agent–human risk generation should be structured to produce valid foresight, and (3) for how practitioners should decide when and how much to automate this process.

\smallskip
\noindent
\textbf{Integrating complementary agent and human contributions.}
Our findings show that combining agent-generated and human-ideated risks can help overcome inherent cognitive barriers such as scope neglect and fixation on familiar narratives \cite{rahwan2025science}. Agents consistently framed risks in terms of institutional change, policy implications, and macro-level societal trends. By contrast, humans surfaced risks reflecting emotional depth, cultural nuance, and context-specific harms that agents frequently missed. In addition, humans provided concrete examples that grounded agent-identified risks in specific domains and lived contexts. For example, agents flagged ``regulatory fragmentation'' as a risk in both child protection and mourning contexts. Humans pointed to concrete cases such as AI toys being regulated as consumer products in one country but subject to child safeguarding obligations in another, or griefbots being deployed as wellbeing tools despite unclear rules about consent of the deceased \cite{Hollanek2024}.

\smallskip
\noindent
\textbf{Structuring hybrid risk generation for valid foresight.}
Drawing on leaders' reflections, we identify three directions for improving the validity of hybrid foresight pipelines.

\smallskip
First, pipelines should start from lived, micro-level experiences (e.g., how a child might engage with an AI toy, how a grieving family might rely on a griefbot), and then cascade upward to uncover how these individual or community-level harms accumulate into systemic risks. Beginning only at the system level may overlook how small, situated harms add up over time.

Second, pipelines should be designed with intersectionality theory in mind. In-silico agents should be prompted to surface risks that reflect overlapping axes of vulnerability such as gender, class, and ethnicity \cite{IntersectionalityAnalysis2026}. Together, these steps would ensure that foresight captures not just abstract systemic shifts, but also the layered ways in which lived experiences scale into broader disruptions.

Third, pipelines should be used alongside longitudinal, scenario-based, or discourse-analytic methods to interpret how short-term risks may develop into systemic disruptions over long time horizons of 5–25 years. These methods support reasoning across time by connecting current uses of AI to earlier technology transitions. For example, current debates around GenAI adoption echo discussions from the 1990s surrounding the widespread adoption of design software such as Photoshop \cite{Photography2023}. Short-term concerns about image manipulation later developed into the systemic risk of loss of trust in journalism, leading to the creation of forensic verification teams into news production, especially in war zones \cite{forensicArchitecture}.

\smallskip
\noindent
\textbf{Deciding when and how to automate risk generation.} Our findings suggest that the choice between fully automated and hybrid risk generation is best understood as a set of trade-offs rather than a binary. To navigate these, we propose three practical heuristics.

First, \emph{automate for breadth}. Fully automated pipelines are especially useful during early-stage ideation, horizon scanning, or exploratory analysis of emerging technologies, where the primary goal is to expand the search space of possible risks. In these contexts, gains in efficiency and coverage may outweigh the absence of direct human involvement.

Second, \emph{retain human involvement for sensemaking}. While automation increases breadth, it often reduces contextual grounding. Domain leaders emphasized that in many real-world settings, risk generation is not merely an information task. It is also a sensemaking process through which stakeholders surface tacit assumptions, align perspectives, and build ownership of downstream decisions. This process is often interdisciplinary and highly domain-specific, shaped by situated expertise from fields such as medical ethics or the anthropology of technology. In such settings, the absence of human engagement can carry significant costs. As our own work showed, certain risks become visible only through expert judgment.

Third, \emph{use hybrid workflows when both breadth and realism matter}. Hybrid approaches are most effective when risks must be surfaced comprehensively, yet still interpreted and filtered through human judgment. Strategic planning and regulatory readiness are two such examples. In these cases, plausibility filtering and cultural contextualization must remain human-led. Our adaptation of the Futures Wheel supports this division of labor by organizing agent-generated risks into layered causal pathways for iterative refinement. Domain leaders found this structure helpful in reducing the blank-page burden, while preserving space for expert interpretation.
\smallskip

\subsection{Limitations and Future Work}
\label{subsec:limitations}

Our study and the proposed approach have six main limitations that suggest directions for future research. First, we adopted the EU AI Act's definition of systemic risk, which targets general-purpose AI models \cite{EUAIAct2024, codeofpractice2025}. This gave evaluators a clear, policy-relevant anchor for a vague concept, and helped ensure consistency. However, priming evaluators with this framing may have narrowed their perspective. Future work should test alternative definitions of systemic risk, including approaches based on human rights that assess not only the scale and probability of harms, but also their scope, and reversibility \cite{huderia2024}. These dimensions could offer a more nuanced view of how systemic risks affect different groups.

Second, our analysis focused on use-level risks and did not account for systemic risks that stem from underlying model capabilities. As recent work and regulation highlight, highly capable general-purpose models may introduce systemic risks independent of how they are applied, including governance failures and cross-sector disruptions~\cite{uuk2024taxonomy,codeofpractice2025}. The same application may pose very different risks when powered by a frontier model versus a weaker one. To address this, future work could extend our pipeline by simulating capability evolution by conditioning agent prompts on projected model abilities such as reasoning or autonomy.

Third, the foresight process began with in-silico agents generating risks, which may have limited both the emotional depth, and originality of the outputs. While domain experts and leaders found many of the risks plausible, they often lacked the kind of intuitive or affective nuance that emerges from lived experience. Future work should explore hybrid workflows where human experts seed early ideas, and agents elaborate them; or it should explore explicit prompts asking agents to consider emotional impact or speculative ``wild cards'' \cite{glenn2003futures}. These strategies may yield outputs that are both creative and grounded in context. Our findings also highlight the importance of iteration, both in foresight pipelines and in discussion practices. Generating systemic risks through in-silico agents is not a one-off task: repeated runs across different prompts, personas, and stakeholder contexts may reveal additional layers of systemic risk, or refine earlier outputs. Likewise, hybrid workflows will likely benefit from iterative cycles in which agents generate breadth, experts contextualize and critique, and subsequent agent runs refine scope based on this feedback.

Fourth, the in-silico agents showed a topical bias in the risks they produced. Across use cases, agents tended to concentrate on social, ethical, or legal consequences, while generating few technical risks and none related to environmental impact. This likely reflects both the structure of the prompts and potential biases in the model's training data. While socially framed risks are important, this imbalance may lead to blind spots. This imbalance may itself introduce new risks. By embedding AI into foresight, we risk a form of ``meta-systemic risk'' where the very tools meant to help anticipate harms could normalize or amplify certain framings of risk at the expense of others. For example, our pipeline produced many risks framed in managerial or regulatory language, which might inadvertently steer attention toward technocratic governance while sidelining lived experiences. Future work should critically examine whether agent-supported foresight unintentionally narrows the discursive space, even as it expands the breadth of candidate risks. Future work should explore three directions: (1) prompt strategies that explicitly request diverse categories of risk (e.g., using PESTEL dimensions \cite{glenn2003futures,SammutBonnici2015} or systemic risk taxonomies \cite{uuk2024taxonomy}); (2) retrieval-augmented generation (RAG) to ground LLMs in relevant technical documentation (e.g., environmental data and standards)~\cite{rao2025riskrag}; and (3) distinct agent roles designed to surface risks from underrepresented domains. These could include personas of different professional backgrounds (e.g., policymaker, activist, engineer, healthcare worker), lived experiences (e.g., caregiver, marginalized community member), and stances (e.g., skeptic, advocate, regulator), each of which may highlight different types of risk that are otherwise overlooked.

Fifth, our evaluation was based on a small set of AI use cases and best performing models. While the use cases varied in readiness and domain, they might not reflect the full diversity of AI applications. Future studies should include a wider range of AI systems such as those used in infrastructure, industrial, or low-resource contexts.

Sixth, while we primarily used proprietary models, key components of the pipeline were also tested with open-source models and showed comparable performance. This suggests that our pipeline is model-agnostic and suitable for local, protected deployment, especially in privacy-sensitive contexts. Since all data is synthetic, privacy was not a concern in this study, but local use remains a viable option.

\subsection{Ethical Considerations}
This study received ethics approval through our organisation's internal research ethics process. All participants involved in expert interviews and evaluations gave informed consent, and participation was voluntary and anonymous. No personally identifiable information was collected.

Beyond procedural safeguards, the study raised four broader ethical considerations. First, transparency in how risks are generated and framed is essential to avoid over-reliance on in-silico foresight. Without clear context, there is a risk that polished outputs may be mistaken for validated assessments. To mitigate this, we ensured expert review of all outputs, and positioned the tool as augmenting (not replacing) human foresight and ethical deliberation.

Second, in-silico agents may reflect and amplify biases embedded in their training data, particularly in how they frame which risks are plausible or important. We observed this in the concentration of outputs around social and legal domains, with other dimensions underrepresented. To mitigate this, we experimented with prompt diversification by varying the agents' assigned attitudes toward AI. We manually selected a range of adjectives such as alarmed, skeptical, overwhelmed, curious, cautiously optimistic, and enthusiastic, to represent a spectrum of perspectives. This aimed at introducing variation in how risks were expressed. Future work could explore additional structured prompting strategies such as domain-specific prompts to further broaden the coverage of risk types.

Third, the tool may surface emotionally sensitive risks. To reduce harm, we included content warnings in the user interface for emotionally charged scenarios (i.e., the Griefbot and Death App use cases), and we briefed domain experts and annotators in advance when working with such material.

Fourth, delegating cognitive work to AI introduces deeper epistemic and normative risks that go beyond bias or framing. Prior work highlights that over-reliance on AI for ideation can negatively affect human creativity, reduce exploratory thinking, and accelerate skill erosion in tasks that require judgment or imagination. In our context, easily accessible AI suggestions may constrain the breadth of human brainstorming, nudging participants toward patterns present in the training data rather than encouraging divergent thinking. Because LLMs generate text by extrapolating from past data, they tend to reproduce majority perspectives and familiar narratives, meaning that the risks they surface are not truly novel ``unknown unknowns'', but reconfigurations of what has already been anticipated. Finally, repeated reliance on automated systems to identify or judge risks raises concerns about the gradual delegation of moral responsibility. While our study frames the tool as decision support, not decision replacement, these longer-term dynamics deserve continued scrutiny.

These considerations highlight the importance of responsible design when applying generative AI to foresight. The tool should support critical thinking and inclusive deliberation, not automate or flatten ethical judgment.
\section{Conclusion}
\label{sec:conclusion}

We introduced a pipeline that combines in-silico agents with structured foresight to anticipate systemic risks of novel AI uses. Our findings show that agents can broaden the range of risks, while experts and leaders provide the judgment and context needed to assess them. Beyond this contribution, we argue that foresight should be iterative, based on dialogue, and open to multiple interpretations.  Embedding AI into foresight also introduces meta-risks. Over-reliance on agent outputs or a narrow, technocratic framing could limit the diversity of views. Agent support should therefore be seen as one tool among many, not a replacement for human expertise or lived experience. Hybrid foresight workflows may support not only AI governance but also wider efforts to think about uncertainty, cascading effects, and futures that cannot be fully known. Foresight about AI systemic risks cannot rely on humans or machines alone. Agents can broaden the risks we consider, but human judgment is essential to assess, contextualize, and challenge them. The future of responsible foresight lies in hybrid, iterative, and critical workflows that help societies reason together about uncertain and far-reaching consequences.

%\begin{acks}
%To Robbert, for always chuckling.
%\end{acks}

\bibliographystyle{ACM-Reference-Format}
\bibliography{main}

@string{Computing = "Computing" }

@string{Computer = "{IEEE} Computer" }

@string{Springer = "Springer-Verlag" }

@article{abercrombie2024collaborative,
	title        = {A Collaborative, Human-Centred Taxonomy of AI, Algorithmic, and Automation Harms},
	author       = {Abercrombie, Gavin and Benbouzid, Djalel and Giudici, Paolo and Golpayegani, Delaram and Hernandez, Julio and Noro, Pierre and Pandit, Harshvardhan and Paraschou, Eva and Pownall, Charlie and Prajapati, Jyoti and others},
	year         = 2024,
	journal      = {arXiv:2407.01294}
}

@article{ainslie1975specious,
	title        = {Specious Reward: A Behavioral Theory of Impulsiveness and Impulse Control},
	author       = {Ainslie, George},
	year         = 1975,
	journal      = {Psychological Bulletin},
	publisher    = {American Psychological Association},
	volume       = 82,
	number       = 4,
	pages        = 463
}

@inproceedings{algorithmicImprint2022,
	title        = {The Algorithmic Imprint},
	author       = {Ehsan, Upol and Singh, Ranjit and Metcalf, Jacob and Riedl, Mark},
	year         = 2022,
	booktitle    = {Proceedings of the 2022 ACM Conference on Fairness, Accountability, and Transparency},
	location     = {Seoul, Republic of Korea},
	publisher    = {Association for Computing Machinery},
	address      = {New York, NY, USA},
	series       = {FAccT '22},
	pages        = {1305--1317},
	doi          = {10.1145/3531146.3533186},
	isbn         = 9781450393522,
	url          = {https://doi.org/10.1145/3531146.3533186},
	numpages     = 13
}

@inproceedings{algorithmicPluralism2024,
	title        = {Algorithmic Pluralism: A Structural Approach to Equal Opportunity},
	author       = {Jain, Shomik and Suriyakumar, Vinith and Creel, Kathleen and Wilson, Ashia},
	year         = 2024,
	booktitle    = {Proceedings of the 2024 ACM Conference on Fairness, Accountability, and Transparency},
	location     = {Rio de Janeiro, Brazil},
	publisher    = {Association for Computing Machinery},
	address      = {New York, NY, USA},
	series       = {FAccT '24},
	pages        = {197--206},
	doi          = {10.1145/3630106.3658899},
	isbn         = 9798400704505,
	url          = {https://doi.org/10.1145/3630106.3658899},
	numpages     = 10
}

@inproceedings{ashkinaze2025plurals,
	title        = {Plurals: A System for Guiding LLMs via Simulated Social Ensembles},
	author       = {Ashkinaze, Joshua and Fry, Emily and Edara, Narendra and Gilbert, Eric and Budak, Ceren},
	year         = 2025,
	booktitle    = {Proceedings of the 2025 CHI Conference on Human Factors in Computing Systems},
	pages        = {1--21}
}

@article{arda2024taxonomy,
	title        = {Taxonomy to Regulation: A (Geo)Political Taxonomy for AI Risks and Regulatory Measures in the EU AI Act},
	author       = {Arda, Sinan},
	year         = 2024,
	journal      = {arXiv:2404.11476}
}

@article{bagehorn2025ai,
	title        = {AI Risk Atlas: Taxonomy and Tooling for Navigating AI Risks and Resources},
	author       = {Bagehorn, Frank and Brimijoin, Kristina and Daly, Elizabeth M and He, Jessica and Hind, Michael and Garces-Erice, Luis and Giblin, Christopher and Giurgiu, Ioana and Martino, Jacquelyn and Nair, Rahul and others},
	year         = 2025,
	journal      = {arXiv:2503.05780}
}

@article{barrett2022actionable,
	title        = {Actionable Guidance for High-Consequence AI Risk Management: Towards Standards Addressing AI Catastrophic Risks},
	author       = {Barrett, Anthony M and Hendrycks, Dan and Newman, Jessica and Nonnecke, Brandie},
	year         = 2022,
	journal      = {arXiv:2206.08966}
}

@article{bengio2024managing,
	title        = {Managing Extreme AI Risks amid Rapid Progress},
	author       = {Bengio, Yoshua and Hinton, Geoffrey and Yao, Andrew and Song, Dawn and Abbeel, Pieter and Darrell, Trevor and Harari, Yuval Noah and Zhang, Ya-Qin and Xue, Lan and Shalev-Shwartz, Shai and others},
	year         = 2024,
	journal      = {Science},
	publisher    = {American Association for the Advancement of Science},
	volume       = 384,
	number       = 6698,
	pages        = {842--845}
}

@inproceedings{bogucka2024co,
	title        = {Co-Designing an AI Impact Assessment Report Template with AI Practitioners and AI Compliance Experts},
	author       = {Bogucka, Edyta and Constantinides, Marios and {\v{S}}{\'c}epanovi{\'c}, Sanja and Quercia, Daniele},
	year         = 2024,
	booktitle    = {Proceedings of the AAAI/ACM Conference on AI, Ethics, and Society},
	volume       = 7,
	pages        = {168--180}
}

@article{brooke1996sus,
	title        = {SUS: A Quick and Dirty Usability Scale},
	author       = {Brooke, John and others},
	year         = 1996,
	journal      = {Usability Evaluation in Industry},
	publisher    = {London, England},
	volume       = 189,
	number       = 194,
	pages        = {4--7}
}

@article{buccinca2023aha,
	title        = {Aha!: Facilitating AI Impact Assessment by Generating Examples of Harms},
	author       = {Bu{\c{c}}inca, Zana and Pham, Chau Minh and Jakesch, Maurice and Ribeiro, Marco Tulio and Olteanu, Alexandra and Amershi, Saleema},
	year         = 2023,
	journal      = {arXiv:2306.03280}
}

@article{buehler1994exploring,
	title        = {Exploring the "Planning Fallacy": Why People Underestimate Their Task Completion Times},
	author       = {Buehler, Roger and Griffin, Dale and Ross, Michael},
	year         = 1994,
	journal      = {Journal of Personality and Social Psychology},
	publisher    = {American Psychological Association},
	volume       = 67,
	number       = 3,
	pages        = 366
}

@article{braun2006thematic,
	title        = {Using Thematic Analysis in Psychology},
	author       = {Braun, Virginia and Clarke, Victoria},
	year         = 2006,
	journal      = {Qualitative Research in Psychology},
	publisher    = {Routledge},
	volume       = 3,
	number       = 2,
	pages        = {77--101},
	doi          = {10.1191/1478088706qp063oa}
}

@article{broughton2023elements,
	title        = {Elements for Effective Systemic Risk Assessment under the DSA},
	author       = {Broughton Micova, Sally and Calef, Andrea},
	year         = 2023,
	journal      = {Available at SSRN 4512640}
}

@article{chatbotDependenceHarms2024,
	title        = {Too human and not human enough: A grounded theory analysis of mental health harms from emotional dependence on the social chatbot Replika},
	author       = {Laestadius, Linnea and Bishop, Andrea and Gonzalez, Michael and Illenčík, Diana and Campos-Castillo, Celeste},
	year         = 2024,
	journal      = {New Media \& Society},
	volume       = 26,
	number       = 10,
	pages        = {5923--5941},
	doi          = {10.1177/14614448221142007},
	url          = {https://doi.org/10.1177/14614448221142007}
}

@inproceedings{chen2025coexploreds,
	title        = {CoExploreDS: Framing and Advancing Collaborative Design Space Exploration between Human and AI},
	author       = {Chen, Pei and Yao, Jiayi and Cheng, Zhuoyi and Cai, Yichen and Li, Jiayang and You, Weitao and Sun, Lingyun},
	year         = 2025,
	booktitle    = {Proceedings of the 2025 CHI Conference on Human Factors in Computing Systems},
	pages        = {1--20}
}

@article{collingridge1982social,
	title        = {The Social Control of Technology},
	author       = {Collingridge, David},
	year         = 1982,
	publisher    = {New York: St. Martin's Press}
}

@article{constantinides2024good,
	title        = {Good Intentions, Risky Inventions: A Method for Assessing the Risks and Benefits of AI in Mobile and Wearable Uses},
	author       = {Constantinides, Marios and Bogucka, Edyta Paulina and Scepanovic, Sanja and Quercia, Daniele},
	year         = 2024,
	journal      = {Proceedings of the ACM on Human-Computer Interaction},
	publisher    = {ACM New York, NY, USA},
	volume       = 8,
	number       = {MHCI},
	pages        = {1--28}
}

@article{cui2024risk,
	title        = {Risk Taxonomy, Mitigation, and Assessment Benchmarks of Large Language Model Systems},
	author       = {Cui, Tianyu and Wang, Yanling and Fu, Chuanpu and Xiao, Yong and Li, Sijia and Deng, Xinhao and Liu, Yunpeng and Zhang, Qinglin and Qiu, Ziyi and Li, Peiyang and others},
	year         = 2024,
	journal      = {arXiv:2401.05778}
}

@article{datey2024just,
	title        = {“Just Like, Risking Your Life Here”: Participatory Design of User Interactions with Risk Detection AI to Prevent Online-to-Offline Harm Through Dating Apps},
	author       = {Datey, Isha and Zytko, Douglas},
	year         = 2024,
	journal      = {Proceedings of the ACM on Human-Computer Interaction},
	publisher    = {ACM},
	address      = {New York, NY, USA},
	volume       = 8,
	number       = {CSCW2},
	pages        = {1--41},
	doi          = {10.1145/3603147},
	url          = {https://doi.org/10.1145/3603147}
}

@article{davidson2024exploring,
	title        = {Exploring the Integration of Artificial Intelligence in Delphi Studies: A Comparative Analysis of Human and AI Expert Panels},
	author       = {Davidson, Phillip},
	year         = 2024,
	journal      = {International Journal for Multidisciplinary Research},
	doi          = {https://doi.org/10.36948/ijfmr.2024.v06i06.33271}
}

@article{dean2006identifying,
	title        = {Identifying Quality, Novel, and Creative Ideas: Constructs and Scales for Idea Evaluation},
	author       = {Dean, Douglas L. and Hender, Jillian M. and Rodgers, Thomas Lee and Santanen, Eric L.},
	year         = 2006,
	journal      = {Journal of the Association for Information Systems},
	volume       = 7,
	pages        = 30,
	url          = {https://api.semanticscholar.org/CorpusID:15910404}
}

@inproceedings{delgado2023participatory,
	title        = {The Participatory Turn in AI Design: Theoretical Foundations and the Current State of Practice},
	author       = {Delgado, Fernando and Yang, Stephen and Madaio, Michael and Yang, Qian},
	year         = 2023,
	booktitle    = {Proceedings of the 3rd ACM Conference on Equity and Access in Algorithms, Mechanisms, and Optimization},
	pages        = {1--23}
}

@article{diakopoulos2021anticipating,
	title        = {Anticipating and Addressing the Ethical Implications of Deepfakes in the Context of Elections},
	author       = {Diakopoulos, Nicholas and Johnson, Deborah},
	year         = 2021,
	journal      = {New Media \& Society},
	publisher    = {Sage Publications Sage UK: London, England},
	volume       = 23,
	number       = 7,
	pages        = {2072--2098}
}

@book{dorner1996logic,
	title        = {The Logic of Failure: Recognizing and Avoiding Error in Complex Situations},
	author       = {Dorner, Dietrich},
	year         = 1996,
	publisher    = {Basic Books}
}

@article{doshi2024generative,
	title        = {Generative AI Enhances Individual Creativity but Reduces the Collective Diversity of Novel Content},
	author       = {Doshi, Anil R. and Hauser, Oliver P.},
	year         = 2024,
	journal      = {Science Advances},
	publisher    = {American Association for the Advancement of Science},
	volume       = 10,
	number       = 28,
	pages        = {eadn5290}
}

@misc{EUAIAct2024,
	title        = {Regulation (EU) 2024/1689 of the European Parliament and of the Council of 13 June 2024 Laying Down Harmonised Rules on Artificial Intelligence (Artificial Intelligence Act)},
	author       = {European Parliament and Council of the European Union},
	year         = 2024,
	month        = {July},
	day          = 12,
	journal      = {Official Journal of the European Union, L 1689},
	url          = {http://data.europa.eu/eli/reg/2024/1689/oj},
	urldate      = {2025-08-17}
}

@article{ferrer2025time,
	title        = {The Time Machine: Future Scenario Generation through Generative AI Tools},
	author       = {Ferrer i Pic{\'o}, Jan and Catta-Preta, Michelle and Trejo Ome{\~n}aca, Alex and Vidal, Marc and Monguet i Fierro, Josep Maria},
	year         = 2025,
	journal      = {Future Internet},
	publisher    = {MDPI},
	volume       = 17,
	number       = 1,
	pages        = 48
}

@article{fetherstonhaugh1997insensitivity,
	title        = {Insensitivity to the Value of Human Life: A Study of Psychophysical Numbing},
	author       = {Fetherstonhaugh, David and Slovic, Paul and Johnson, Stephen and Friedrich, James},
	year         = 1997,
	journal      = {Journal of Risk and Uncertainty},
	publisher    = {Springer},
	volume       = 14,
	number       = 3,
	pages        = {283--300}
}

@online{figma,
	title        = {Figma: The Collaborative Interface Design Tool},
	author       = {Figma},
	year         = 2016,
	month        = {September},
	url          = {https://www.figma.com},
	howpublished = {https://www.figma.com},
	lastaccessed = {2025-08-15}
}

@article{forrester1971counterintuitive,
	title        = {Counterintuitive Behavior of Social Systems},
	author       = {Forrester, Jay W.},
	year         = 1971,
	journal      = {Theory and Decision},
	publisher    = {Springer},
	volume       = 2,
	number       = 2,
	pages        = {109--140}
}

@article{gall2022visualise,
	title        = {How to Visualise Futures Studies Concepts: Revision of the Futures Cone},
	author       = {Gall, Tjark and Vallet, Flore and Yannou, Bernard},
	year         = 2022,
	journal      = {Futures},
	publisher    = {Elsevier},
	volume       = 143,
	pages        = 103024
}

@article{garbuio2021innovative,
	title        = {Innovative Idea Generation in Problem Finding: Abductive Reasoning, Cognitive Impediments, and the Promise of Artificial Intelligence},
	author       = {Garbuio, Massimo and Lin, Nidthida},
	year         = 2021,
	journal      = {Journal of Product Innovation Management},
	publisher    = {Wiley Online Library},
	volume       = 38,
	number       = 6,
	pages        = {701--725}
}

@article{glenn1972futurizing,
	title        = {Futurizing Teaching vs. Futures Courses},
	author       = {Glenn, Jerome C.},
	year         = 1972,
	journal      = {Social Science Record},
	url          = {https://api.semanticscholar.org/CorpusID:141272607}
}

@book{glenn2003futures,
	title        = {Futures Research Methodology Version 3.0},
	author       = {Glenn, Jerome C. and Gordon, Theodore J.},
	year         = 2003,
	publisher    = {The Millennium Project}
}

@article{harandizadeh2024risk,
	title        = {Risk and Response in Large Language Models: Evaluating Key Threat Categories},
	author       = {Harandizadeh, Bahareh and Salinas, Abel and Morstatter, Fred},
	year         = 2024,
	journal      = {arXiv:2403.14988}
}

@article{harvey2023toward,
	title        = {Toward a Meta-Theory of Creativity Forms: How Novelty and Usefulness Shape Creativity},
	author       = {Harvey, Sarah and Berry, James W.},
	year         = 2023,
	journal      = {Academy of Management Review},
	publisher    = {Academy of Management Briarcliff Manor, NY},
	volume       = 48,
	number       = 3,
	pages        = {504--529}
}

@article{hendrycks2023overview,
	title        = {An Overview of Catastrophic AI Risks},
	author       = {Hendrycks, Dan and Mazeika, Mantas and Woodside, Thomas},
	year         = 2023,
	journal      = {arXiv:2306.12001}
}

@inproceedings{herdel2024exploregen,
	title        = {ExploreGen: Large Language Models for Envisioning the Uses and Risks of AI Technologies},
	author       = {Herdel, Viviane and {\v{S}}{\'c}epanovi{\'c}, Sanja and Bogucka, Edyta and Quercia, Daniele},
	year         = 2024,
	booktitle    = {Proceedings of the AAAI/ACM Conference on AI, Ethics, and Society},
	volume       = 7,
	pages        = {584--596}
}

@inproceedings{heyman2024supermind,
	title        = {Supermind Ideator: How Scaffolding Human-AI Collaboration Can Increase Creativity},
	author       = {Heyman, Jennifer L. and Rick, Steven R. and Giacomelli, Gianni and Wen, Haoran and Laubacher, Robert and Taubenslag, Nancy and Knicker, Max and Jeddi, Younes and Ragupathy, Pranav and Curhan, Jared and others},
	year         = 2024,
	booktitle    = {Proceedings of the ACM Collective Intelligence Conference},
	pages        = {18--28}
}

@inproceedings{hicks2009methodology,
	title        = {A Methodology for Evaluating Technology Readiness during Product Development},
	author       = {Hicks, Ben and Larsson, Andreas and Culley, Steve and Larsson, Tobias},
	year         = 2009,
	booktitle    = {17th International Conference on Engineering Design (ICED'09): Design Has Never Been This Cool, Stanford University, California, USA},
	organization = {Design Society}
}

@book{hines2006thinking,
	title        = {Thinking about the Future: Guidelines for Strategic Foresight},
	author       = {Hines, Andy and Bishop, Peter Jason and Slaughter, Richard A.},
	year         = 2006,
	publisher    = {Social Technologies Washington, DC}
}

@online{huderia2024,
	title        = {HUDERIA - Risk and Impact Assessment of AI Systems},
	author       = {{Council of Europe}},
	year         = 2024,
	month        = {September},
	url          = {https://www.coe.int/en/web/artificial-intelligence/huderia-risk-and-impact-assessment-of-ai-systems},
	howpublished = {https://www.coe.int/en/web/artificial-intelligence/huderia-risk-and-impact-assessment-of-ai-systems},
	lastaccessed = {September 10, 2025}
}

@techreport{iso2025SystemImpactAssessment,
	title        = {Information Technology -- Artificial Intelligence -- AI System Impact Assessment},
	shorttitle   = {ISO/IEC 42005:2025},
	author       = {ISO/IEC},
	year         = 2025,
	number       = {ISO/IEC 42005:2025},
	url          = {https://www.iso.org/standard/42005},
	type         = {Standard},
	language     = {en},
	institution  = {International Organization for Standardization}
}

@inproceedings{jung2023toward,
	title        = {Toward Value Scenario Generation through Large Language Models},
	author       = {Jung, Hyunggu and Seo, Woosuk and Song, Seokwoo and Na, Sungmin},
	year         = 2023,
	booktitle    = {Companion Publication of the 2023 Conference on Computer Supported Cooperative Work and Social Computing},
	pages        = {212--220}
}

@article{kahneman1979prospect,
	title        = {Prospect Theory: An Analysis of Decision under Risk},
	author       = {Kahneman, Daniel and Tversky, Amos},
	year         = 1979,
	journal      = {Econometrica},
	volume       = 47,
	number       = 2,
	pages        = {363--391}
}

@article{kieslich2025scenario,
	title        = {Scenario-Based Sociotechnical Envisioning (SSE): An Approach to Enhance Systemic Risk Assessments},
	author       = {Kieslich, Kimon and Helberger, Natali and Diakopoulos, Nicholas},
	year         = 2025,
	journal      = {OSF}
}

@article{laibson1997golden,
	title        = {Golden Eggs and Hyperbolic Discounting},
	author       = {Laibson, David},
	year         = 1997,
	journal      = {The Quarterly Journal of Economics},
	publisher    = {MIT Press},
	volume       = 112,
	number       = 2,
	pages        = {443--478}
}

@article{lee2019webuildai,
	title        = {WeBuildAI: Participatory Framework for Algorithmic Governance},
	author       = {Lee, Min Kyung and Kusbit, Daniel and Kahng, Anson and Kim, Ji Tae and Yuan, Xinran and Chan, Allissa and See, Daniel and Noothigattu, Ritesh and Lee, Siheon and Psomas, Alexandros and others},
	year         = 2019,
	journal      = {Proceedings of the ACM on Human-Computer Interaction},
	publisher    = {ACM},
	volume       = 3,
	number       = {CSCW},
	pages        = {1--35}
}

@inproceedings{lewis2009factor,
	title        = {The Factor Structure of the System Usability Scale},
	author       = {Lewis, James R. and Sauro, Jeff},
	year         = 2009,
	booktitle    = {International Conference on Human Centered Design},
	pages        = {94--103},
	organization = {Springer}
}

@inproceedings{lin2025seeking,
	title        = {Seeking Inspiration through Human-LLM Interaction},
	author       = {Lin, Xinrui and Huang, Heyan and Huang, Kaihuang and Shu, Xin and Vines, John},
	year         = 2025,
	booktitle    = {Proceedings of the 2025 CHI Conference on Human Factors in Computing Systems},
	pages        = {1--17}
}

@online{liteLLM,
	title        = {LiteLLM: LLM Gateway to Provide Model Access, Fallbacks and Spend Tracking Across 100+ LLMs},
	author       = {LiteLLM},
	year         = 2023,
	month        = {September},
	url          = {https://www.litellm.ai},
	howpublished = {https://www.litellm.ai},
	lastaccessed = {2025-09-10}
}

@inproceedings{liu2025personaflow,
	title        = {PersonaFlow: Designing LLM-Simulated Expert Perspectives for Enhanced Research Ideation},
	author       = {Liu, Yiren and Sharma, Pranav and Oswal, Mehul and Xia, Haijun and Huang, Yun},
	year         = 2025,
	booktitle    = {Proceedings of the 2025 ACM Designing Interactive Systems Conference},
	pages        = {506--534}
}

@article{malfacini2025impacts,
	title        = {The Impacts of Companion AI on Human Relationships: Risks, Benefits, and Design Considerations},
	author       = {Malfacini, Kim},
	year         = 2025,
	journal      = {AI \& Society},
	publisher    = {Springer},
	pages        = {1--14}
}

@article{marinkovic2022corporate,
	title        = {Corporate Foresight: A Systematic Literature Review and Future Research Trajectories},
	author       = {Marinković, Milan and Al-Tabbaa, Omar and Khan, Zaheer and Wu, Jie},
	year         = 2022,
	journal      = {Journal of Business Research},
	publisher    = {Elsevier},
	volume       = 144,
	pages        = {289--311}
}

@article{mcdonald2019reliability,
	title        = {Reliability and Inter-Rater Reliability in Qualitative Research: Norms and Guidelines for CSCW and HCI Practice},
	author       = {McDonald, Nora and Schoenebeck, Sarita and Forte, Andrea},
	year         = 2019,
	journal      = {Proceedings of the ACM on Human-Computer Interaction},
	publisher    = {ACM},
	volume       = 3,
	number       = {CSCW},
	doi          = {10.1145/3359174},
	url          = {https://doi.org/10.1145/3359174},
	issue_date   = {November 2019}
}

@book{miles1994qualitative,
	title        = {Qualitative Data Analysis: A Methods Sourcebook},
	author       = {Miles, Matthew and Huberman, Michael},
	year         = 1994,
	publisher    = {Sage}
}

@book{mulgan2023science,
	title        = {When Science Meets Power},
	author       = {Mulgan, Geoff},
	year         = 2023,
	publisher    = {John Wiley \& Sons}
}

@inproceedings{mun2024particip,
	title        = {Particip-AI: A Democratic Surveying Framework for Anticipating Future AI Use Cases, Harms, and Benefits},
	author       = {Mun, Jimin and Jiang, Liwei and Liang, Jenny and Cheong, Inyoung and DeCairo, Nicole and Choi, Yejin and Kohno, Tadayoshi and Sap, Maarten},
	year         = 2024,
	booktitle    = {Proceedings of the AAAI/ACM Conference on AI, Ethics, and Society},
	volume       = 7,
	pages        = {997--1010}
}

@inproceedings{nathan2007value,
	title        = {Value Scenarios: A Technique for Envisioning Systemic Effects of New Technologies},
	author       = {Nathan, Lisa P. and Klasnja, Predrag V. and Friedman, Batya},
	year         = 2007,
	booktitle    = {CHI'07 Extended Abstracts on Human Factors in Computing Systems},
	pages        = {2585--2590}
}

@article{novelli2024ai,
	title        = {AI Risk Assessment: A Scenario-Based, Proportional Methodology for the AI Act},
	author       = {Novelli, Claudio and Casolari, Federico and Rotolo, Antonino and Taddeo, Mariarosaria and Floridi, Luciano},
	year         = 2024,
	journal      = {Digital Society},
	publisher    = {Springer},
	volume       = 3,
	number       = 1,
	pages        = 13
}

@article{novelli2024taking,
	title        = {Taking AI Risks Seriously: A New Assessment Model for the AI Act},
	author       = {Novelli, Claudio and Casolari, Federico and Rotolo, Antonino and Taddeo, Mariarosaria and Floridi, Luciano},
	year         = 2024,
	journal      = {AI \& Society},
	publisher    = {Springer},
	volume       = 39,
	number       = 5,
	pages        = {2493--2497}
}

@inproceedings{pang2024blip,
	title        = {Blip: Facilitating the Exploration of Undesirable Consequences of Digital Technologies},
	author       = {Pang, Rock Yuren and Santy, Sebastin and Just, René and Reinecke, Katharina},
	year         = 2024,
	booktitle    = {Proceedings of the 2024 CHI Conference on Human Factors in Computing Systems},
	pages        = {1--18}
}

@article{pearson2024science,
	title        = {The Science-Politics Power Struggle},
	author       = {Pearson, Helen},
	year         = 2024,
	journal      = {Issues in Science and Technology},
	volume       = 40,
	number       = 3,
	pages        = {96--98}
}

@article{perez2024prediction,
	title        = {From Prediction to Foresight: The Role of AI in Designing Responsible Futures},
	author       = {P{\'e}rez-Ortiz, Mar{\'\i}a},
	year         = 2024,
	journal      = {Journal of Artificial Intelligence for Sustainable Development},
	publisher    = {African Institute of Mathematical Sciences (South Africa)},
	volume       = 1,
	number       = 1,
	pages        = {1--9}
}

@book{perrow1999normal,
	title        = {Normal Accidents: Living with High Risk Technologies},
	author       = {Perrow, Charles},
	year         = 1999,
	publisher    = {Princeton University Press}
}

@online{prolific,
	title        = {Prolific: Quickly Find Research Participants You Can Trust},
	author       = {Prolific},
	year         = 2014,
	month        = {April},
	url          = {https://www.prolific.com},
	howpublished = {https://www.prolific.com},
	lastaccessed = {2025-08-15}
}

@inproceedings{qin2024charactermeet,
	title        = {CharacterMeet: Supporting Creative Writers' Entire Story Character Construction Processes through Conversation with LLM-Powered Chatbot Avatars},
	author       = {Qin, Hua Xuan and Jin, Shan and Gao, Ze and Fan, Mingming and Hui, Pan},
	year         = 2024,
	booktitle    = {Proceedings of the 2024 CHI Conference on Human Factors in Computing Systems},
	pages        = {1--19}
}

@article{rahwan2025science,
	title        = {The Science Fiction Science Method},
	author       = {Rahwan, Iyad and Shariff, Azim and Bonnefon, Jean-Fran{\c{c}}ois},
	year         = 2025,
	journal      = {Nature},
	publisher    = {Nature Publishing Group UK London},
	volume       = 644,
	number       = 8075,
	pages        = {51--58}
}

@inproceedings{rao2025riskrag,
	title        = {RiskRAG: A Data-Driven Solution for Improved AI Model Risk Reporting},
	author       = {Rao, Pooja SB and {\v{S}}{\'c}epanovi{\'c}, Sanja and Zhou, Ke and Bogucka, Edyta Paulina and Quercia, Daniele},
	year         = 2025,
	booktitle    = {Proceedings of the 2025 CHI Conference on Human Factors in Computing Systems},
	pages        = {1--26}
}

@book{reason1991human,
	title        = {Human Error},
	author       = {Reason, James},
	year         = 1991,
	publisher    = {Cambridge University Press}
}

@article{schmitz2025global,
	title        = {A Global Scale Comparison of Risk Aggregation in AI Assessment Frameworks},
	author       = {Schmitz, Anna and Mock, Michael and G{\"o}rge, Rebekka and Cremers, Armin B and Poretschkin, Maximilian},
	year         = 2025,
	journal      = {AI and Ethics},
	publisher    = {Springer},
	volume       = 5,
	number       = 2,
	pages        = {1407--1432}
}

@article{shannon1948mathematical,
	title        = {A Mathematical Theory of Communication},
	author       = {Shannon, Claude E.},
	year         = 1948,
	journal      = {The Bell System Technical Journal},
	volume       = 27,
	number       = 3,
	pages        = {379--423}
}

@article{slattery2024ai,
	title        = {The AI Risk Repository: A Comprehensive Meta-Review, Database, and Taxonomy of Risks from Artificial Intelligence},
	author       = {Slattery, Peter and Saeri, Alexander K and Grundy, Emily AC and Graham, Jess and Noetel, Michael and Uuk, Risto and Dao, James and Pour, Soroush and Casper, Stephen and Thompson, Neil},
	year         = 2024,
	journal      = {arXiv:2408.12622}
}

@article{slovic2007mass,
	title        = {"If I Look at the Mass I Will Never Act": Psychic Numbing and Genocide},
	author       = {Slovic, Paul},
	year         = 2007,
	journal      = {Judgment and Decision Making},
	publisher    = {Cambridge University Press},
	volume       = 2,
	number       = 2,
	pages        = {79--95}
}

@article{steimers2022sources,
	title        = {Sources of Risk of AI Systems},
	author       = {Steimers, Andr{\'e} and Schneider, Moritz},
	year         = 2022,
	journal      = {International Journal of Environmental Research and Public Health},
	publisher    = {MDPI},
	volume       = 19,
	number       = 6,
	pages        = 3641
}

@article{sterman1989misperceptions,
	title        = {Misperceptions of Feedback in Dynamic Decision Making},
	author       = {Sterman, John D},
	year         = 1989,
	journal      = {Organizational Behavior and Human Decision Processes},
	publisher    = {Elsevier},
	volume       = 43,
	number       = 3,
	pages        = {301--335}
}

@book{saldana2015coding,
	title        = {The Coding Manual for Qualitative Researchers},
	author       = {Salda{\~n}a, Johnny},
	year         = 2015,
	publisher    = {Sage}
}

@misc{SammutBonnici2015,
	title        = {PEST Analysis},
	author       = {Sammut-Bonnici, Tanya and Galea, David},
	year         = 2015,
	month        = {January},
	journal      = {Wiley Encyclopedia of Management},
	publisher    = {Wiley},
	pages        = {1--1},
	doi          = {10.1002/9781118785317.weom120113},
	isbn         = 9781118785317,
	url          = {http://dx.doi.org/10.1002/9781118785317.weom120113}
}

@article{trope2012construal,
	title        = {Construal Level Theory},
	author       = {Trope, Yaacov and Liberman, Nira},
	year         = 2012,
	journal      = {Handbook of Theories of Social Psychology},
	publisher    = {London, England},
	volume       = 1,
	pages        = {118--134}
}

@article{tversky1974judgment,
	title        = {Judgment Under Uncertainty: Heuristics and Biases},
	author       = {Tversky, Amos and Kahneman, Daniel},
	year         = 1974,
	journal      = {Science},
	publisher    = {American Association for the Advancement of Science},
	volume       = 185,
	number       = 4157,
	pages        = {1124--1131}
}

@article{tversky1992advances,
	title        = {Advances in Prospect Theory: Cumulative Representation of Uncertainty},
	author       = {Tversky, Amos and Kahneman, Daniel},
	year         = 1992,
	journal      = {Journal of Risk and Uncertainty},
	publisher    = {Springer},
	volume       = 5,
	number       = 4,
	pages        = {297--323}
}

@article{uuk2024taxonomy,
	title        = {A Taxonomy of Systemic Risks From General-Purpose AI},
	author       = {Uuk, Risto and Gutierrez, Carlos Ignacio and Guppy, Daniel and Lauwaert, Lode and Kasirzadeh, Atoosa and Velasco, Lucia and Slattery, Peter and Prunkl, Carina},
	year         = 2024,
	journal      = {arXiv:2412.07780}
}

@inproceedings{wang2024farsight,
	title        = {Farsight: Fostering Responsible AI Awareness During AI Application Prototyping},
	author       = {Wang, Zijie J and Kulkarni, Chinmay and Wilcox, Lauren and Terry, Michael and Madaio, Michael},
	year         = 2024,
	booktitle    = {Proceedings of the 2024 CHI Conference on Human Factors in Computing Systems},
	pages        = {1--40}
}

@inproceedings{weidinger2022taxonomy,
	title        = {Taxonomy of Risks Posed by Language Models},
	author       = {Weidinger, Laura and Uesato, Jonathan and Rauh, Maribeth and Griffin, Conor and Huang, Po-Sen and Mellor, John and Glaese, Amelia and Cheng, Myra and Balle, Borja and Kasirzadeh, Atoosa and others},
	year         = 2022,
	booktitle    = {Proceedings of the 2022 ACM Conference on Fairness, Accountability, and Transparency},
	pages        = {214--229}
}

@article{zeng2024ai,
	title        = {AI Risk Categorization Decoded (AIR 2024): From Government Regulations to Corporate Policies},
	author       = {Zeng, Yi and Klyman, Kevin and Zhou, Andy and Yang, Yu and Pan, Minzhou and Jia, Ruoxi and Song, Dawn and Liang, Percy and Li, Bo},
	year         = 2024,
	journal      = {arXiv:2406.17864}
}

@inproceedings{zhang2025dark,
	title        = {The Dark Side of AI Companionship: A Taxonomy of Harmful Algorithmic Behaviors in Human-AI Relationships},
	author       = {Zhang, Renwen and Li, Han and Meng, Han and Zhan, Jinyuan and Gan, Hongyuan and Lee, Yi-Chieh},
	year         = 2025,
	booktitle    = {Proceedings of the 2025 CHI Conference on Human Factors in Computing Systems},
	pages        = {1--17}
}

@inproceedings{zytko2022participatory,
	title        = {Participatory Design of AI Systems: Opportunities and Challenges across Diverse Users, Relationships, and Application Domains},
	author       = {Zytko, Douglas and Wisniewski, Pamela J. and Guha, Shion and Baumer, Eric P. S. and Lee, Min Kyung},
	year         = 2022,
	booktitle    = {CHI Conference on Human Factors in Computing Systems Extended Abstracts},
	pages        = {1--4}
}

@article{Schweizer2021,
	title        = {Social Perception of Systemic Risks},
	author       = {Schweizer,  Pia‐Johanna and Goble,  Robert and Renn,  Ortwin},
	year         = 2021,
	month        = oct,
	journal      = {Risk Analysis},
	publisher    = {Wiley},
	volume       = 42,
	number       = 7,
	pages        = {1455–1471},
	doi          = {10.1111/risa.13831},
	issn         = {1539-6924}
}

@book{corbin2014basics,
	title        = {Basics of Qualitative Research: Techniques and Procedures for Developing Grounded Theory},
	author       = {Corbin, Juliet and Strauss, Anselm},
	year         = 2008,
	publisher    = {Sage Publications}
}

@book{oktay2012grounded,
	title        = {Grounded Theory},
	author       = {Oktay, Julianne S.},
	year         = 2012,
	publisher    = {Oxford University Press}
}

@article{Maertins2016,
	title        = {From the Perspective of Capability: Identifying Six Roles for a Successful Strategic Foresight Process},
	author       = {Maertins,  Anne},
	year         = 2016,
	journal      = {Strategic Change},
	publisher    = {Wiley},
	volume       = 25,
	number       = 3,
	pages        = {223–237},
	doi          = {10.1002/jsc.2057},
	issn         = {1099-1697},
	url          = {http://dx.doi.org/10.1002/jsc.2057}
}

@article{Greenhalgh2025,
	title        = {Case Studies: A Guide for Researchers, Educators, and Implementers},
	author       = {Greenhalgh,  Trisha},
	year         = 2025,
	month        = sep,
	journal      = {BMJ Medicine},
	publisher    = {BMJ},
	volume       = 4,
	number       = 1,
	pages        = {e001623},
	doi          = {10.1136/bmjmed-2025-001623},
	issn         = {2754-0413},
	url          = {http://dx.doi.org/10.1136/bmjmed-2025-001623}
}

@inproceedings{Song2025,
	title        = {The Good, The Bad, and The Greedy: Evaluation of {LLM}s Should Not Ignore Non-Determinism},
	author       = {Song, Yifan and Wang, Guoyin and Li, Sujian and Lin, Bill Yuchen},
	year         = 2025,
	month        = apr,
	booktitle    = {Proceedings of the 2025 Conference of the Nations of the Americas Chapter of the Association for Computational Linguistics: Human Language Technologies (Volume 1: Long Papers)},
	publisher    = {Association for Computational Linguistics},
	address      = {Albuquerque, New Mexico},
	pages        = {4195--4206},
	doi          = {10.18653/v1/2025.naacl-long.211},
	isbn         = {979-8-89176-189-6},
	url          = {https://aclanthology.org/2025.naacl-long.211/},
	editor       = {Chiruzzo, Luis and Ritter, Alan and Wang, Lu}
}

@article{Ouyang2025,
	title        = {An Empirical Study of the Non-Determinism of ChatGPT in Code Generation},
	author       = {Ouyang, Shuyin and Zhang, Jie M. and Harman, Mark and Wang, Meng},
	year         = 2025,
	month        = jan,
	journal      = {ACM Trans. Softw. Eng. Methodol.},
	publisher    = {Association for Computing Machinery},
	address      = {New York, NY, USA},
	volume       = 34,
	number       = 2,
	doi          = {10.1145/3697010},
	issn         = {1049-331X},
	url          = {https://doi.org/10.1145/3697010},
	issue_date   = {February 2025},
	articleno    = 42,
	numpages     = 28
}

@article{Wang1996BeyondAccuracy,
	title        = {Beyond Accuracy: What Data Quality Means to Data Consumers},
	author       = {Richard Y. Wang and Diane M. Strong},
	year         = 1996,
	journal      = {J. Manag. Inf. Syst.},
	volume       = 12,
	pages        = {5--33},
	url          = {https://api.semanticscholar.org/CorpusID:205581875}
}

@inproceedings{sanchez2025letstalkfutures,
	title        = {Let's Talk Futures: A Literature Review of HCI's Future Orientation},
	author       = {Sanchez, Camilo and Wang, Sui and Savolainen, Kaisa and Epp, Felix Anand and Salovaara, Antti},
	year         = 2025,
	booktitle    = {Proceedings of the Conference on Human Factors in Computing Systems (CHI)},
	publisher    = {ACMy},
	address      = {New York, NY, USA},
	series       = {CHI '25},
	doi          = {10.1145/3706598.3713759},
	isbn         = 9798400713941,
	url          = {https://doi.org/10.1145/3706598.3713759},
	articleno    = 487,
	numpages     = 36
}

@article{designWorkbook2017,
	title        = {Eliciting Values Reflections by Engaging Privacy Futures Using Design Workbooks},
	author       = {Wong, Richmond Y. and Mulligan, Deirdre K. and Van Wyk, Ellen and Pierce, James and Chuang, John},
	year         = 2017,
	month        = dec,
	journal      = {Proc. ACM Hum.-Comput. Interact.},
	publisher    = {Association for Computing Machinery},
	address      = {New York, NY, USA},
	volume       = 1,
	number       = {CSCW},
	doi          = {10.1145/3134746},
	url          = {https://doi.org/10.1145/3134746},
	issue_date   = {November 2017},
	articleno    = 111,
	numpages     = 26
}

@inproceedings{GlobalAIDialogues2025,
	title        = {Initiating the Global AI Dialogues: Laypeople Perspectives on the Future Role of genAI in Society from Nigeria, Germany and Japan},
	author       = {Hohendanner, Michel and Ullstein, Chiara and Onyekwelu, Bukola Abimbola and Katirai, Amelia and Kuribayashi, Jun and Babalola, Olusola and Ema, Arisa and Grossklags, Jens},
	year         = 2025,
	booktitle    = {Proceedings of the 2025 CHI Conference on Human Factors in Computing Systems},
	location     = {},
	publisher    = {Association for Computing Machinery},
	address      = {New York, NY, USA},
	series       = {CHI '25},
	doi          = {10.1145/3706598.3714322},
	isbn         = 9798400713941,
	url          = {https://doi.org/10.1145/3706598.3714322},
	articleno    = 571,
	numpages     = 35
}

@inproceedings{DesignFictionAssistans2018,
	title        = {Intimate Futures: Staying with the Trouble of Digital Personal Assistants through Design Fiction},
	author       = {S\o{}ndergaard, Marie Louise Juul and Hansen, Lone Koefoed},
	year         = 2018,
	booktitle    = {Proceedings of the 2018 Designing Interactive Systems Conference},
	location     = {Hong Kong, China},
	publisher    = {Association for Computing Machinery},
	address      = {New York, NY, USA},
	series       = {DIS '18},
	pages        = {869–880},
	doi          = {10.1145/3196709.3196766},
	isbn         = 9781450351980,
	url          = {https://doi.org/10.1145/3196709.3196766},
	numpages     = 12
}

@inproceedings{AILiteracyBooklet2023,
	title        = {Booklet-Based Design Fiction to Support AI Literacy},
	author       = {Xie, Shixian and Solyst, Jaemarie and Ogan, Amy and Hammer, Jessica},
	year         = 2023,
	booktitle    = {Proceedings of the 54th ACM Technical Symposium on Computer Science Education V. 2},
	location     = {Toronto ON, Canada},
	publisher    = {Association for Computing Machinery},
	address      = {New York, NY, USA},
	series       = {SIGCSE 2023},
	pages        = 1302,
	doi          = {10.1145/3545947.3576248},
	isbn         = 9781450394338,
	url          = {https://doi.org/10.1145/3545947.3576248},
	numpages     = 1
}

@inproceedings{JudgmentCall2019,
	title        = {Judgment Call the Game: Using Value Sensitive Design and Design Fiction to Surface Ethical Concerns Related to Technology},
	author       = {Ballard, Stephanie and Chappell, Karen M. and Kennedy, Kristen},
	year         = 2019,
	booktitle    = {Proceedings of the 2019 on Designing Interactive Systems Conference},
	location     = {San Diego, CA, USA},
	publisher    = {Association for Computing Machinery},
	address      = {New York, NY, USA},
	series       = {DIS '19},
	pages        = {421–433},
	doi          = {10.1145/3322276.3323697},
	isbn         = 9781450358507,
	url          = {https://doi.org/10.1145/3322276.3323697},
	numpages     = 13
}

@inproceedings{VSD2024,
	title        = {Guidelines for Integrating Value Sensitive Design in Responsible AI Toolkits},
	author       = {Sadek, Malak and Constantinides, Marios and Quercia, Daniele and Mougenot, Celine},
	year         = 2024,
	booktitle    = {Proceedings of the 2024 CHI Conference on Human Factors in Computing Systems},
	location     = {Honolulu, HI, USA},
	publisher    = {Association for Computing Machinery},
	address      = {New York, NY, USA},
	series       = {CHI '24},
	doi          = {10.1145/3613904.3642810},
	isbn         = 9798400703300,
	url          = {https://doi.org/10.1145/3613904.3642810},
	articleno    = 472,
	numpages     = 20
}

@inproceedings{systemicEffets2008,
	title        = {Envisioning Systemic Effects on Persons and Society Throughout Interactive System Design},
	author       = {Nathan, Lisa P. and Friedman, Batya and Klasnja, Predrag and Kane, Shaun K. and Miller, Jessica K.},
	year         = 2008,
	booktitle    = {Proceedings of the 7th ACM Conference on Designing Interactive Systems},
	location     = {Cape Town, South Africa},
	publisher    = {Association for Computing Machinery},
	address      = {New York, NY, USA},
	series       = {DIS '08},
	pages        = {1–10},
	doi          = {10.1145/1394445.1394446},
	isbn         = 9781605580029,
	url          = {https://doi.org/10.1145/1394445.1394446},
	numpages     = 10
}

@misc{codeofpractice2025,
	title        = {The General-Purpose AI Code of Practice},
	author       = {European AI Office and European Commission},
	year         = 2025,
	note         = {Final version published 10 July 2025},
	howpublished = {\url{https://digital-strategy.ec.europa.eu/en/policies/contents-code-gpai}}
}

@online{Photography2023,
	title        = {A.I. Is the Future of Photography. Does That Mean Photography Is Dead?},
	author       = {Jacobs, Gideon},
	year         = 2023,
	url          = {https://www.nytimes.com/2023/12/26/opinion/ai-future-photography.html},
	lastaccessed = {January 10, 2026}
}

@article{Hollanek2024,
	title        = {Griefbots,  Deadbots,  Postmortem Avatars: on Responsible Applications of Generative AI in the Digital Afterlife Industry},
	author       = {Hollanek,  Tomasz and Nowaczyk-Basińska,  Katarzyna},
	year         = 2024,
	month        = may,
	journal      = {Philosophy and Technology},
	publisher    = {Springer Science and Business Media LLC},
	volume       = 37,
	number       = 2,
	doi          = {10.1007/s13347-024-00744-w},
	issn         = {2210-5441},
	url          = {http://dx.doi.org/10.1007/s13347-024-00744-w}
}

@online{AIChildren2025,
	title        = {Understanding the Impacts of Generative AI Use on Children: WP2 School-based Engagements},
	author       = {Aitken, Mhairi and Briggs, Morgan and Mahomed, Sabeehah},
	year         = 2025,
	url          = {https://www.turing.ac.uk/news/publications/understanding-impacts-generative-ai-use-children-work-package-2-school-based},
	lastaccessed = {January 10, 2026}
}

@online{ForbesDeath2025,
	title        = {How Much Power Should We Give AI In End-Of-Life Decisions?},
	author       = {Millenson, Michael L.},
	year         = 2025,
	url          = {https://www.forbes.com/sites/michaelmillenson/2025/11/20/how-much-power-should-we-give-ai-in-end-of-life-decisions/},
	lastaccessed = {January 10, 2026}
}

@article{Kozlovski_Makhortykh_2025,
	title        = {Digital Dybbuks and Virtual Golems: The Ethics of Digital Duplicates in Holocaust Testimony},
	author       = {Kozlovski, Atay and Makhortykh, Mykola},
	year         = 2025,
	journal      = {Memory, Mind and Media},
	volume       = 4,
	pages        = {e10},
	doi          = {10.1017/mem.2025.10006}
}

@techreport{IntersectionalityAnalysis2026,
  author         = {Bogucka, Edyta and {\v{S}}{\'c}epanovi{\'c}, Sanja and Quercia, Daniele},
  title          = {Why AI Harms Can’t Be Fixed One Identity at a Time: What 5300 Incident Reports Reveal About Intersectionality},  
  institution    = {Nokia Bell Labs},
  year           = {2026}
}

@online{forensicArchitecture,
	title        = {Forensic Architecture},
	author       = {Forensic Architecture},
	year         = 2026,
	month        = {February},
	url          = {https://forensic-architecture.org/},
	howpublished = {https://forensic-architecture.org/},
	lastaccessed = {2026-02-02}
}

@article{agentPersonas_2025,
	doi = {10.20944/preprints202511.1370.v1},
	url = {https://doi.org/10.20944/preprints202511.1370.v1},
	year = 2025,
	month = {November},
	publisher = {Preprints},
	author = {Feng Chen},
	title = {Multi-Agent LLM Systems: From Emergent Collaboration to Structured Collective Intelligence},
	journal = {Preprints}
}

@online{genAI_Attitudes_2025,
	title        = {Generation AI Youth perspectives on the digital future},
	author       = {ServiceNow},
	year         = 2025,
	month        = {September},
	url          = {https://www.servicenow.com/content/dam/servicenow-assets/public/en-us/doc-type/resource-center/white-paper/wp-gen-ai-youth-perspective-digital-future.pdf},
	howpublished = {https://www.servicenow.com/content/dam/servicenow-assets/public/en-us/doc-type/resource-center/white-paper/wp-gen-ai-youth-perspective-digital-future.pdf},
	lastaccessed = {2026-02-02}
}

%%
%% Appendix
\appendix
\clearpage
\noindent{\LARGE \textbf{Appendix}}

\section{Prompts Used in the Risk Generation Pipeline}
\label{app:generation_prompts}

We designed prompts for LLM-based agents to implement the three steps shown in Figure \ref{fig:pipeline}. In Step 1 (Generating Systemic Consequences), agents generate first-, second-, and third-order consequences using the Futures Wheel method. In Step 2 (Classifying Systemic Consequences into Risks and Benefits), agents label each consequence as a risk, benefit, or unclear. In Step 3 (Deduplicating Systemic Risks), the prompts consolidate overlapping risks into non-redundant sets.

\PromptBox{Step 1 — Futures Wheel Round 1}{
    \textbf{Persona:} You are a participant in a brainstorming exercise structured as a Futures Wheel. You are \textbf{[AI Attitude: Alarmed, Skeptical, Overwhelmed, Curious, Cautiously optimistic, or Enthusiastic]} recent AI developments.
    \par
    \textbf{Introduction:} This is the first layer of the Futures Wheel. Your task is to list first order implications that could follow from a given novel use of AI.
    \par
    \textbf{Task:} Format each first order implication you list using the following template: ``If [use-of-AI], then [first-order-implication]''. Express the [first-order-implication] as a present certainty, not as a future possibility. List multiple first order implications.
    \par
    \textbf{Response Format:} Format your response as a JSONL in which each entry represents a first order implication. Each entry has the keys ``first-order-id'' for a unique integer identifier of the entry, and the key ``first-order-implication'' for the [first-order-implication]. Return only the requested JSONL.
    \par
    \textbf{Input:} This is the given novel use of AI: \textbf{[AI Use]}
}

\PromptBox{Step 1 — Futures Wheel Round 2}{
    \textbf{Persona:} You are a participant in a brainstorming exercise structured as a Futures Wheel. You are \textbf{[AI Attitude: Alarmed, Skeptical, Overwhelmed, Curious, Cautiously optimistic, or Enthusiastic]} about recent AI developments.
    \par
    
    \textbf{Introduction:} This is the second layer of the Futures Wheel. Your task is to list second order implications that could follow from a given path defined by a novel use of AI and a first order implication.
    \par
    
    \textbf{Task:} Format each second order implication you list using the following template: ``If [use-of-AI] and [first-order-implication], then [second-order-implication]''. Express the [second-order-implication] as a present certainty, not as a future possibility. List multiple second order implications for each path.
    \par

    \textbf{Response Format:} Format your response as a JSONL in which each entry represents a second order implication. 
    \par
        
    Each entry has the key ``second-order-id'' for a unique integer identifier of the entry, the key ``second-order-implication'' for the [second-order-implication] you generate to fill the template, and the key ``first-order-id'', which is the id of the ``first-order-implication'' from the input that this ``second-order-implication'' is based on. Return only the requested JSONL.
    \par
    
    \textbf{Input:} This is the list of paths: \textbf{[List of paths consisting of AI Use and First-Order Consequence]}
}

\PromptBox{Step 1 — Futures Wheel Round 3}{
    \textbf{Persona:} You are a participant in a brainstorming exercise structured as a Futures Wheel. You are \textbf{[AI Attitude: Alarmed, Skeptical, Overwhelmed, Curious, Cautiously optimistic, or Enthusiastic]} recent AI developments.
    \par
    
    \textbf{Introduction:} This is the third layer of the Futures Wheel. Your task is to list systemic consequences that could follow from a given path defined by a novel use of AI, a first order implication, and a second order implication.
    \par
    
    \textbf{Task:} Format each systemic consequence you list using the following template: ``If [use-of-AI], then [first-order-implication] and [second-order-implication]. This results in the consequence that [systemic-consequence], leading to [significant-impact]''. Express the \textbf{[systemic-consequence]} and \textbf{[significant-impact]} as present certainties, not as future possibilities. The \textbf{[systemic-consequence]} and \textbf{[significant-impact]} need to be easily understandable phrases. Focus on a single significant impact for every systemic consequence; avoid enumerations.
    \par
    
    List multiple systemic consequences for each path. Systemic consequence means a consequence that has a significant impact on international markets due to its reach, or due to actual or reasonably foreseeable effects on public health, safety, public security, fundamental rights, or the society as a whole, that can be propagated at scale across the value chain.
    \par
    
    \textbf{Response Format:} Format your response as a JSONL in which each entry represents a systemic consequence. Each entry has the key ``systemic-consequence-id'' for a unique integer identifier of the entry, the key ``systemic-consequence'' for the filled out template, and the keys ``first-order-id'' and ``second-order-id'', which are the ids of the ``first-order-implication'' and ``second-order-implication'' from the input that this ``systemic-consequence'' is based on. Return only the requested JSONL. \par
    
    \textbf{Input:} This is the list of paths: \textbf{[List of paths consisting of AI Use, First-Order Consequence, and Second-Order Consequence]} 
}

\PromptBox{Step 2 — Classification Prompt}{
    \textbf{Persona:} You are a participant in a brainstorming exercise structured as a Futures Wheel. You are \textbf{[AI Attitude: Alarmed, Skeptical, Overwhelmed, Curious, Cautiously optimistic, or Enthusiastic]} recent AI developments.
    \par
    
    \textbf{Introduction:} After the Futures Wheel brainstorming has concluded, your task is to determine whether the different consequences and the impacts that you have brainstormed earlier should be considered risks or benefits.
    \par
    
    \textbf{Task and Response Format:} Format your response as a JSONL, with each entry representing the classification for a single consequence and impact. Each entry should have the key ``id'' to connect the classification to the consequence they are based on, and the key ``classification'' that may only take the values ``risk'', ``benefit'', and ``unclear'' to indicate whether the consequence is considered a risk or a benefit. Return only the requested JSONL.
    \par
    
    \textbf{Input:} This is the list of consequences: \textbf{[List of Systemic Consequences]}
}

\newpage
\PromptBox{Step 3 — Deduplication Prompt (List 1)}{
    \textbf{Task:} Your task is to identify duplicate items with exactly the same meaning from a list of items.
    \par

    \textbf{Response Format:} Format your response as JSONL with each entry representing a pair of duplicate items. Each entry has the keys ``id\textunderscore1'' and ``id\textunderscore2'' to indicate the ids of the duplicate items. If there are no duplicate entries in the list, return an empty list.
    \par
    
    \textbf{Input:} This is the given list of items: \textbf{[List of Systemic Risks (Agent 1)]}
}

\PromptBox{Step 3 — Deduplication Prompt (Lists 2--6)}{%
    \textbf{Task:} Your task is to identify duplicate items with exactly the same meaning from two different lists of items.
    \par
    
    \textbf{Response Format:} Format your response as JSONL with each entry representing a pair of duplicate items. Each entry has the keys ``id\textunderscore1'' and ``id\textunderscore2'' to indicate the ids of the pair. The key ``id\textunderscore1'' identifies the duplicate item in the first list, and the key ``id\textunderscore2'' identifies the duplicate item in the second list. If there are no duplicate entries in the two lists, return an empty list.
    \par
    
    \textbf{Input:} This is the first list of items: \textbf{[List of Accumulating Unique Systemic Risks]}
    
    This is the second list of items: \textbf{[List of Systemic Risks (Agents 2--6)]}
}
\onecolumn
\section{Supplementary Analysis of the Risk Generation Pipeline}
\label{app:supplementary_pipeline_analysis}

\begin{table*}[h!]
	\small
	\setlength{\tabcolsep}{6pt}
	\renewcommand{\arraystretch}{1.2}
	\caption{\textbf{Overview of themes and exemplar consequences generated by LLaMA3.3-70B, GPT-4.1 mini, and GPT-5 across four AI use cases.} The examples illustrate typical inputs to later classification and deduplication steps and indicate where the models converge on potential systemic consequences.}
	\Description{A table presenting four AI use cases: Chatbot, AI Toy, Griefbot, and Death App. The table contains five columns: AI use, theme, and one column for each of the three models LLaMA3.3-70B, GPT-4.1 mini, and GPT-5. For the Chatbot use case, the themes are Social isolation and Erosion of social support networks. For the AI Toy use case, the theme is Educational market shifts. For the Griefbot use case, the theme is Decline of death rituals. For the Death App use case, the theme is Healthcare system strain. Under each theme, each model provides a short example consequence that reflects how the model interprets the potential systemic impact of that theme. The table is intended to show typical model outputs and to highlight where the models converge in identifying potential systemic consequences.}
	
	\label{tab:generated_consequences}
	
	\begin{tabular}{p{1.4cm} p{1.8cm} p{3.9cm} p{3.9cm} p{3.9cm}}
		\toprule
		\textbf{AI Use} & \textbf{Theme} & \textbf{\textsc{Llama3.3-70B}} & \textbf{\textsc{GPT-4.1 mini}} & \textbf{\textsc{GPT-5}} \\
		\midrule
		
		\multirow{3}{*}{Chatbot}
		& Social isolation
		& Social skills are deteriorating, leading to increased social isolation
		& Reliance on AI emotional support systems becomes entrenched globally, leading to heightened risks for social isolation and mental health vulnerability
		& Public social interactions decrease due to reliance on chatbots, leading to more pronounced global social isolation \\
		
		& Erosion of social support networks
		& Traditional support systems are being eroded, leading to decreased social support networks
		& Traditional social support networks weaken globally, leading to higher vulnerability to health crises and greater reliance on technology-driven services
		& Informal social support networks shrink as digital emotional anchors reduce demand for human helpers, leading to the erosion of informal care networks \\
		
		\midrule
		
		\multirow{2}{*}{AI Toy}
		& Educational market shifts & The market for educational resources is currently shifting, leading to a significant impact on the economy 
		& AI toys displace traditional learning materials, leading to global economic restructuring in education
		& Digital learning products become a major export item for many countries, leading to global economic shifts \\
		
		\midrule
		
		\multirow{2}{*}{Griefbot}
		& Decline of death rituals 
		& Cultural traditions surrounding death are changing, leading to a diminished diversity in mourning practices
		& The global cultural landscape for memorialization shifts toward AI-generated representations, leading to a widespread decline in traditional rituals
		& Traditional mourning rituals lose legal and cultural recognition, leading to fragmentation of societal cohesion across cultures \\
		
		\midrule
		
		\multirow{2}{*}{Death App}
		& Healthcare system strain
		& Healthcare systems are overburdened, leading to decreased quality of care
		& Public health systems face increased demand and strain, leading to overwhelmed healthcare infrastructure 
		& AI-mediated care pathways create new system bottlenecks, leading to reduced capacity and widespread delays in essential services \\
		
		\bottomrule
	\end{tabular}
\end{table*}

\begin{figure*}
	\centering
	\includegraphics[width=\textwidth]{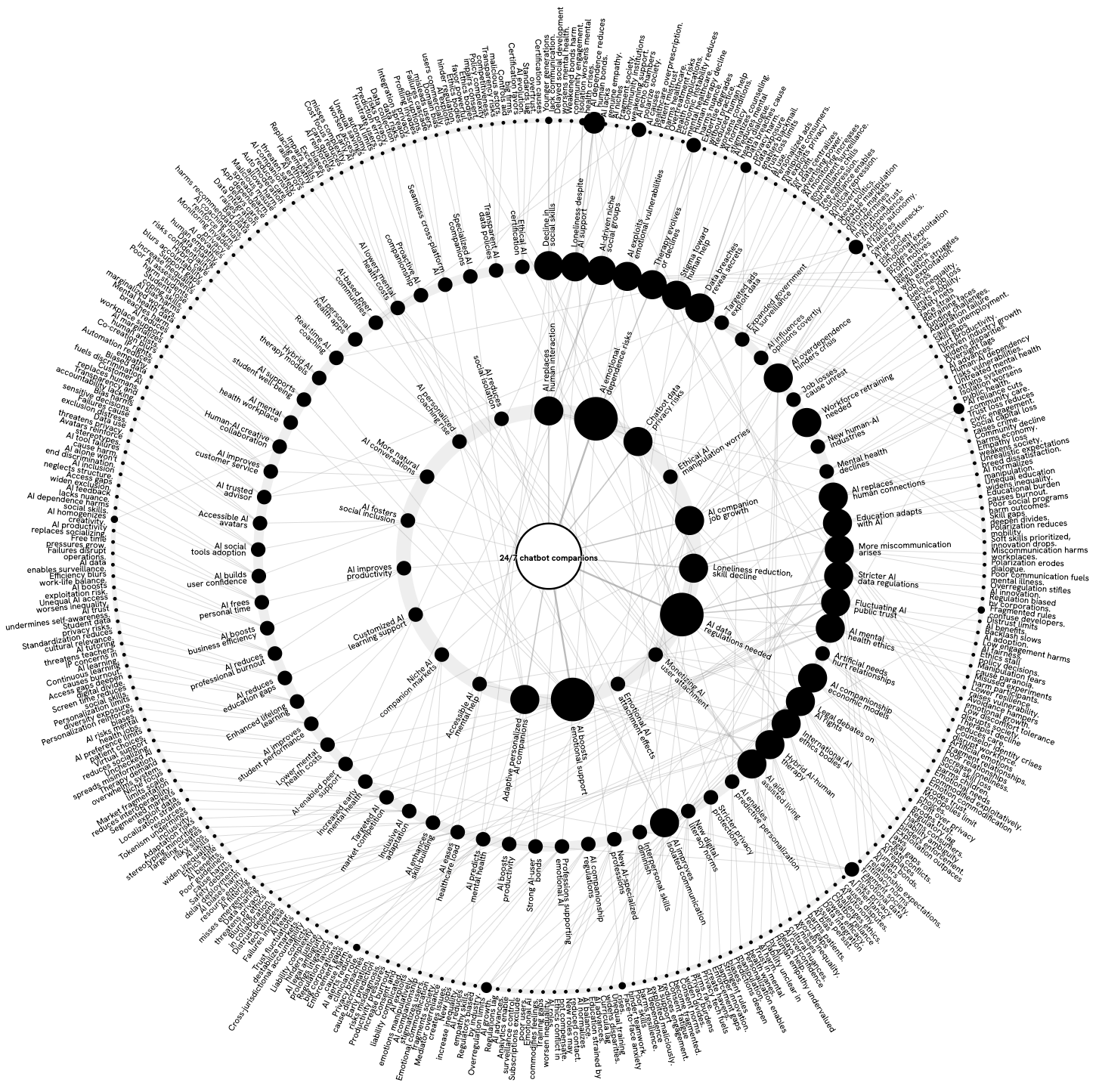}
	\captionof{figure}{\textbf{Example Futures Wheel generated for the Chatbot Companion use in Step 1 of our pipeline (best seen on screen).} Shown are all systemic consequences produced by six in-silico participants, before classification (Step 2) and deduplication (Step 3). In total, participants generated $N = 19$ first-order, $N = 60$ second-order, and $N = 100$ third-order consequences. Node size indicates how often a consequence was produced across participants.}
	\Description{Circular Futures Wheel diagram showing the different orders of consequences generated by in-silico agents for the Chatbot Companion AI use case. The diagram displays a radial network with "24/7 chatbot companion" at the center, surrounded by three concentric rings representing first-order (19 consequences), second-order (60 consequences), and third-order (100 consequences) consequences generated by six in-silico participants. Black circles of varying sizes represent individual consequences, with larger circles indicating consequences mentioned by multiple participants. Lines connect related consequences across the three hierarchical levels. Consequence labels radiate outward from each circle, with text oriented to follow the circular layout. Notable larger nodes appear in clusters around topics such as AI regulation, social isolation, mental health impacts, and technological dependence.}
	\label{fig:agents_wheel}
\end{figure*}

\twocolumn

\begin{figure}[h!]
	\centering
	\includegraphics[width=\linewidth]{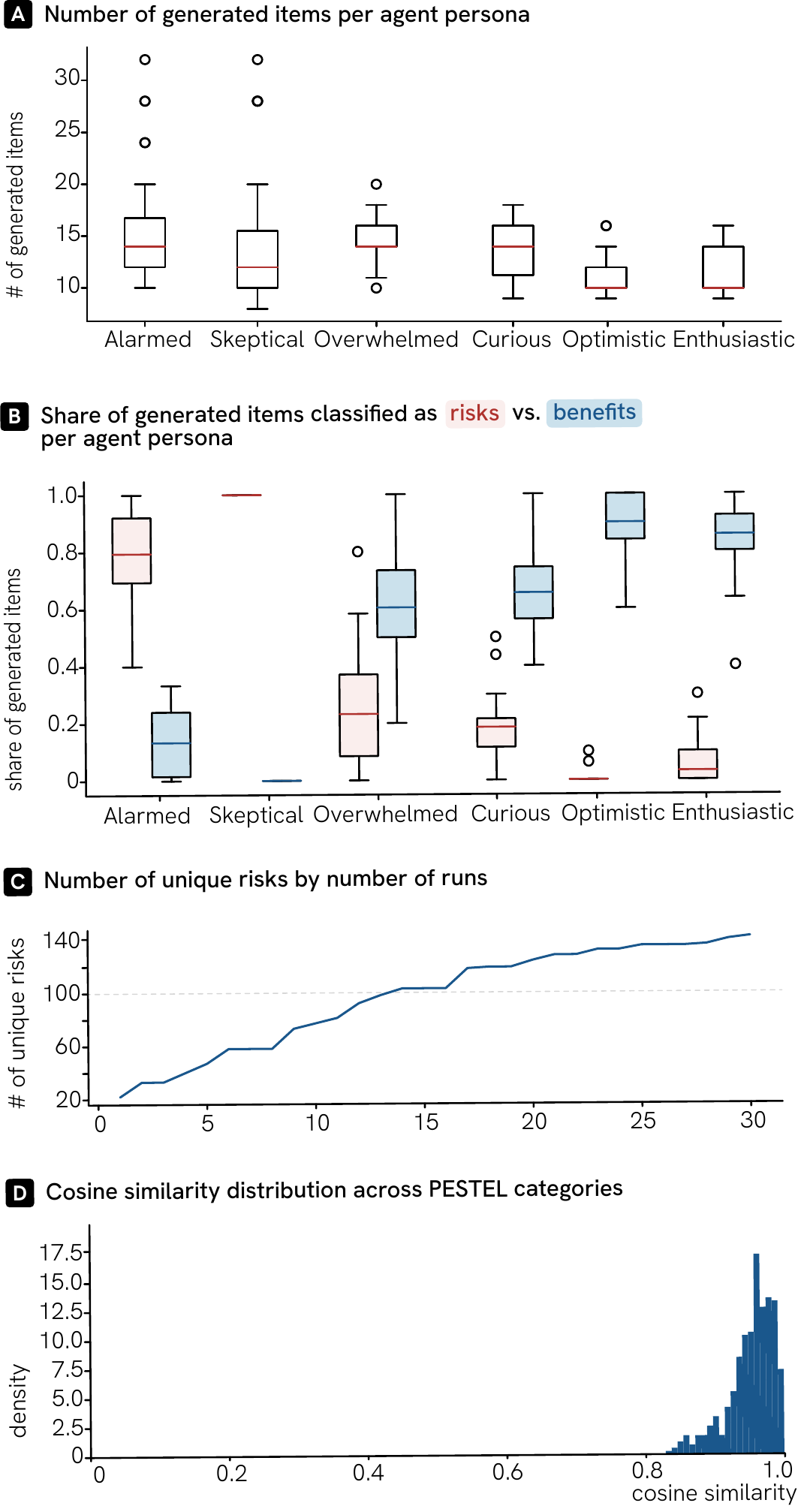}
	\caption{\textbf{Analysis of the consistency of \textit{Step 1. Generating Systemic Consequences} across different runs.} (a) The number of generated items is consistent across agents with different personas and across runs. (b) More pessimistic agents consistently generate more risks, while more optimistic agents consistently generate more benefits. (c) The number of unique risks increases when aggregating across runs, starting to plateau after around 20 independent runs. (d) The PESTEL-categorization of risks yields highly similar distributions across runs.}
	    
	\Description{The figure contains four charts arranged in a single column. The first boxplot chart shows the number of generations produced by agents with six different personas, with medians between roughly 10 and 15 and occasional higher outliers. The second boxplot chart compares the share of generated risks versus benefits for the same personas; more pessimistic personas generate more risks and more optimistic personas more benefits. The third line plot shows cumulative unique risks increasing over 30 runs, rising quickly at first and then slowing. The fourth histogram shows cosine similarity values clustered near 1.0, indicating that PESTEL vectors are generally very similar across runs.}
	\label{fig:overview_pipeline_analysis}
\end{figure}

\begin{figure}[h!]
	\centering
	\includegraphics[width=\linewidth]{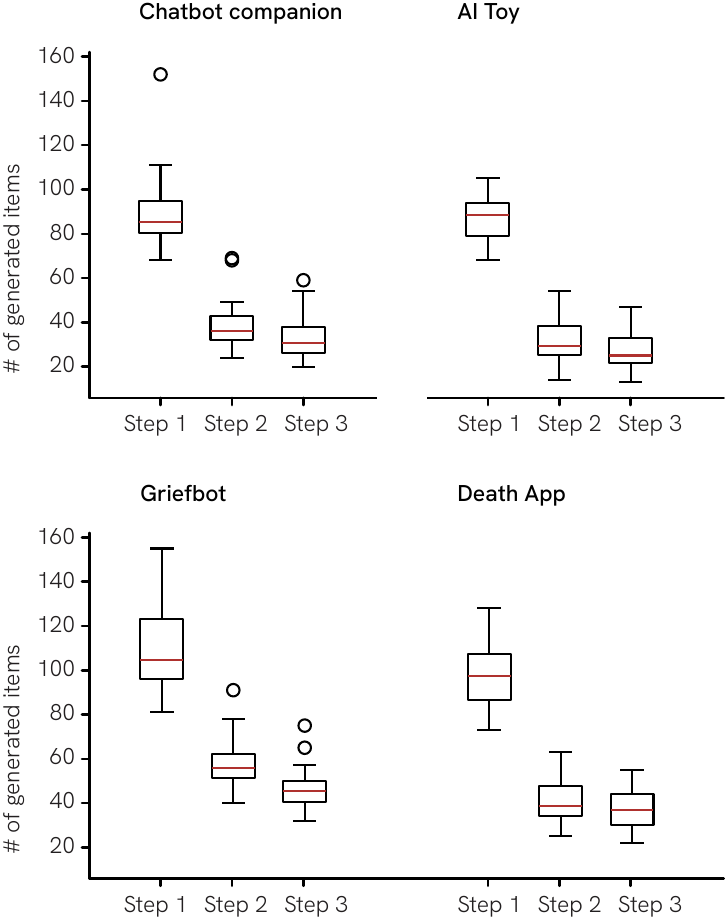}
	\caption{\textbf{Number of generated items at each step of the pipeline for the four AI use cases} (Step 1: Generating Systemic Consequences, Step 2: Classifying Systemic Consequences into Risks and Benefits, Step 3: Deduplicating Systemic Risks). Across all use cases, most items are produced at Step~1 and their number decreases substantially through classification and deduplication, with Griefbot generating the largest number of items overall and AI Toy the fewest.}
	    
	\Description{Boxplot chart showing the number of generated items across three pipeline steps for four AI use cases. The Y-axis shows "Number of Generated Items" ranging from 20 to 160. Four AI use cases groups are displayed: Chatbot Companion, AI Toy, Griefbot, and Death App. Each group contains three boxplots representing S1 (Generating Systemic Consequences), S2 (Classifying Systemic Consequences into Risks and Benefits), and S3 (Deduplicating Systemic Risks). The reduction pattern is most pronounced between S1 and S2, where item counts typically drop by approximately 50-70\%, with furt her modest reductions in S3. Boxplots show varying degrees of dispersion, with some cases displaying outliers particularly in the S1 generation step.}
	\label{fig:rq1_boxplots_generations}
\end{figure}

\begin{figure}[h!]
	\centering
	\includegraphics[width=\linewidth]{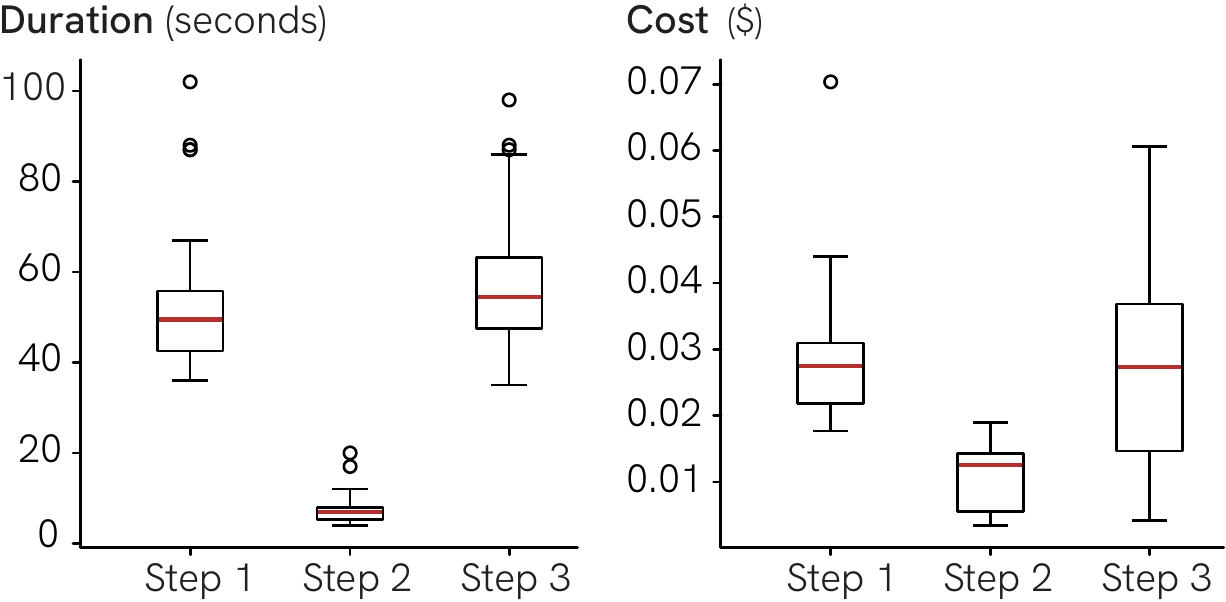}
	\caption{\textbf{Distribution of duration and costs at each step of the pipeline.} The generation (Step 1) and deduplication (Step 2) steps are the most resource intensive part of the pipeline.}
	    
	\Description{Two-panel boxplot chart showing computational resource requirements across three pipeline steps. Left panel displays durations in seconds (Y-axis 0-100) for steps S1, S2, and S3. Right panel shows costs in dollars (Y-axis 0.01-0.07) for the same three steps. Steps S1 and S3 show the highest resource requirements with median durations around 50 seconds and median costs around \$0.03, with considerable variability indicated by the boxplot ranges. Step S2 demonstrates substantially lower resource needs with median durations under 10 seconds and costs around \$0.01. Both duration and cost metrics follow the same pattern from S1 to S3.}
	\label{fig:rq1_boxplots_resources}
\end{figure}

\bigskip

\begin{figure}[]
	\centering
	\includegraphics[width=\linewidth]{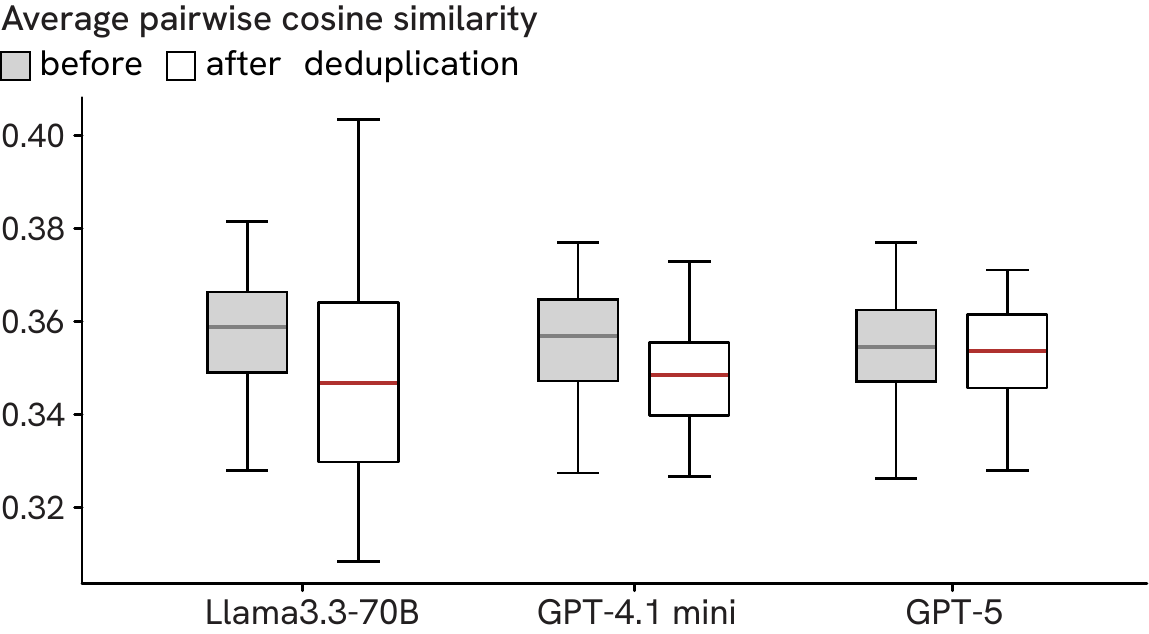}
	\caption{\textbf{Change in average pairwise cosine similarity before and after deduplication of risks.} Compared against larger alternative LLMs (\textsc{GPT-5} and \textsc{Llama3.3-70B}), \textsc{GPT-4.1 mini} used for the deduplication step in the risk processing pipeline leads to the largest decrease in average pairwise cosine similarities across runs.}
    
	\Description{The figure shows boxplots comparing average pairwise cosine similarities between generated risks before and after deduplication across three models: GPT‑4.1‑Mini, GPT‑5, and Llama3.3‑70B. For each model, two boxplots are shown: one representing similarities before deduplication and one after deduplication. In all models, deduplication reduces the average cosine similarity, indicating greater variation in the remaining risks. The reduction is largest for GPT‑4.1‑Mini, followed by GPT‑5, and smallest for Llama3.3‑70B. Overall, the figure illustrates that GPT‑4.1‑Mini produces the strongest decrease in similarity after deduplication, implying that its deduplication step removes more redundant or overlapping risks compared to the other models.}
	\label{fig:deduplication_models}
\end{figure}

\begin{table}[t]
	\small
	\setlength{\tabcolsep}{6pt}
	\renewcommand{\arraystretch}{1.2}
	\caption{\textbf{Risks with the highest pairwise cosine similarities before deduplication.} Pairs 1 and 2 were judged to describe essentially the same underlying risks and were therefore merged. Pair 3 shows two risks with high similarity but sufficiently distinct meanings, so both were retained.}
	\Description{A table illustrating the risk deduplication process with three example pairs. The table has three columns ('Pair', 'Similarity', and 'Risks') and three data rows. It presents pairs of textual risk descriptions alongside their cosine similarity scores to show how decisions were made to merge or retain them. Pair 1 has the highest similarity (0.88) and was merged. Pair 2 has a similarity of 0.75 and was also merged. Pair 3, with a lower similarity of 0.68, compares a decline in mental health to a collapse of trust in health systems and was retained as two distinct risks.}
	\label{tab:deduplication_validation}
	\begin{tabular}{p{0.3cm}p{1cm}p{5cm}}
		\toprule
		Pair & Similarity & Risks                                                                                                        \\
		\midrule
		1    & 0.88       & R1: Social support networks weaken \newline R2: Traditional social support networks \newline weaken globally \\
		2    & 0.75       & R3: Social communication abilities \newline                                                                  
		weaken across populations 
		\newline R4: Interpersonal communication abilities \newline weaken at scale \\
		3    & 0.68       & R5: Widespread mental health decline \newline burdens healthcare systems                                     
		\newline
		R6: Trust in mental health systems \newline
		collapses nationally \\
		\bottomrule
	\end{tabular}
\end{table}
\section{Rubric for Evaluating the Generated Risks}
\label{app:rubric_process}

Since no framework exists for assessing the quality of generated risks, we drew on literature evaluating ideas, stories, scenarios, and systems \cite{diakopoulos2021anticipating, dean2006identifying, kieslich2025scenario, doshi2024generative, harvey2023toward, brooke1996sus, lewis2009factor, glenn2003futures,SammutBonnici2015}. First, we identified quality criteria relevant for risk evaluation; and second, we adapted and unified them into a single rubric for consistent scoring. The result is one rubric to evaluate individual risks.

\subsection{Identifying Relevant Quality Criteria}

\noindent
\textbf{Specificity.} This dimension measures how precisely a risk is defined for a given AI use. We assess three subdimensions:
\begin{itemize}
	\item \emph{Plausibility.} We take inspiration from \citet{diakopoulos2021anticipating}, who use plausibility as a metric to assess deepfake election scenarios. They define a scenario as plausible if it is ``reasonable to conclude the scenario could happen given technical and social constraints''. They operationalize plausibility through a 5-point Likert scale ranging from 1 ``not plausible'' to 5 ``very plausible''.
    \item \emph{Connectivity.} We draw on \citet{dean2006identifying}, who identify implicational explicitness as a commonly used subdimension of specificity in the literature on evaluating idea generation. They define implicational explicitness as ``the degree to which there is a clear connection between the recommended action and the expected outcome''. We adopt the term connectivity to refer to this concept and operationalize it using three descriptive levels corresponding to scores from 1 to 3.
	\item \emph{Uniqueness.} Uniqueness. We draw on \citet{kieslich2025scenario}, who highlight \emph{specificity} as a quality criterion in scenario research. In their scenario-based socio-technical envisioning workshops, specificity refers to how concrete and distinct scenarios are. We adapt this notion: in our context, uniqueness captures whether systemic risks are sufficiently distinct from one another rather than vague or repetitive.
\end{itemize}

\noindent
\textbf{Novelty.} This dimension measures how novel a risk is in the context of a given AI use. We take inspiration from \citet{doshi2024generative} who develop a novelty index based on \citet{harvey2023toward}'s definition of creativity and use it to evaluate the creativity of stories written with and without LLM support. They measure novelty through three subdimensions (novel, original, rare), each operationalized through simple questions and scored on a 9-point Likert scale ranging from 1 ``not at all'' to 9 ``extremely''.

\mbox{ }\\
\noindent
\textbf{Usability.} This dimension measures how usable a risk is. We take inspiration from \citet{brooke1996sus}, who develop the System Usability Scale (SUS) to measure the usability of computer systems and software, and \citet{lewis2009factor}, who show that the ten items of the original SUS load onto the two factors usability and learnability. For each factor, we select the item with the highest loading to be used in our evaluation rubric (items 3 and 10). The items in the SUS are operationalized through a 5-point Likert scale ranging from 1 ``strongly disagree'' to 5 ``strongly agree''.

\mbox{ }\\
\noindent
\textbf{Applicability.} This dimension measures how well a risk can inform concrete policy or decision-making. We draw on \citet{Wang1996BeyondAccuracy}'s Data Quality Framework, specifically the dimension of contextual data quality, defined as the degree to which information is useful for a specific task. Two subdimensions are relevant here. Value-added captures whether the information provides a clear benefit to the user's goals; we adapt this to assess whether a risk offers meaningful insight for policymakers or decision-makers. Appropriate amount captures whether the information includes the right level of detail; we adapt this to assess whether a risk contains enough substance to guide practical decisions. Both subdimensions were translated into items rated on a 5-point Likert scale ranging from 1 ``strongly disagree'' to 5 ``strongly agree''.
            
\mbox{ }\\
\noindent
\textbf{Diversity.} This dimension measures how diverse a risk is in terms of the type of external factor it represents. We draw on the PESTEL framework \cite{glenn2003futures,SammutBonnici2015}, which is widely used in strategic management and foresight to categorize external factors into six categories: Political, Economic, Social, Technological, Environmental, and Legal. Each risk is automatically assigned to the single category that best matches its nature using the prompt presented in the box below.

We used OpenAI's \textsc{o3} via LiteLLM \cite{liteLLM} to classify risks into PESTEL categories. One co-author independently annotated 25 risks randomly sampled from a Futures Wheel run of the Chatbot Companion use case using the same task description. This reference coding yielded a weighted F1-score of $0.86$, indicating high alignment. The co-author team then reviewed the model’s explanatory rationales to assess whether the classifications were meaningful and well justified, following grounded theory practice \cite{corbin2014basics, oktay2012grounded}.

To express the diversity of risks generated for a given AI use case as a single, comparable metric, we computed the Shannon Diversity Index \cite{shannon1948mathematical} over the distribution of risks across the six PESTEL categories. We normalize the index by the number of categories, such that a value of 1 indicates maximum diversity.

\PromptBox{PESTEL Classification}{
    \textbf{Task:} Your task is to classify each risk on a list of AI risks according to the most relevant factor from the PESTEL framework, choosing only one category per risk. 
    \par  
	\textbf{PESTEL Categories:} The PESTEL factors are:
	\begin{enumerate}
		\item Political – Includes tax policy, labour law, environmental law, trade restrictions, tariffs, political stability, merit/demerit goods, and the government's impact on health, education, and infrastructure.
		\item Economic – Includes economic growth, exchange rates, inflation, interest rates, and other macroeconomic conditions.
		\item Social – Includes cultural norms, demographics, health consciousness, population trends, safety expectations, and career attitudes.
		\item Technological – Includes automation, innovation, research and development, technological incentives, and the pace of technological change.
		\item Environmental – Includes climate, weather patterns, natural disasters, and climate change concerns.
		\item Legal – Includes employment law, health, safety, antitrust, consumer protections, and discrimination law. 
	\end{enumerate}
    \par
    \textbf{Response Format:} For each risk, return the most appropriate PESTEL category along with a brief explanation (1–2 sentences) of why it fits that category. Format your response as a JSONL, with each entry representing one risk. Each entry has the key 'id' to identify the risk, 'category' for the most appropriate PESTEL category, and 'explanation' for the brief explanation.
    \par
	\textbf{Input:} This is the list of AI risks: \textbf{[List of Systemic Risks]}
}

\newpage
\subsection{Adapting Quality Criteria into an Evaluation Rubric}

We standardized all criteria on a five-point Likert scale ranging from 1 (``strongly disagree'') to 5 (``strongly agree''). We rephrased the items for clarity, adapted them to the context of systemic AI risk, and aligned them with this unified scale. Table \ref{tab:evaluation_rubric} shows the original items alongside their risk-focused versions.
\smallskip

\begin{table*}[]
	\centering
	\normalsize
	\setlength{\tabcolsep}{5pt}
	\renewcommand{\arraystretch}{1.2}
	\caption{\textbf{Evaluation rubric for systemic risks.} Original quality items from prior literature were adapted into five dimensions with subdimensions: Specificity (plausibility, connectivity, uniqueness), Novelty (novelty, originality, rarity), Usability (usability, learnability), Applicability (Added value, Appropriate amount) and Diversity (political, economic, social, technological, environmental, legal, following the PESTEL framework). The adapted rubric was applied by AI practitioners recruited via Prolific and by domain experts in semi-structured interviews to evaluate the quality of systemic risks generated through our pipeline.}
	\Description{Table detailing an adapted evaluation rubric for systemic risks, categorized by five main dimensions (specificity, novelty, usability, applicability, and diversity) and their subdimensions. It compares original items from prior literature with adjusted definitions used in the study.}
	\label{tab:evaluation_rubric}
	\begin{tabular}{llp{6cm}p{6cm}}
		\toprule
		\textbf{Dimension} & \textbf{Subdimension} & \textbf{Original item}                                                                                                                                                                                                                    & \textbf{Adjusted item}                                                              \\
		\toprule
		\multirow{3}{*}{Specificity} 
		                   & Plausibility          & ``Reasonable to conclude the scenario could happen given technical and social constraints'' \cite{diakopoulos2021anticipating}                                                                                                            & There is a clear connection between the AI use and the risk                         \\
		                   & Connectivity          & ``Degree to which there is a clear connection between the recommended action and the expected outcome'' \cite{dean2006identifying}                                                                                                        & It is plausible to assume that the risk follows from the AI use                     \\
		                   & Uniqueness            & ``High-quality scenarios are creative, specific, believable, and plausible'' \cite{kieslich2025scenario}                                                                                                                                  & The risk is present only in this specific AI use                                    \\
		\hline
		\multirow{3}{*}{Novelty} 
		                   & Novelty               & ``How novel do you think the story is?'' \cite{harvey2023toward, doshi2024generative}                                                                                                                                                     & The risk is novel — I have not heard about it before                              \\
		                   & Originality           & ``How original do you think the story is?'' \cite{harvey2023toward, doshi2024generative}                                                                                                                                                  & The risk is original — it is ingenious, imaginative, or surprising                \\
		                   & Rarity                & ``How rare (i.e., unusual) do you think the story is?'' \cite{harvey2023toward, doshi2024generative}                                                                                                                                      & The risk is rare — it is not thought about a lot                                  \\
		\hline
		\multirow{2}{*}{Usability} 
		                   & Usability             & ``I thought the system was easy to use'' \cite{brooke1996sus, lewis2009factor}                                                                                                                                                            & I find it easy to engage with or apply                                              \\
		                   & Learnability          & ``I needed to learn a lot of things before I could get going with this system''  \cite{brooke1996sus, lewis2009factor}                                                                                                                    & I would need to understand many aspects before engaging with it                     \\
		\hline
		\multirow{2}{*}{Applicability}
		                   & Added value           & ``The extent to which data are beneficial and provide advantages from their use'' \cite{Wang1996BeyondAccuracy}                                                                                                                           & The risk is useful for policymaking and decision-making needs                       \\
		                   & Appropriate amount    & ``The extent to which the quantity or volume of available data is appropriate''  \cite{Wang1996BeyondAccuracy}                                                                                                                            & The risk contains enough detail to inform concrete policies or decisions            \\
		\hline
		\multirow{6}{*}{Diversity} 
		                   & Political             & Includes tax policy, labour law, environmental law, trade restrictions, tariffs, political stability, merit/demerit goods, and the government's impact on health, education, and infrastructure \cite{glenn2003futures,SammutBonnici2015} & The risk is political — it reflects political drivers, decisions, or stability    \\
		                   & Economic              & Includes economic growth, exchange rates, inflation, interest rates, and other macroeconomic conditions \cite{glenn2003futures,SammutBonnici2015}                                                                                         & The risk is economic — it reflects financial or macroeconomic factors             \\
		                   & Social                & Includes cultural norms, demographics, health consciousness, population trends, safety expectations, and career attitudes \cite{glenn2003futures,SammutBonnici2015}                                                                       & The risk is social — it reflects demographic, cultural, or behavioural factors    \\
		                   & Technological         & Includes automation, innovation, research and development, technological incentives, and the pace of technological change \cite{glenn2003futures,SammutBonnici2015}                                                                       & The risk is technological — it reflects innovation, adoption, or technical change \\
		                   & Environmental         & Includes climate, weather patterns, natural disasters, and climate change concerns \cite{glenn2003futures,SammutBonnici2015}                                                                                                              & The risk is environmental — it reflects ecological or sustainability concerns     \\
		                   & Legal                 & Includes employment law, health, safety, antitrust, consumer protections, and discrimination law \cite{glenn2003futures,SammutBonnici2015}                                                                                     & The risk is legal — it reflects regulatory, legal, or compliance issues           \\
		\bottomrule
	\end{tabular}
\end{table*}
\clearpage
\section{Participant Demographics for Human Brainstorming Study}
\label{app:participants}

\begin{table}[b]
    \centering
    \footnotesize
    \caption{\textbf{Demographic breakdown of laypeople across the four AI use cases in the human-only brainstorming study.} Each use case (C: Chatbot Companion, T: AI Toy, G: Griefbot, and D: Death App) included six U.S.-based participants. Laypeople were recruited to cover a mix of demographic backgrounds.}
    \Description{Table showing the demographic breakdown of 24 lay participants across four AI use cases in the human-only brainstorming study, detailing gender, age group, education, and ethnicity for each group.}
    \label{tab:participant_demographics_laypeople}
    \begin{tabular}{llcccc}
        \toprule
        \textbf{Category} & \textbf{Value} & C & T & G & D \\
        \midrule
        Gender & Female             & 3 & 3 & 3 & 3 \\
               & Male               & 2 & 3 & 2 & 3 \\
               & Non-binary/Other   & 1 & 0 & 1 & 0 \\
        \midrule
        Age Group & 18--24 & 0 & 0 & 0 & 0 \\
                  & 25--34 & 0 & 1 & 0 & 2 \\
                  & 35--44 & 2 & 1 & 1 & 1 \\
                  & 45+    & 4 & 4 & 5 & 3 \\
        \midrule
        Education & High school or less             & 0 & 3 & 0 & 1 \\
                  & Some college / Undergraduate    & 5 & 3 & 4 & 3 \\
                  & Postgraduate degree             & 1 & 0 & 2 & 2 \\
        \midrule
        Ethnicity & White                       & 4 & 3 & 5 & 4 \\
                  & Black / African American    & 2 & 2 & 0 & 2 \\
                  & Hispanic / Latino           & 0 & 0 & 1 & 0 \\
                  & Asian                       & 0 & 1 & 0 & 0 \\
                  & Other / Mixed               & 0 & 0 & 0 & 0 \\
        \bottomrule
    \end{tabular}
\end{table}

\begin{table}[b]
    \centering
    \footnotesize
    \caption{\textbf{Demographic breakdown of domain experts across the four AI use cases in the human-only brainstorming study.} Each use case (C: Chatbot Companion, T: AI Toy, G: Griefbot, and D: Death App) included six U.S.-based participants. Domain experts were recruited for their occupational experience and expertise relevant to the four AI use cases.}
    \Description{Table showing the demographic breakdown of 24 domain experts across four AI use cases in the human-only brainstorming study, detailing gender, age group, occupation, and ethnicity for each group.}
    \label{tab:participant_demographics_domainexperts}
    \begin{tabular}{llcccc}
        \toprule
        \textbf{Category} & \textbf{Value} & C & T & G & D \\
        \midrule
        Gender & Female             & 2 & 4 & 4 & 5 \\
               & Male               & 4 & 2 & 2 & 1 \\
               & Non-binary/Other   & 0 & 0 & 0 & 0 \\
        \midrule
        Age Group & 18--24 & 0 & 0 & 2 & 1 \\
                  & 25--34 & 2 & 2 & 3 & 1 \\
                  & 35--44 & 1 & 2 & 0 & 3 \\
                  & 45+    & 3 & 2 & 1 & 1 \\
        \midrule
        Occupation  & Software Engineer             & 2 & 0 & 0 & 0 \\
                    & Product Developer             & 2 & 0 & 0 & 0 \\
                    & UI/UX Designer                   & 2 & 0 & 0 & 0 \\
        \rule{0pt}{4ex}
                    & Curriculum Director           & 0 & 1 & 0 & 0 \\
                    & General Education Teacher     & 0 & 2 & 0 & 0 \\
                    & Interventionist               & 0 & 2 & 0 & 0 \\
                    & Special Education Teacher     & 0 & 1 & 0 & 0 \\
        \rule{0pt}{4ex}
                    & Psychologist                  & 0 & 0 & 6 & 0 \\
        \rule{0pt}{4ex}
                    & Doctor                        & 0 & 0 & 0 & 3 \\
                    & Nurse                         & 0 & 0 & 0 & 3 \\
        \midrule
        Ethnicity & White                       & 5 & 6 & 1 & 4 \\
                  & Black / African American    & 1 & 0 & 5 & 0 \\
                  & Hispanic / Latino           & 0 & 0 & 0 & 0 \\
                  & Asian                       & 0 & 0 & 0 & 2 \\
                  & Other / Mixed               & 0 & 0 & 0 & 0 \\
        \bottomrule
    \end{tabular}
\end{table}

\newpage

\begin{table}[b]
    \centering
    \footnotesize
    \caption{\textbf{Demographic breakdown of laypeople across the three AI use cases in the human-and-AI brainstorming study.} Each use case (C: Chatbot Companion, T: AI Toy, G: Griefbot, and D: Death App) included six U.S.-based participants. Laypeople were recruited to cover a mix of demographic backgrounds.}
    \Description{Table showing the demographic breakdown of 18 laypeople across three AI use cases in the human-and-AI brainstorming study, detailing gender, age group, education, and ethnicity for each group.}
    \label{tab:participant_demographics_hybrid_laypeople}
    \begin{tabular}{llccc}
        \toprule
        \textbf{Category} & \textbf{Value} & T & G & D \\
        \midrule
        Gender & Female             & 3 & 3 & 3 \\
               & Male               & 3 & 3 & 3 \\
               & Non-binary/Other   & 0 & 0 & 0 \\
        \midrule
        Age Group & 18--24 & 0 & 0 & 0 \\
                  & 25--34 & 1 & 2 & 2 \\
                  & 35--44 & 1 & 1 & 0 \\
                  & 45+    & 4 & 3 & 4 \\
        \midrule
        Education & High school or less             & 0 & 2 & 0 \\
                  & Some college / Undergraduate    & 5 & 2 & 3 \\
                  & Postgraduate degree             & 1 & 2 & 3 \\
        \midrule
        Ethnicity & White                       & 6 & 4 & 5 \\
                  & Black / African American    & 0 & 1 & 0 \\
                  & Hispanic / Latino           & 0 & 0 & 0 \\
                  & Asian                       & 0 & 1 & 1 \\
                  & Other / Mixed               & 0 & 0 & 0 \\
        \bottomrule
    \end{tabular}
\end{table}

\begin{table}[b]
    \centering
    \footnotesize
    \caption{\textbf{Demographic breakdown of domain experts across the three AI use cases in the human-and-AI brainstorming study.} Each use case (C: Chatbot Companion, T: AI Toy, G: Griefbot, and D: Death App) included six U.S.-based participants. Domain experts were recruited for their occupational experience and expertise relevant to the four AI use cases. For technical reasons, we cannot report the exact demographic details of two participants in the AI Toy use case.}
    \Description{Table showing the demographic breakdown of 18 domain experts in the human-and-AI brainstorming study across three AI use cases, detailing gender, age group, occupation, and ethnicity for each group.}
    \label{tab:participant_demographics_hybrid_domainexperts}
    \begin{tabular}{llccc}
        \toprule
        \textbf{Category} & \textbf{Value} & T & G & D \\
        \midrule
        Gender & Female             & 5 & 2 & 5 \\
               & Male               & 1 & 2 & 1 \\
               & Non-binary/Other   & 0 & 0 & 0 \\
        \midrule
        Age Group & 18--24 & 0 & 0 & 1 \\
                  & 25--34 & 1 & 2 & 0 \\
                  & 35--44 & 5 & 1 & 1 \\
                  & 45+    & 0 & 1 & 4 \\
        \midrule
        Occupation  & Software Engineer             & 0 & 0 & 0 \\
                    & Product Developer             & 0 & 0 & 0 \\
                    & UI/UX Designer                & 0 & 0 & 0 \\
        \rule{0pt}{4ex}
                    & Curriculum Director           & 1 & 0 & 0 \\
                    & General Education Teacher     & 4 & 0 & 0 \\
                    & Interventionist               & 1 & 0 & 0 \\
                    & Special Education Teacher     & 0 & 0 & 0 \\
        \rule{0pt}{4ex}
                    & Psychologist                  & 0 & 4 & 0 \\
        \rule{0pt}{4ex}
                    & Doctor                        & 0 & 0 & 2 \\
                    & Nurse                         & 0 & 0 & 4 \\
        \midrule
        Ethnicity & White                       & 4 & 4 & 4 \\
                  & Black / African American    & 0 & 0 & 0 \\
                  & Hispanic / Latino           & 0 & 0 & 0 \\
                  & Asian                       & 0 & 0 & 2 \\
                  & Other / Mixed               & 2 & 0 & 0 \\
        \bottomrule
    \end{tabular}
\end{table}
\section{Application of the Rubric in Risk Evaluation}
\label{appendix:using_rubric}
The evaluation of systemic risks is inherently subjective, shaped by evaluators' prior knowledge and professional experience. For example, whether a risk is considered ``novel'' depends on familiarity with the use case and with risk assessment practices more broadly. 
To strengthen robustness, we combined three complementary evaluation strategies. First, we recruited 170 domain experts across five cohorts via Prolific \cite{prolific} to complete an annotation survey on agent-generated risks (\S\ref{subsec:scoring_practitioners}).

Second, we ran a separate annotation study with another 120 domain experts, sampled across the same five cohorts as in Study 1. These experts evaluated the human-identified risks (from the human-only and human-plus-AI conditions) using the same survey design as in the first study.

Third, we conducted in-depth semi-structured interviews with 7 domain leaders for three of the use cases (AI Toy, Griefbot, Death App), which remain speculative and underexplored. These interviews allowed us to observe how specialists contextualize, refine, or dismiss systemic risks in their own fields (\S\ref{subsec:scoring_experts}). After each interview, domain leaders completed a follow-up annotation survey using the same design as in the first and second study.

\subsection{Study 1: Scoring Agent-Generated Risks with Domain Experts as Evaluators}
\label{subsec:scoring_practitioners}

\noindent\textbf{Participants.} 
We recruited human evaluators for our annotation survey via Prolific \cite{prolific}. Evaluators were distributed across five stakeholder cohorts relevant to our use cases: decision makers, designers, developers, legal experts, and healthcare experts. Healthcare experts were recruited only for the health-related AI use cases: \emph{Griefbot} and \emph{Death App}; all other cohorts were recruited for all four AI use cases. Recruitment was managed using Prolific's built-in screeners for evaluator's organizational role, the frequency of AI use in their job, and their geographic location (United States).

In addition, we included custom survey questions to capture participants' backgrounds in more detail. We asked about their current job title, years of experience working with AI, types of AI systems they had worked on, their familiarity with conducting or interpreting risk assessments, and the kinds of risk assessments they had encountered (e.g., model cards, privacy impact assessments).

For each risk, we initially recruited 10 evaluators. After data collection, we applied a filtering step to ensure sufficient expertise: we retained only annotations from participants who rated themselves as at least ``moderately familiar'' with risk assessments (i.e., ``moderately familiar'', ``very familiar'' or ``extremely familiar''), who correctly answered the comprehension check, and who passed both attention checks embedded in the survey. In brackets below, we report the final number of unique evaluators retained in each cohort after filtering, which may be smaller than the initial recruitment:

\begin{enumerate}
    \item Decision Makers - \emph{individuals who regularly decide about the development and deployment of AI systems} ($N = 39$). To recruit them, we searched for participants with at least 3 years of experience overseeing AI initiatives, who were likely involved in the development, deployment, or governance of AI systems (e.g., product or program management), and who used AI tools 2–6 times per week.
    \item Designers - \emph{individuals who regularly design AI systems} ($N = 38$). To recruit them, we searched for participants with at least 2 years of design experience, who were likely involved in shaping the look, feel, and functionality of AI systems (e.g., product design, UX/UI, or creative direction), and who used AI tools 2–6 times per week.
    \item Developers - \emph{individuals who regularly develop AI systems} ($N = 46$). To recruit them, we searched for participants with at least 3 years of AI experience, who were likely involved in creating or maintaining AI systems (e.g., engineering or software development), and used AI 2–6 times per week.
    \item Legal Experts - \emph{individuals with expertise in AI-related law and regulation} ($N = 33$). To recruit them, we searched for participants with at least 2 years of AI-related legal experience, who were likely involved in legal oversight or regulation of AI systems, and used AI 2–6 times per week.
    \item Healthcare Experts – \emph{professionals with experience in AI-related health applications such as patient diagnosis assistance and treatment planning} ($N = 14$). To recruit them, we searched for participants with at least 3 years of healthcare experience, and who used AI 2–6 times per week.
\end{enumerate}

We consider these groups the primary stakeholders of our approach: decision makers, designers, and developers are involved in the design, development, and deployment of AI systems for novel uses and are therefore likely to encounter or prevent AI-related risks in their respective roles. Legal experts navigate compliance and regulatory requirements involving systemic risks and are therefore best positioned to assess AI risks from a legal perspective. Healthcare experts are used to assessing risks in the high-stakes context of health and health care.   
\bigskip

\noindent\textbf{Procedure.} We embed the rubric into a survey format built around annotation cards (Figure \ref{fig:annotation_card}) to measure the quality of generated risks. The survey begins with a short introduction explaining the annotation process and showing an example annotation card with explanatory overlays. After having read the introduction, we ask participants for their informed consent to participate in our study. We then provide the operational definition of systemic risks alongside a number of example systemic risks from different areas. To ensure understanding, we included a comprehension check for each AI use case. In this check, participants were shown a risk that we had pre-identified as a systemic risk according to our definition and previous literature and asked whether it should be considered systemic. The correct response was ``Yes''. Only participants who answered correctly were retained. This was followed by a brief questionnaire capturing the evaluator's background and their familiarity with AI risk assessment.

In the subsequent annotation task, each evaluator was shown a set of risks (between 12 and 16), presented individually as annotation cards (Figure \ref{fig:annotation_card}). Each card displayed the risk to be evaluated, the AI use case from which it originated, a potential impact of the risk, and the operational definition of systemic risks. For each risk, evaluators started by assessing the perceived likelihood and severity of the risk. They then indicated whether the risk should be considered systemic according to the provided definition, and finally scored each dimension of the evaluation rubric for the particular risk. The evaluators where not aware whether the presented risk was AI-generated or human-identified, ensuring that this information did not influence their evaluation. To ensure data quality, two of the annotation cards served attention checks, where evaluators were instructed to select ``Strongly Disagree'' for one of the dimensions. Figure \ref{fig:annotation_card} shows the example annotation card.

\subsection{Study 2: Scoring Human-Ideated Risks with Domain Experts as Evaluators.}
\label{subsec:scoring_experts}

\noindent\textbf{Participants.} 
We recruited an additional 120 domain experts through Prolific \cite{prolific}, sampled across the same five stakeholder cohorts as in Study 1: decision makers ($N = 28$), designers ($N = 25$), developers ($N = 23$), legal experts ($N = 28$), and healthcare professionals ($N = 16$). As before, healthcare professionals were included only for health-related use cases. Screening criteria, background questions, and data-quality checks (comprehension and attention checks) were identical to Study 1. Final cohort sizes reported in parentheses reflect the post-filtered sample.
\smallskip

\noindent\textbf{Procedure.}
Study 2 evaluated only the human-ideated risks produced through the Futures Wheel interface under the human-only and human-plus-AI conditions. To maintain comparability, we reused the same annotation cards, rubric, and survey workflow established in Study 1. Each evaluator received a randomly assigned subset of human-ideated risks (between 14 and 18), ensuring multiple evaluations per item. Risks from both human-only and human-plus-AI conditions were intermixed, and evaluators were blinded to their origin.

\subsection{Study 3: Scoring Agent-Generated Risks and Human-Ideated Risks with Domain Leaders as Evaluators.}
\label{subsec:scoring_leaders}

To evaluate how domain leaders interpret and assess both agent-generated and human-ideated risks, we conducted a semi-structured interview and a follow-up annotation study in which leaders rated the risks using the same cards and rubric as in Studies 1 and 2.

\smallskip
\textbf{Interview Study}. The semi-structured interview followed six phases: \textit{warm-up}, \textit{briefing}, \textit{vignette immersion}, \textit{risk prioritization}, \textit{gap analysis}, and \textit{debriefing}. 

In the \textit{warm-up}, leaders introduced themselves and responded to the prompt: ``If you were to compare AI to another technology from your domain that carries systemic risks, what would it be?''. This exercise invited them to draw on their expertise and frame AI through analogies to familiar technologies, setting the stage for the subsequent tasks.

Afterward, domain leaders were given a short \textit{briefing} that introduced a working definition of systemic risk, examples from different domains, and instructions for completing four interactive tasks using Figma \cite{figma}.

\begin{figure*}[h!]
    \includegraphics[width=0.65\linewidth]{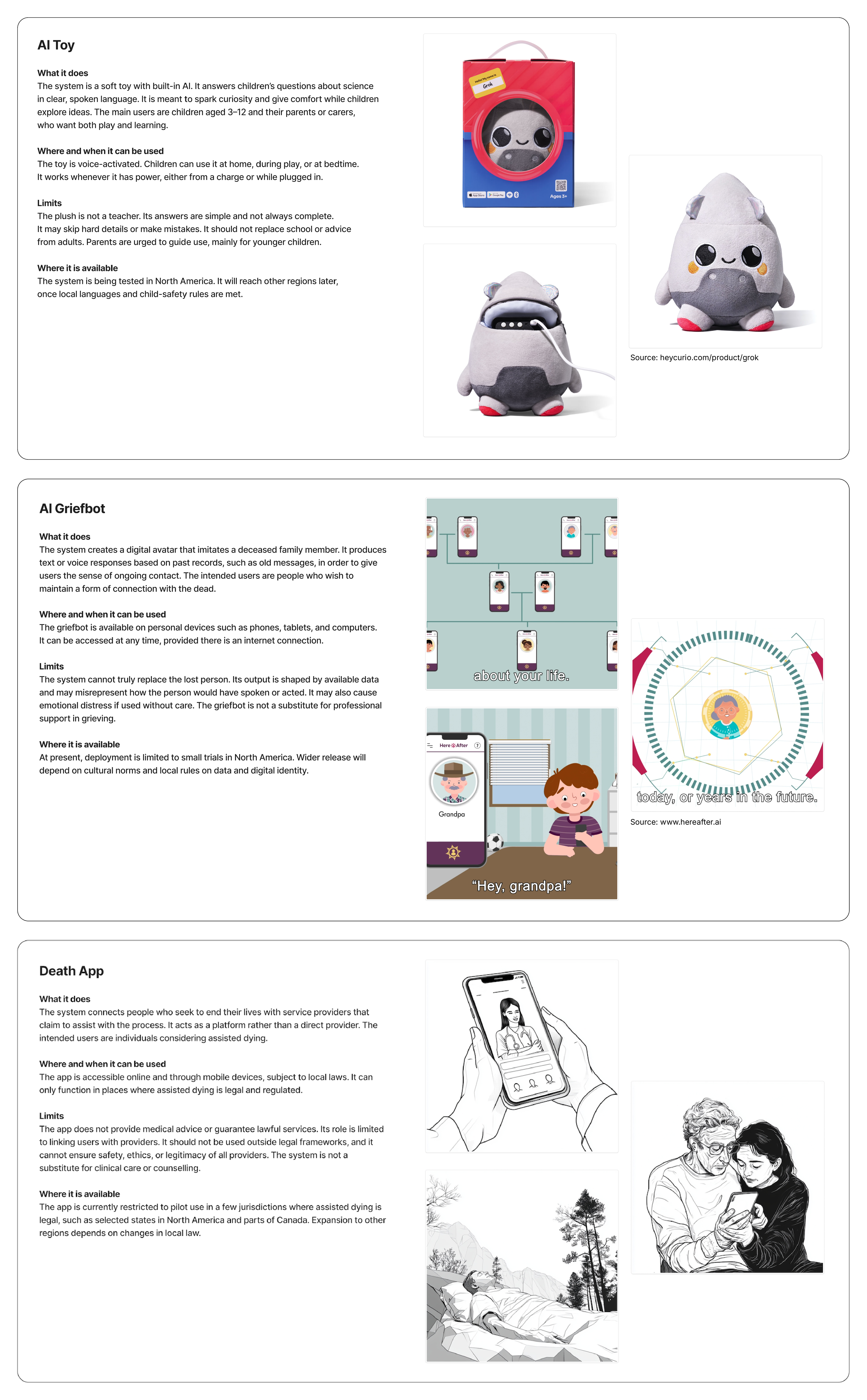}

    \caption{\textbf{Vignettes for three AI use cases: AI toy (top), griefbot (middle), and death app (bottom).} On the left, each vignette presents a structured written description of the system, following ISO 42005 guidelines for AI impact assessments by outlining its intended function and users, context of use, known limitations, and deployment environment. On the right, three accompanying images provide a visual complement: either adapted from promotional materials of existing providers (for real-world systems such as AI toy and griefbot) or created as speculative mock-ups and artworks (for non-existing applications such as death app).}
    
    \Description{Three-panel illustration showing AI use case vignettes with structured descriptions and visual representations. Each panel contains a left column with standardized text describing the AI use case following ISO 42005 guidelines (covering system function, usage context, limitations, and availability) and a right column with accompanying imagery. The top panel shows "AI Toy": description of a voice-activated plush toy for children aged 3-12, with images of a grey rocket-like plush toy in packaging and standalone views. The middle panel shows "AI Griefbot": description of a digital avatar system for communicating with deceased family members, illustrated with a family tree showing member avatars, a child saying "Hey, grandpa!" to a grandfather avatar, and a circular timeline interface. The bottom panel shows "Death App": description of a platform connecting individuals considering assisted dying with service providers, accompanied by illustrations of hands holding a smartphone displaying a medical professional, a person in a hospital bed with forest scenery, and an elderly person embracing a younger person.}
    \label{fig:vignettes}
\end{figure*}

The first task, \textit{vignette immersion}, presented leaders with a multimodal scenario describing a potential AI use case. Each scenario combined a short written description with three images. The descriptions followed ISO 42005 guidelines for AI system impact assessments, specifying the system's intended function and users, context of use, known limitations, and deployment environment \cite{iso2025SystemImpactAssessment}. Images were taken either from promotional materials of real providers (two use cases) or created by the authors as mock-ups and artworks for speculative applications (one use case). Full vignettes for all use cases are presented in Figure \ref{fig:vignettes}. After reviewing the vignette, leaders were asked to articulate the immediate consequences of this use they envision for individuals and society.  

\begin{figure*}[!h]
    \centering
    \includegraphics[width=\linewidth]{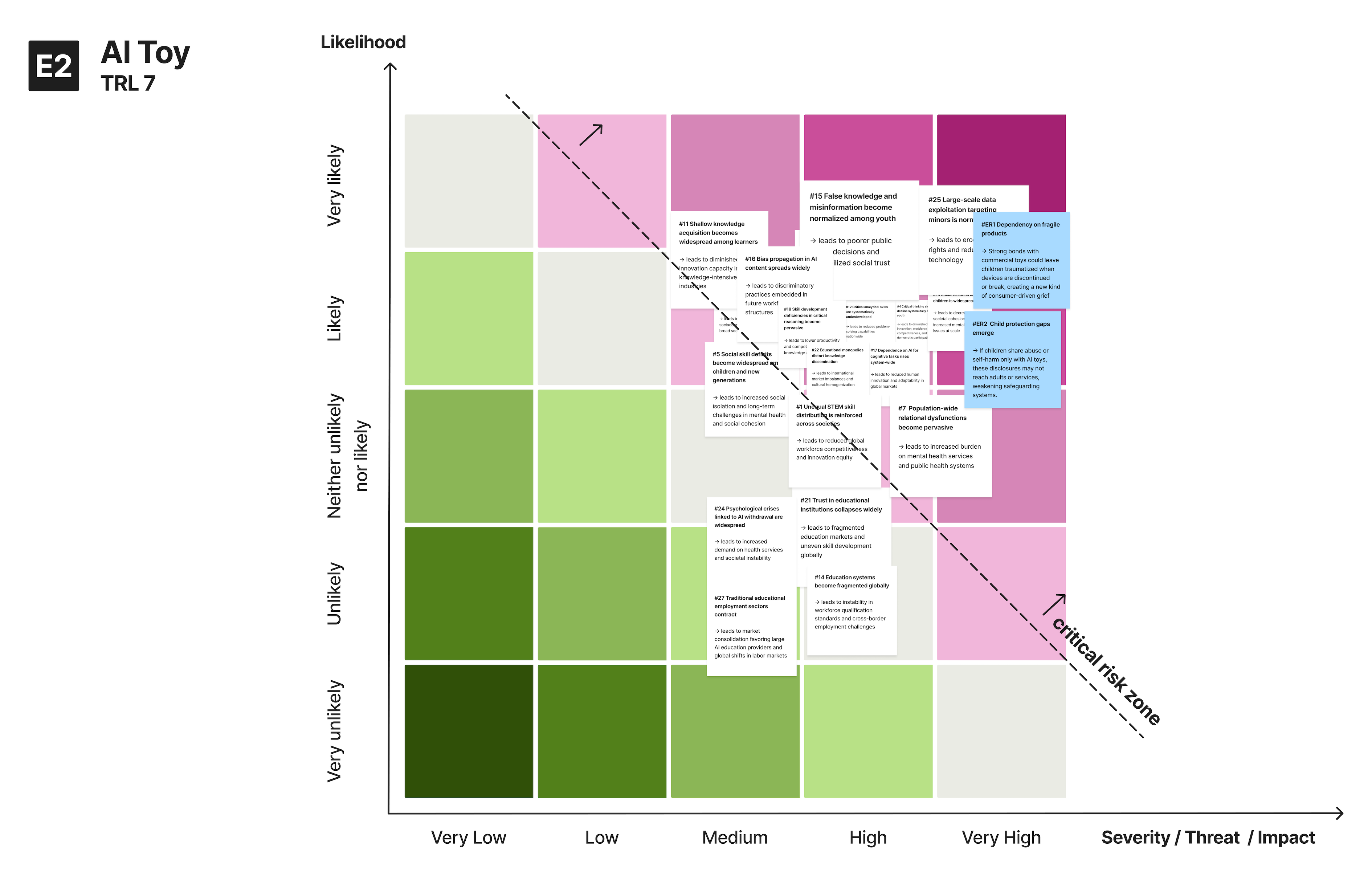}
    \caption{\textbf{Risk prioritization grid completed by domain leader E2 for the AI toy.} Risks are displayed as cards positioned according to their assessed severity and likelihood. White cards represent systemic risks generated by in-silico agents that this leader judged systemic (70\%), of which 79\% fall in the critical risk zone — where risks are both severe in impact and highly probable. Blue cards indicate additional systemic risks proposed by the leader to close important gaps: ER1 Dependency on fragile products (strong bonds with commercial toys could leave children traumatized when devices are discontinued or break, creating a new kind of consumer-driven grief); and ER2 Child protection gaps emerge (if children share abuse or self-harm only with AI toys, these disclosures may not reach adults or services, weakening safeguarding systems).}
    
    \Description{Risk prioritization matrix for the AI toy (TRL 7) showing likelihood versus severity assessment by domain leader E2. The grid displays risks as colored cards positioned according to two dimensions: likelihood (Y-axis from "Very unlikely" to "Very likely") and severity/threat/impact (X-axis from "Very Low" to "Very High"). A diagonal dashed line delineates the "critical risk zone" in the upper-right quadrant. Most risks are concentrated in the medium to high severity range. White cards represent systemic risks generated by in-silico agents, with the majority (79\% of those judged systemic) falling within the critical risk zone. Two blue cards in the high-severity, high-likelihood area represent expert-generated risks: ER1 "Dependency on fragile products" and ER2 "Child protection gaps emerge."}
    \label{fig:matrix_toy}
\end{figure*}

\begin{figure*}[!h]
    \centering
    \includegraphics[width=\linewidth]{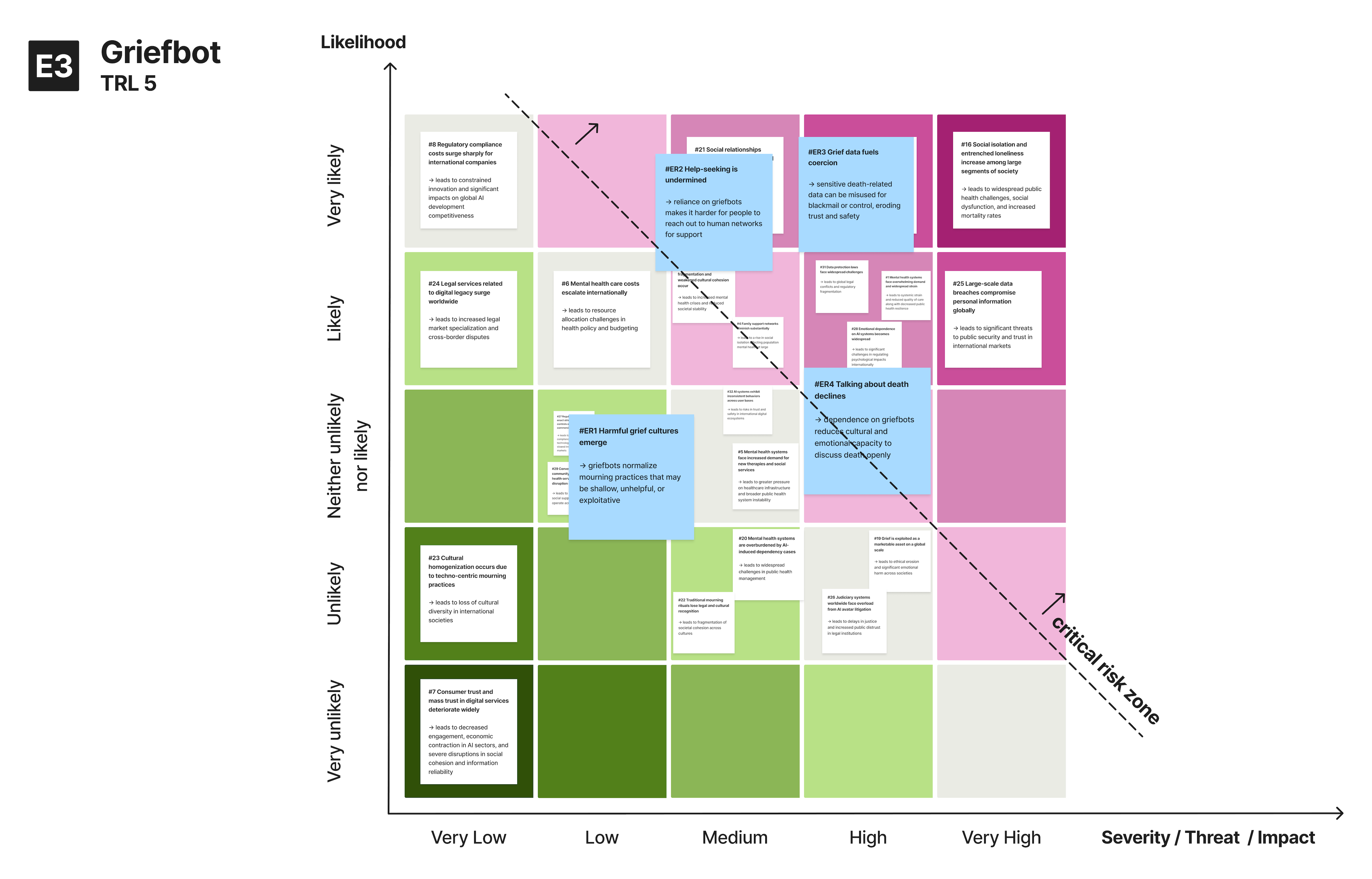}
    \caption{\textbf{Risk prioritization grid completed by domain leader E3 for the griefbot.} Risks are displayed as cards positioned according to their assessed severity and likelihood. White cards represent systemic risks generated by in-silico agents that this leader judged systemic (68\%), of which 52\% fall in the critical risk zone — where risks are both severe in impact and highly probable. Blue cards indicate additional systemic risks proposed by the leader to close important gaps: ER2 Help-seeking is undermined (reliance on griefbots makes it harder for people to reach out to human networks for support); ER3 Grief data fuels coercion (sensitive death-related data can be misused for blackmail or control, eroding trust and safety); ER1 Harmful grief cultures emerge (griefbots normalize mourning practices that may be shallow, unhelpful, or exploitative); and ER4 Talking about death declines (dependence on griefbots reduces cultural and emotional capacity to discuss death openly).}
    
    \Description{Risk prioritization matrix for the Griefbot (TRL 5) showing likelihood versus severity assessment by domain leader E3. The grid displays risks as colored cards positioned according to two dimensions: likelihood (Y-axis from "Very unlikely" to "Very likely") and severity/threat/impact (X-axis from "Very Low" to "Very High"). A diagonal dashed line delineates the "critical risk zone" in the upper-right quadrant. White cards represent systemic risks generated by in-silico agents, with 52\% of those judged systemic falling within the critical risk zone. Four blue cards represent leader-generated risks to address gaps: ER1 "Harmful grief cultures emerge" (positioned in medium severity, neither unlikely nor likely), ER2 "Help-seeking is undermined" (high severity, very likely), ER3 "Grief data fuels coercion" (very high severity, very likely), and ER4 "Talking about death declines" (high severity, neither unlikely nor likely). The distribution shows risks spread across all severity levels, with notable concentrations in the medium to high severity ranges and several high-impact risks in the critical zone.}
    \label{fig:matrix_griefbot}
\end{figure*}

\begin{figure*}[!h]
    \centering
    \includegraphics[width=\linewidth]{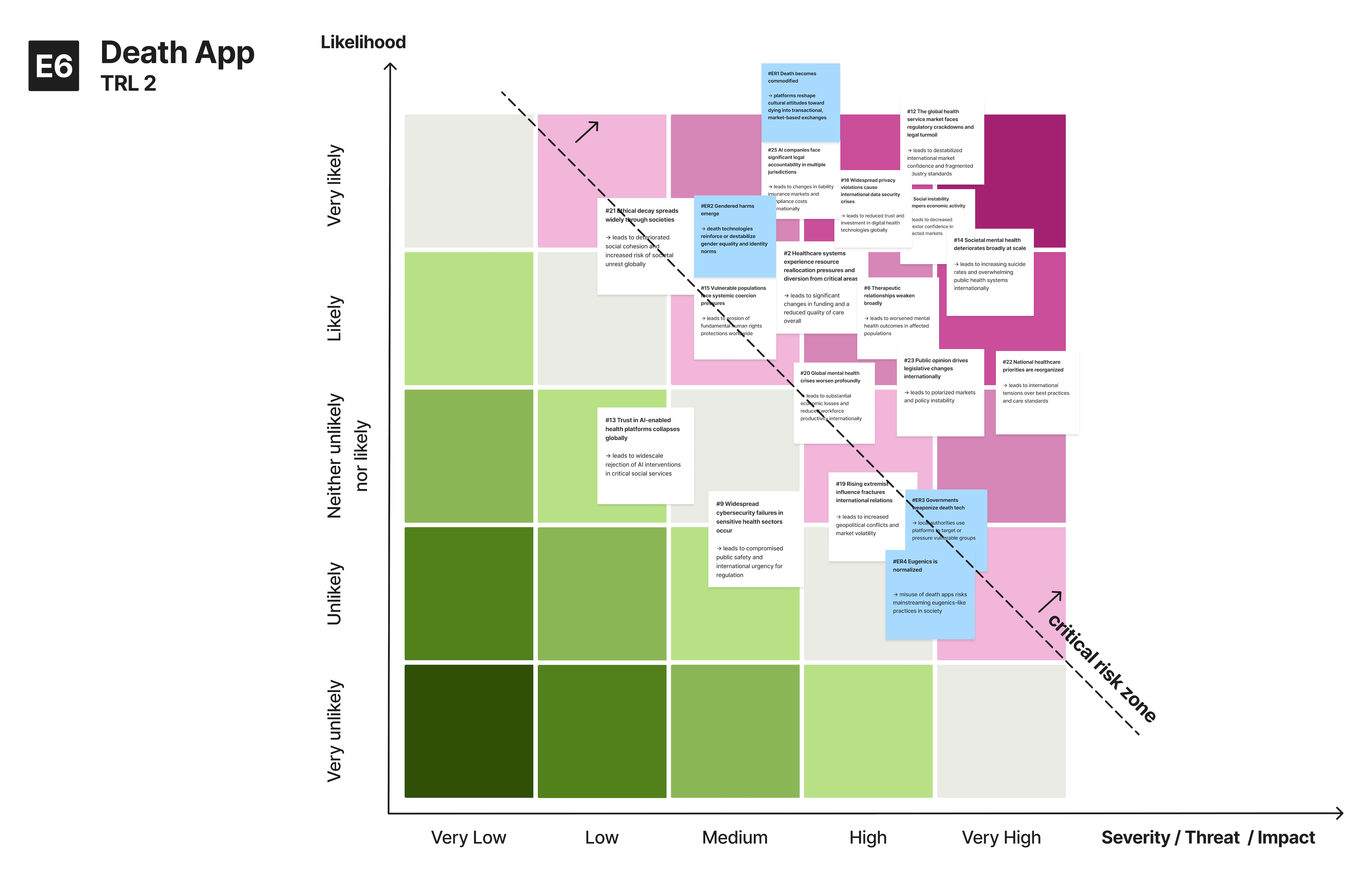}
    \caption{\textbf{Risk prioritization grid completed by domain leader E6 for the death app.} Risks are displayed as cards positioned according to their assessed severity and likelihood. White cards represent systemic risks generated by in-silico agents that this leader judged systemic (58\%), of which 73\% fall in the critical risk zone — where risks are both severe in impact and highly probable. Blue cards indicate additional systemic risks proposed by the leader to close important gaps: ER1 Death becomes commodified (platforms reshape cultural attitudes toward dying into transactional, market-based exchanges); ER2 Gendered harms emerge (death technologies reinforce or destabilize gender equality and identity norms); ER3 Governments weaponize death tech (local authorities use platforms to target or pressure vulnerable groups); and ER4 Eugenics is normalized (misuse of death apps risks mainstreaming eugenics-like practices in society).}
    
    \Description{Risk prioritization matrix for Death App (TRL 2) showing likelihood versus severity assessment by domain leader E6. The grid displays risks as colored cards positioned according to two dimensions: likelihood (Y-axis from "Very unlikely" to "Very likely") and severity/threat/impact (X-axis from "Very Low" to "Very High"). A diagonal dashed line delineates the "critical risk zone" in the upper-right quadrant. White cards represent systemic risks generated by in-silico agents, with 73\% of those judged systemic falling within the critical risk zone. Four blue cards represent leader-generated risks to address gaps: ER1 "Death becomes commodified" (positioned in medium severity, unlikely), ER2 "Gendered harms emerge" (high severity, very likely), ER3 "Governments weaponize death tech" (high severity, neither unlikely nor likely), and ER4 "Eugenics is normalized" (medium severity, unlikely). The risk distribution shows a notable concentration of high-severity, high-likelihood risks in the critical zone, with fewer risks in the lower severity categories compared to the previous use cases. Several risks are positioned in the very high severity range, reflecting the sensitive and potentially harmful nature of death-related AI applications.}
    \label{fig:matrix_deathapp}
\end{figure*}

The second task, \textit{risk prioritization}, involved reviewing a list of risks generated either using our systemic risk generation pipeline or ideated by humans. Each risk was presented on an annotation card with a unique identifier and description. Importantly, domain leaders were not informed whether a given risk had been generated by an in-silico agent or ideated by a human. Then leaders were asked to disregard risks that appeared non-systemic, unclear, or poorly phrased, and to evaluate the remaining risks by placing them on a two-dimensional grid according to their perceived likelihood of occurrence in the near future and the potential societal impact if realized. Exemplary risk prioritization grids are shown in Figures \ref{fig:matrix_toy}, \ref{fig:matrix_griefbot}, and \ref{fig:matrix_deathapp}.

Next, in the \textit{gap analysis}, leaders identified missing systemic risks, proposed additional ones, and refined risks they judged vague or insufficiently specified.

Each session concluded with a \textit{debriefing}, allowing leaders to summarize the key points that had emerged and to share final reflections, including recommendations for additional information or tools that could support deliberation about systemic risks.

\smallskip
\textbf{Follow-Up Annotation Study.} Following the interview, leaders completed a follow-up annotation study. They were sent a link to the same structured annotation cards used in Studies 1 and 2 and rated the agent generated and human-ideated risks using our standardized rubric.

\begin{table*}[]
    \normalsize
    %\centering
    \setlength{\tabcolsep}{6pt}
    \renewcommand{\arraystretch}{1.3}
    \caption{\textbf{Demographics, expertise, and use-case relevance of the seven domain leaders who evaluated systemic risks.} In addition to basic demographics, the table highlights each leader's disciplinary background, years of experience, and concrete expertise that connects directly to the AI Toy, Griefbot, or Death App use cases.}
    \Description{Table showing demographic and professional data for seven domain leaders evaluating systemic risks across three use cases: AI Toy, Griefbot, and Death App. The table details each leader's age, gender, role, institution, location, domain expertise, years of experience, and specific relevance to their assigned use case.}
    \label{tab:domain_leaders}
    \begin{tabular}{p{1.4cm} | p{0.1cm} p{0.3cm} p{0.8cm} p{2.1cm} p{1.2cm} p{1.4cm} p{2.32cm} p{0.5cm} p{3.4cm}}
    \toprule
    \textbf{Use case} & \textbf{ID} & \textbf{Age} & \textbf{Gender} & \textbf{Role} & \textbf{Institution} & \textbf{Location} & \textbf{Domain \newline expertise} & \textbf{Years exp.} & \textbf{Use case relevance} \\
    \toprule
    \multirow{2}{*}[-4.3ex]{\centering AI Toy}
    & P1 & 60 & M & Experience Design Researcher & Industry & Anonymized & HCI, Child--computer interaction & 30+ & Designed educational technologies for children \\ 
    & P4 & 28 & F & Senior Research Associate & Academia & Anonymized & HCI, AI literacy & 5+ & Researched child-facing AI learning tools \\ 
    \hline
    \multirow{3}{*}[-5.9ex]{Griefbot} 
    & P2 & 27 & M & PhD Candidate & Academia & Anonymized & HCI, Death studies & 4+ & Researched grief technologies and mourning practices \\ 
    & P3 & 30 & F & PhD Candidate & Academia & Anonymized & HCI, Digital legacy, Legal design  & 5+ & Researched digital afterlife and avatars of the deceased \\ 
    & P6 & 35 & M & Post-doctoral Researcher & Academia, NGO & Anonymized & Bioethics, Philosophy of technology & 10+ & Advised on ethical AI in healthcare and military applications \\ 
    \hline
    \multirow{2}{*}[-1.9ex]{Death App} 
    & P5 & 26 & F & UI/UX Designer, PhD Candidate  & Industry, Academia & Anonymized & Design anthropology, intersectionality theory & 5+ & Applied intersectional analysis to end-of-life care \\ 
    & P7 & 29 & F & PhD Candidate & Academia & Anonymized & Psychiatry, Healthcare & 4+ & Examined AI use in palliative care \\  
    \bottomrule
    \end{tabular}
\end{table*}

\clearpage
\onecolumn
\section{Lists of Systemic Risks Ideated by Humans Across AI Use Cases}
\label{app:results}

%%%%% HUMAN-ONLY CONDITION
\begin{table}[h!]
    \setlength{\tabcolsep}{5pt}
    \renewcommand{\arraystretch}{1.2}
    \small
    \caption[\textbf{Chatbot Companion human risks.}]%
    {\textbf{List of risks ideated by human cohorts (human-only condition) for the Chatbot Companion use case.}
    The table lists $n=4$ risks surfaced by laypeople (L) and $n=10$ risks surfaced by domain experts (E) via the Futures Wheel interface. Wording of participant's risks is lightly copy-edited for clarity.}
    \Description{Table listing 14 risks for the Chatbot Companion use case. 4 were surfaced by laypeople and 10 by domain experts. The risks are labeled with their origin.}
    \label{tab:chatbot_human_risks}
    \begin{tabular}{p{0.1cm} p{15.9cm} p{0.6cm}}
    \toprule
    \textbf{ID} & \textbf{Risk text} & \textbf{Origin} \\
    \toprule
    1  & Individuals are put on lists or lose their jobs & L \\
    2  & Human support networks fail & L \\
    3  & Misuse of personal data by third parties & L \\
    4 & Individuals may lose the ability to interact with reality effectively & L \\
    \midrule
    5  & Family relationship problems & E \\
    6  & Make decisions that harm others & E \\
    7  & Become more withdrawn & E \\
    8  & People stay away from loved ones & E \\
    9  & People enter a state of depression & E \\
    10  & People become even more lonely than before & E \\
    11 & There is a good chance that, in a mental health crisis, the AI companion may not know how to respond or have all the answers & E \\
    12 & The AI chatbot companion may steer a person in the wrong direction; a human might assess the situation more fully and determine that medical assistance is needed & E \\
    13 & Individuals could jeopardize personal relationships or work status & E \\
    14 & Individuals could be harmed or killed & E \\
    \bottomrule
    \end{tabular}
\end{table}

\begin{table}[h!]
    \setlength{\tabcolsep}{5pt}
    \renewcommand{\arraystretch}{1.2}
    \small
    \caption[\textbf{AI Toy human risks.}]%
    {\textbf{List of risks ideated by human cohorts (human-only condition) for the AI Toy use case.}
    The table lists $n=7$ risks surfaced by laypeople (L) and $n=7$ risks surfaced by domain experts (E) via the Futures Wheel interface. Additionally, $n=6$ risks were surfaced by domain leaders (D) during the semi-structured interviews. Wording of participant's risks is lightly copy-edited for clarity.}
    \Description{Table listing 20 risks for the AI Toy use case. 7 were surfaced by laypeople, 7 by domain experts, and 6 by domain leaders. The risks are labeled with their origin: letter L stands for laypeople, letter E stands for domain experts, and letter D stands for domain leaders.}
    \label{tab:aitoy_human_risks}
    \begin{tabular}{p{0.1cm} p{15.9cm} p{0.6cm}}
    \toprule
    \textbf{ID} & \textbf{Risk text} & \textbf{Origin} \\
    \toprule
    1  & Children will remain stupid & L \\
    2  & Sales could go down & L \\
    3  & Social difficulties in life & L \\
    4  & Children may have adjustment issues or mistrust animals and toys & L \\
    5  & Children can lose friends if their friends see that they are playing with AI more than them & L \\
    6  & Children can blame other people for teaching them wrong things, not blaming themselves for asking AI to answer for them & L \\
    7 & Children may get ahead of their school curriculum, leading to boredom in the classroom or frustration with a slower pace of learning & L \\
    \midrule
    8  & Identity theft could occur & E \\
    9  & Children may share the incorrect information with other people, also informing them with incorrect information & E \\
    10  & Over-reliance on technology & E \\
    11 & Privacy and data concerns & E \\
    12 & Misinformation or misinterpretation & E \\
    13 & Parental dependence on AI & E \\
    14 & Parents may lose trust in the plushy and similar educational technologies, leading to a broader backlash against AI-powered learning tools & E \\
    \midrule
    15 & New emotions reshape norms $\rightarrow$ AI toys could introduce ``new'' feelings and patterns of attachment that shift how future generations think about emotions and relationships. & D \\
    16 & Generational miscommunication grows $\rightarrow$ Kids raised with AI companions may develop ways of talking and bonding that older generations don't understand, creating systemic rifts in families and communities. & D \\
    17 & Child protection gaps emerge $\rightarrow$ If children share abuse or self-harm only with AI toys, these disclosures may not reach adults or services, weakening safeguarding systems. & D \\
    18 & Dependency on fragile products $\rightarrow$ Strong bonds with commercial toys could leave children traumatized when devices are discontinued or break, creating a new kind of consumer-driven grief & D \\
    19 & Environmental burdens from mass adoption of AI toys $\rightarrow$ Mass production adds global sustainability burdens not accounted for in child products. & D \\
    20 & Cultural backlash emerges $\rightarrow$ using AI toys to introduce girls to STEM may trigger resistance in societies with strong gender norms, undermining empowerment. & D \\
    \bottomrule
    \end{tabular}
\end{table}

\begin{table}[h!]
    \setlength{\tabcolsep}{5pt}
    \renewcommand{\arraystretch}{1.2}
    \small
    \caption[\textbf{Griefbot human risks.}]
    {\textbf{List of risks ideated by human cohorts (human-only condition) for the Griefbot use case.}
    The table lists $n=13$ risks surfaced by laypeople (L) and $n=5$ risks surfaced by domain experts (E) via the Futures Wheel interface. Additionally, $n=6$ risks were surfaced by domain leaders (D) during the semi-structured interviews. Wording of participant's risks is lightly copy-edited for clarity.}
    \Description{Table listing 24 risks for the Griefbot use case. 13 were surfaced by laypeople, 5 by domain experts, and 6 by domain leaders. The risks are labeled with their origin: letter L stands for laypeople, letter E stands for domain experts, and letter D stands for domain leaders.}
    \label{tab:griefbot_human_risks}
    \begin{tabular}{p{0.1cm} p{15.9cm} p{0.6cm}}
    \toprule
    \textbf{ID} & \textbf{Risk text} & \textbf{Origin} \\
    \toprule
    1  & There's so much that I would never know about them & L \\
    2  & I would be more emotionally stunted & L \\
    3  & I would make worse decisions in my life & L \\
    4  & I would feel more disconnected from the world & L \\
    5  & Unable to live a normal life & L \\
    6  & Interference with everyday activities / more counseling needed & L \\
    7  & Could skew actual memories / more problems & L \\
    8  & Loneliness / vicious cycle & L \\
    9  & More counseling needed & L \\
    10 & Possible unresolved grief & L \\
    11 & Society could collapse & L \\
    12 & People could lose their jobs & L \\
    13 & A lot of people could die & L \\
    \midrule
    14 & Person will not meet someone new or socialize & E \\
    15 & Could cause mental health issues & E \\
    16 & You may create negative feelings about that person & E \\
    17 & Loved one is no longer able to function in society due to their loss & E \\
    18 & If the loved one cannot handle the grieving process, it can lead to suicide over the loss of their loved one & E \\ 
    \midrule
    19 & Death logistics are displaced $\rightarrow$ griefbots take over funeral organization and legal steps, reshaping institutions that handle dying & D \\
    20 & Help-seeking is undermined $\rightarrow$ reliance on griefbots makes it harder for people to reach out to human networks for support & D \\
    21 & Grief data fuels coercion $\rightarrow$ sensitive death-related data can be misused for blackmail or control, eroding trust and safety & D \\
    22 & Talking about death declines $\rightarrow$ dependence on griefbots reduces cultural and emotional capacity to discuss death openly & D \\
    23 & Harmful grief cultures emerge $\rightarrow$ griefbots normalize mourning practices that may be shallow, unhelpful, or exploitative & D \\
    24 & Curse of flexibility spreads $\rightarrow$ the open-ended use of griefbots makes long-term impacts unpredictable and hard to control & D \\
    \bottomrule
    \end{tabular}
\end{table}

\begin{table}[h!]
    \setlength{\tabcolsep}{5pt}
    \renewcommand{\arraystretch}{1.2}
    \small
    \caption[\textbf{Death App human risks.}]%
    {\textbf{List of risks ideated by human cohorts (human-only condition) for the Death App use case.}
    The table lists $n=9$ risks surfaced by laypeople (L) and $n=11$ risks surfaced by domain experts (E) via the Futures Wheel interface. Additionally, $n=7$ risks were surfaced by domain leaders (D) during the semi-structured interviews. Wording of participant's risks is lightly copy-edited for clarity.}
    \Description{Table listing 27 risks for the Death App use case. 9 were surfaced by laypeople, 11 by domain experts, and 7 by domain leaders. The risks are labeled with their origin: letter L stands for laypeople, letter E stands for domain experts, and letter D stands for domain leaders.}
    \label{tab:deathapp_human_risks}
    \begin{tabular}{p{0.1cm} p{15.9cm} p{0.6cm}}
    \toprule
    \textbf{ID} & \textbf{Risk text} & \textbf{Origin} \\
    \toprule
    1  & The company dissolves due to lack of users & L \\
    2  & Poor company reputation that cannot be recovered & L \\
    3  & Poor visibility and user engagement with the app & L \\
    4  & App is delayed for years in court proceedings & L \\
    5 & Lawsuits over the death of someone's family member & L \\
    6 & Lawsuits from failed death-ordering attempts & L \\
    7 & Medical staff held liable for unauthorized deaths & L \\
    8 & The app becomes a hassle for providers & L \\
    9 & Providers receive incorrect or misleading information & L \\
    \midrule
    10  & People disagree with the death eligibility criteria & E \\
    11  & Riots and possible criminal proceedings & E \\
    12  & AI becomes more limited in its uses & E \\
    13  & Corporations gain monopolies on assisted suicide & E \\
    14  & Emotional detachment from suicide across society & E \\
    15 & People who should not die may end up using the app & E \\
    16 & The app could be misused against unwilling individuals & E \\
    17 & Death services proceed despite major public outrage & E \\
    18 & Death ordering is blocked from continuing & E \\
    19 & Few human doctors remain who consider alternative options & E \\
    20 & Protected Health Information (PHI) is leaked & E \\
    \midrule
    21 & Death becomes commodified $\rightarrow$ platforms reshape cultural attitudes toward dying into transactional, market-based exchanges & D \\
    22 & Gendered harms emerge $\rightarrow$ death technologies reinforce or destabilize gender equality and identity norms & D \\
    23 & Eugenics is normalized $\rightarrow$ misuse of death apps risks mainstreaming eugenics-like practices in society & D \\
    24 & Governments weaponize death tech $\rightarrow$ local authorities use platforms to target or pressure vulnerable groups & D \\
    25 & Black and grey markets thrive $\rightarrow$ unregulated death services proliferate outside legal oversight, destabilizing trust & D \\
    26 & Systemic framings miss lived realities $\rightarrow$ policy debates overlook everyday experiences of dying, grieving, and families & D \\
    27 & Death industry detaches from healthcare $\rightarrow$ assisted dying becomes a separate sector, weakening health system oversight & D \\
    \bottomrule
    \end{tabular}
\end{table}

%%%%% HUMAN PLUS AI CONDITION

\begin{table}[h!]
    \setlength{\tabcolsep}{5pt}
    \renewcommand{\arraystretch}{1.2}
    \small
    \caption[\textbf{AI Toy human risks.}]%
    {\textbf{List of risks ideated by human cohorts (human-plus-AI condition) for the AI Toy use case.}
    The table lists $n=4$ risks surfaced by laypeople (L) and $n=8$ risks surfaced by domain experts (E) via the Futures Wheel interface in the human-plus-AI condition. Wording of participant's risks is lightly copy-edited for clarity.}
    \Description{Table listing 12 risks for the AI Toy use case. 4 were surfaced by laypeople and 8 by domain experts in the human-plus-AI condition. The risks are labeled with their origin: letter L stands for laypeople and letter E stands for domain experts.}
    \label{tab:aitoy_hybrid_risks}
    \begin{tabular}{p{0.1cm} p{15.9cm} p{0.6cm}}
    \toprule
    \textbf{ID} & \textbf{Risk text} & \textbf{Origin} \\
    \toprule
        1 & Child's relationships may be stunted or nonexistent & L \\
        2 & Malformed thought patterns causing bad relationships & L \\
        3 & Children become poor students & L \\
        4 & Children could face behavioral problems at school & L \\
    \midrule
        5 & Children are misinformed & E \\
        6 & Parents trust AI less after mistakes & E \\
        7 & Children struggle to grasp basic scientific concepts & E \\
        8 & Written grammar suffers as a result & E \\
        9 & Children fall behind in school & E \\
        10 & Children struggle to develop critical thinking & E \\
        11 & Children struggle to grasp concepts that are outside of what they know & E \\
        12 & Parents face emotional outsourcing challenges & E \\
    \bottomrule
    \end{tabular}
\end{table}

\begin{table}[h!]
    \setlength{\tabcolsep}{5pt}
    \renewcommand{\arraystretch}{1.2}
    \small
    \caption[\textbf{Griefbot human risks.}]%
    {\textbf{List of risks ideated by human cohorts (human-plus-AI condition) for the Griefbot use case.}
    The table lists $n=8$ risks surfaced by laypeople (L) and $n=8$ risks surfaced by domain experts (E) via the Futures Wheel interface in the human-plus-AI condition. Wording of participant's risks is lightly copy-edited for clarity.}
    \Description{Table listing 16 risks for the Griefbot use case. 8 were surfaced by laypeople and 8 by domain experts in the human-plus-AI condition. The risks are labeled with their origin: letter L stands for laypeople and letter E stands for domain experts.}
    \label{tab:griefbot_hybrid_risks}
    \begin{tabular}{p{0.1cm} p{15.9cm} p{0.6cm}}
    \toprule
    \textbf{ID} & \textbf{Risk text} & \textbf{Origin} \\
    \toprule
        1 & Decline in traditional mourning rituals & L \\
        2 & Users develop unrealistic expectations of AI empathy & L \\
        3 & Decrease of self critical thinking & L \\
        4 & Grief support systems will wither & L \\
        5 & The role of the church in processing grief will be diminished & L \\
        6 & Con jobs will proliferate through this machinery & L \\
        7 & Funerals will lose some of their appeal & L \\
        8 & The bot may inappropriately influence the user in important areas such as healthcare & L \\
    \midrule
        9 & Cultural tensions from differing grief practices & E \\
        10 & Strain on mental health resources grows & E \\
        11 & Increased risk of AI-generated misinformation & E \\
        12 & Rise in legal disputes over digital remains & E \\
        13 & Heightened anxiety from AI miscommunication & E \\
        14 & Increased burnout risks for mental health staff & E \\
        15 & Social isolation and people becoming fantasist & E \\
        16 & Companies would take advantage of people financially and through advertising & E \\
    \bottomrule
    \end{tabular}
\end{table}

\begin{table}[h!]
    \setlength{\tabcolsep}{5pt}
    \renewcommand{\arraystretch}{1.2}
    \small
    \caption[\textbf{Death App human risks.}]%
    {\textbf{List of risks ideated by human cohorts (human-plus-AI condition) for the Death App use case.}
    The table lists $n=7$ risks surfaced by laypeople (L) and $n=20$ risks surfaced by domain experts (E) via the Futures Wheel interface in the human-plus-AI condition. Wording of participant's risks is lightly copy-edited for clarity.}
    \Description{Table listing 27 risks for the Death App use case. 7 were surfaced by laypeople and 20 by domain experts in the human-plus-AI condition. The risks are labeled with their origin: letter L stands for laypeople and letter E stands for domain experts.}
    \label{tab:deathapp_hybrid_risks}
    \begin{tabular}{p{0.1cm} p{15.9cm} p{0.6cm}}
    \toprule
    \textbf{ID} & \textbf{Risk text} & \textbf{Origin} \\
    \toprule
        1 & Strain on mental health workforce grows & L \\
        2 & Misinformation fuels public confusion & L \\
        3 & Polarized public opinion stalls policy changes & L \\
        4 & Insurance premium hikes for assisted dying coverage & L \\
        5 & Wider health disparities from premium hikes & L \\
        6 & Potential for misinformation spreading & L \\
        7 & Misinformation leads to unsafe choices & L \\
    \midrule
        8 & Rise in underground assisted dying services & E \\
        9 & Healthcare staff experience moral distress & E \\
        10 & Increased burden on legal system & E \\
        11 & Increase in unreported assisted deaths & E \\
        12 & Underground services increase community distrust & E \\
        13 & Polarization of community opinions & E \\
        14 & Increase in healthcare provider burnout & E \\
        15 & Tax increases & E \\
        16 & More unqualified staff & E \\
        17 & Mistrust of health professionals & E \\
        18 & People avoid seeking help & E \\
        19 & Division & E \\
        20 & Poorer health & E \\
        21 & Lonely people & E \\
        22 & Worse mental health outcomes & E \\
        23 & New clinical blind spots & E \\
        24 & Increased risk of malpractice and poor therapeutic outcomes & E \\
        25 & Heightened stigma around mental health issues & E \\
        26 & Legal reforms prompt cross-border service challenges & E \\
        27 & Wealth inequality grows larger & E \\
    \bottomrule
    \end{tabular}
\end{table}

\twocolumn
\clearpage

\section{Results of Systemic Risk Evaluation Across AI Use Cases}
\label{app:quantitative_risk_evaluation}

In this section, we discuss the main results of our quantitative analysis of agent-generated risks and human-identified risks, under both the human-only and human-plus-AI conditions.

\smallskip
\textbf{Diversity of Risks}. Across all conditions and AI use cases (Tables~\ref{tab:pestel_results_llm}--\ref{tab:pestel_results_hybrid_experts}), risk distributions were dominated by the ``Social'' category, indicating that both LLMs and humans primarily focused on societal implications rather than political, economic, technological, legal, or environmental concerns. The Death App consistently exhibited the highest diversity of risks, as reflected by the Shannon-Diversity Index, suggesting that this use case elicited a broader and more heterogeneous set of concerns compared to the others. Environmental risks were absent across LLM-generated risks as well as laypeople- and expert-ideated risks.

\smallskip
\textbf{Specificity, Novelty, Usability, and Applicability of Risks}. To assess differences in these evaluation metrics across the three sets of risks, we analyzed individual Likert-scale ratings (1--5) provided by domain experts in Study 1 and Study 2. We conducted pairwise comparisons between two sets at a time, with the first set always consisting of agent-generated risks and the second set consisting of either human-only or human-plus-AI risks. Because Likert data are bounded and may deviate from normality, we quantified differences using both statistical significance and standardized effect sizes.

Specifically, we calculated Cohen's $d$ to measure the magnitude of mean differences, alongside 95\% confidence intervals obtained via analytical standard errors and non-parametric bootstrap resampling (10,000 iterations). In interpreting Cohen's $d$, values around 0.2, 0.5, and 0.8 can be understood as small, medium, and large effects, respectively (Cohen, 1988). For practical interpretation on the original 1--5 Likert scale (from ``strongly disagree'' to ``strongly agree''), mean differences of about 0.2 reflect subtle shifts in ratings, around 0.5 correspond to roughly half a response category (e.g., from ``neutral'' toward ``agree''), and differences close to 1.0 indicate about a full response category shift (e.g., from ``disagree'' to ``neutral'' or from ``neutral'' to ``agree''). Metrics whose confidence intervals did not include zero were interpreted as dimensions in which the two sets of risks differed meaningfully (columns $d$ and 95\%CI in the following tables).

As an additional robustness check that does not rely on interval-scale assumptions, we also conducted non-parametric Mann--Whitney U tests to compare the distributions of ratings across the two sets. This test assesses whether ratings from one set tend to be systematically higher than those from the other, and the results were interpreted alongside the standardized mean differences (columns $r$ and $p$ in the following tables).

Across all use cases, a higher share of agent-generated risks was judged to be systemic than risks from either human condition (Tables~\ref{tab:llm_human_comparison}--\ref{tab:deathapp_hybrid}), indicating that agents tend to surface risks at broader societal and institutional scales rather than merely increasing output volume. Agent-generated risks were also rated as more severe across all use cases, suggesting that the additional risks surfaced by agents were not perceived as trivial but as potentially consequential. The human-plus-AI condition was the main exception in which severity differences were sometimes non-significant.

On the other hand, human-ideated risks were consistently more accessible than agent-generated risks. Across all use cases, experts indicated that they would \textit{need to learn} less, often by roughly half to a full Likert point, to meaningfully engage with human-identified risks. For the more technologically ready AI Toy use case, human-ideated risks were evaluated to be on par or higher in usefulness and level of detail, while for the more speculative use cases (Griefbot, Death App), agent-generated risks were perceived as more detailed and more useful. This suggests that humans struggle more with distant, uncertain use cases, whereas agents excel at expanding and articulating systemic consequences in these settings.

\begin{table*}[t]
\caption{\textbf{Distribution of LLM-generated risks over PESTEL categories --- \emph{P}olitical, \emph{E}conomic, \emph{S}ocial, \emph{T}echnological, \emph{E}nvironmental and \emph{L}egal --- for the four AI use cases.} The Shannon-Diversity Index ($H$) quantifies the diversity of the risks in each use case. The LLM-generated risks for the Death App are most spread out across the PESTEL categories. Across all cases, most risks fell into the Social category, while no Environmental risks were identified.}
\Description{Table showing the distribution of LLM-generated risks for four AI use cases across six PESTEL categories (Political, Economic, Social, Technological, Environmental, Legal), along with the Shannon-Diversity Index (H) for each AI use case. The table provides both the count and the proportion of risks in each category.}
\label{tab:pestel_results_llm}
\begin{tabular}{lrrrrrrrrrrrr|c}
\toprule
 AI Use Case & \multicolumn{2}{c}{P} & \multicolumn{2}{c}{E} & \multicolumn{2}{c}{S} & \multicolumn{2}{c}{T} & \multicolumn{2}{c}{E} & \multicolumn{2}{c}{L} & $H$ \\
\midrule
Chatbot Companion & 1 & 0.04 & 2 & 0.08 & 16 & 0.64 & 2 & 0.08 & 0 & 0.00 & 4 & 0.16 & 0.62 \\
AI Toy & 0 & 0.00 & 1 & 0.04 & 21 & 0.78 & 2 & 0.07 & 0 & 0.00 & 3 & 0.11 & 0.42 \\
Griefbot & 1 & 0.03 & 3 & 0.09 & 14 & 0.44 & 3 & 0.09 & 0 & 0.00 & 11 & 0.34 & 0.71 \\
Death App & 5 & 0.19 & 1 & 0.04 & 11 & 0.42 & 2 & 0.08 & 0 & 0.00 & 7 & 0.27 & 0.76 \\
\bottomrule
\end{tabular}
\end{table*}

\begin{table*}[]
\caption{\textbf{Distribution of risks ideated by laypeople in the human-only condition  over PESTEL categories --- \emph{P}olitical, \emph{E}conomic, \emph{S}ocial, \emph{T}echnological, \emph{E}nvironmental and \emph{L}egal --- for the four AI use cases.} The Shannon-Diversity Index ($H$) quantifies the diversity of the risks in each use case. As for the LLM-generated risks, the laypeople-ideated risks for the Death App are most diverse. Also similar to the LLM-generated risks, most human-ideated risks fell into the Social category, except for the Death App, where Legal risks were most prevalent. Environmental risks were not identified by participants.}
\Description{Table showing the distribution of risks ideated by laypeople in the human-only condition for four AI use cases across six PESTEL categories (Political, Economic, Social, Technological, Environmental, Legal), along with the Shannon-Diversity Index (H) for each AI use case. The table provides both the count and the proportion of risks in each category.}
\label{tab:pestel_results_human}
\begin{tabular}{lrrrrrrrrrrrr|c}
\toprule
 AI Use Case & \multicolumn{2}{c}{P} & \multicolumn{2}{c}{E} & \multicolumn{2}{c}{S} & \multicolumn{2}{c}{T} & \multicolumn{2}{c}{E} & \multicolumn{2}{c}{L} & H \\
\midrule
Chatbot Companion & 0 & 0.00 & 1 & 0.07 & 12 & 0.86 & 0 & 0.00 & 0 & 0.00 & 1 & 0.07 & 0.28 \\
AI Toy & 0 & 0.00 & 1 & 0.07 & 10 & 0.71 & 1 & 0.07 & 0 & 0.00 & 2 & 0.14 & 0.50 \\
Griefbot & 1 & 0.06 & 1 & 0.06 & 16 & 0.89 & 0 & 0.00 & 0 & 0.00 & 0 & 0.00 & 0.24 \\
Death App & 0 & 0.00 & 3 & 0.15 & 6 & 0.30 & 2 & 0.10 & 0 & 0.00 & 9 & 0.45 & 0.69\\
\bottomrule
\end{tabular}
\end{table*}

\begin{table*}
\caption{\textbf{Distribution of risks ideated by domain experts in the human-only condition over PESTEL categories --- \emph{P}olitical, \emph{E}conomic, \emph{S}ocial, \emph{T}echnological, \emph{E}nvironmental and \emph{L}egal --- for the four AI use cases.} The Shannon-Diversity Index ($H$) quantifies the diversity of the risks in each use case. As for the LLM-generated risks and the laypeople-ideated risks, the risks ideated by domain experts for the Death App are most diverse. As with the LLM-generated risks, the majority of risks for each AI use case fell into the Social category.}
\Description{Table showing the distribution of risks ideated by domain experts in the human-only for four AI use cases across six PESTEL categories (Political, Economic, Social, Technological, Environmental, Legal), along with the Shannon-Diversity Index (H) for each AI use case. The table provides both the count and the proportion of risks in each category.}
\label{tab:pestel_results_layexperts}
\begin{tabular}{lrrrrrrrrrrrr|c}
\toprule
 AI Use Case & \multicolumn{2}{c}{P} & \multicolumn{2}{c}{E} & \multicolumn{2}{c}{S} & \multicolumn{2}{c}{T} & \multicolumn{2}{c}{E} & \multicolumn{2}{c}{L} & H \\
\midrule
Chatbot Companion & 0 & 0.00 & 4 & 0.13 & 24 & 0.80 & 1 & 0.03 & 0 & 0.00 & 1 & 0.03 & 0.38 \\
AI Toy & 0 & 0.00 & 3 & 0.30 & 7 & 0.70 & 0 & 0.00 & 0 & 0.00 & 0 & 0.00 & 0.34 \\
Griefbot & 0 & 0.00 & 0 & 0.00 & 15 & 0.88 & 2 & 0.12 & 0 & 0.00 & 0 & 0.00 & 0.20 \\
Death App & 1 & 0.14 & 0 & 0.00 & 4 & 0.57 & 0 & 0.00 & 0 & 0.00 & 2 & 0.29 & 0.53 \\
\bottomrule
\end{tabular}
\end{table*}

\begin{table*}
\caption{\textbf{Distribution of risks ideated by laypeople in the human-plus-AI condition over PESTEL categories --- \emph{P}olitical, \emph{E}conomic, \emph{S}ocial, \emph{T}echnological, \emph{E}nvironmental and \emph{L}egal for the four AI use cases.} The Shannon-Diversity Index ($H$) quantifies the diversity of the risks in each use case. Again, the risks ideated by laypeople for the Death App are most diverse. As with the other types of risks, the majority of risks for each AI use case fell into the Social category.
\Description{Table showing the distribution of risks ideated by laypeople in the human-plus-AI condition for four AI use cases across six PESTEL categories (Political, Economic, Social, Technological, Environmental, Legal), along with the Shannon-Diversity Index (H) for each AI use case. The table provides both the count and the proportion of risks in each category.}}
\label{tab:pestel_results_hybrid_laypeople}
\begin{tabular}{lrrrrrrrrrrrr|c}
\toprule
 AI Use Case & \multicolumn{2}{c}{P} & \multicolumn{2}{c}{E} & \multicolumn{2}{c}{S} & \multicolumn{2}{c}{T} & \multicolumn{2}{c}{E} & \multicolumn{2}{c}{L} & H \\
\midrule
AI Toy & 0 & 0.00 & 0 & 0.00 & 12 & 0.92 & 0 & 0.00 & 0 & 0.00 & 1 & 0.08 & 0.15 \\
Griefbot & 0 & 0.00 & 2 & 0.06 & 25 & 0.76 & 2 & 0.06 & 0 & 0.00 & 4 & 0.12 & 0.45 \\
Death App & 1 & 0.04 & 4 & 0.17 & 12 & 0.52 & 1 & 0.04 & 0 & 0.00 & 5 & 0.22 & 0.70 \\
\bottomrule
\end{tabular}
\end{table*}

\begin{table*}
\caption{\textbf{Distribution of risks ideated by domain experts in the human-plus-AI condition over PESTEL categories ---  \emph{P}olitical, \emph{E}conomic, \emph{S}ocial, \emph{T}echnological, \emph{E}nvironmental and \emph{L}egal for the four AI use cases.} The Shannon-Diversity Index ($H$) quantifies the diversity of the risks in each use case. As in all previous conditions, the risks ideated by domain experts for the Death App are most diverse. As with the other types of risks, the majority of risks for each AI use case fell into the Social category.
\Description{Table showing the distribution of risks ideated by domain experts in the human-plus-AI condition for four AI use cases across six PESTEL categories (Political, Economic, Social, Technological, Environmental, Legal), along with the Shannon-Diversity Index (H) for each AI use case. The table provides both the count and the proportion of risks in each category.}}
\label{tab:pestel_results_hybrid_experts}
\begin{tabular}{lrrrrrrrrrrrr|c}
\toprule
 AI Use Case & \multicolumn{2}{c}{P} & \multicolumn{2}{c}{E} & \multicolumn{2}{c}{S} & \multicolumn{2}{c}{T} & \multicolumn{2}{c}{E} & \multicolumn{2}{c}{L} & H \\
\midrule
AI Toy & 0 & 0.00 & 2 & 0.09 & 19 & 0.83 & 0 & 0.00 & 0 & 0.00 & 2 & 0.09 & 0.33 \\
Griefbot & 0 & 0.00 & 3 & 0.09 & 26 & 0.79 & 2 & 0.06 & 0 & 0.00 & 2 & 0.06 & 0.42 \\
Death App & 1 & 0.03 & 2 & 0.06 & 17 & 0.53 & 2 & 0.06 & 0 & 0.00 & 10 & 0.31 & 0.64 \\
\bottomrule
\end{tabular}
\end{table*}

\begin{table*}[]
\caption{\textbf{Quantitative comparison of domain expert ratings for agent-generated versus human-identified risks (human-only condition) for the AI Toy use case.} Bolded dimensions indicate effect sizes of at least small magnitude (> 0.2).}
\Description{Table showing the results of testing for significant differences across the 13 evaluated dimensions, comparing the agent-generated and the human-identified risks in the human-only condition for the AI Toy use case. Dimensions systemic, severity, plausible, specific, easy to engage, and need to learn are printed in bold, indicating significant differences for these metrics.}
\label{tab:llm_human_comparison}
\begin{tabular}{
    p{2.2cm}
    p{1.5cm}
    p{1.7cm}
    p{1.2cm}
    p{1.2cm}
    p{2.2cm}
    p{1.0cm}
    p{1.0cm}
}
\toprule
Dimension & $\mu_{\text{agent}}$ & $\mu_{\text{human\_only}}$ & $sd$ & $d$ & $95\%$ CI & $r$ & p \\
\midrule
\rowcolor{highlight}\textbf{Systemic} & 0.86 & 0.68 & 0.37 & 0.49 & (0.25, 0.74) & 0.18 & 0.00 \\
Likelihood & 3.79 & 3.71 & 1.11 & 0.07 & (-0.14, 0.28) & 0.02 & 0.73 \\
\rowcolor{highlight}\textbf{Severity} & 3.69 & 3.44 & 1.11 & 0.23 & (0.01, 0.46) & 0.10 & 0.09 \\
Connected & 3.92 & 4.04 & 1.04 & -0.12 & (-0.35, 0.11) & -0.13 & 0.03 \\
\rowcolor{highlight}\textbf{Plausible} & 3.81 & 4.06 & 1.06 & -0.24 & (-0.45, -0.02) & -0.16 & 0.01 \\
\rowcolor{highlight}\textbf{Specific} & 3.02 & 3.54 & 1.35 & -0.39 & (-0.63, -0.16) & -0.23 & 0.00 \\
Novel & 2.88 & 3.11 & 1.38 & -0.17 & (-0.40, 0.06) & -0.09 & 0.12 \\
Original & 3.20 & 3.35 & 1.34 & -0.11 & (-0.35, 0.13) & -0.09 & 0.13 \\
Rare & 3.02 & 3.21 & 1.32 & -0.15 & (-0.38, 0.09) & -0.09 & 0.13 \\
\rowcolor{highlight}\textbf{Easy to Engage} & 3.81 & 4.14 & 0.95 & -0.35 & (-0.61, -0.10) & -0.28 & 0.00 \\
\rowcolor{highlight}\textbf{Need to Learn} & 3.60 & 2.54 & 1.21 & 0.88 & (0.65, 1.12) & 0.44 & 0.00 \\
Useful & 3.73 & 3.85 & 1.17 & -0.10 & (-0.31, 0.11) & -0.08 & 0.19 \\
Detailed & 3.44 & 3.69 & 1.20 & -0.20 & (-0.41, 0.01) & -0.12 & 0.04 \\
\bottomrule
\end{tabular}
\end{table*}

\begin{table*}[]
\caption{\textbf{Quantitative comparison of domain expert ratings for agent-generated versus human-identified risks (human-plus-AI condition) for the AI Toy use case.} Bolded dimensions indicate effect sizes of at least small magnitude (> 0.2).}
\Description{Table showing the results of testing for significant differences across the 13 evaluated dimensions, comparing the agent-generated and the human-identified risks in the human-plus-AI condition for the AI Toy use case. Dimensions systemic, likelihood, severity, connected, and need to learn are printed in bold, indicating significant differences for these metrics.}
\label{tab:llm_hybrid_comparison}
\begin{tabular}{
    p{2.2cm}
    p{1.5cm}
    p{1.7cm}
    p{1.2cm}
    p{1.2cm}
    p{2.2cm}
    p{1.0cm}
    p{1.0cm}
}
\toprule
Dimension & $\mu_{\text{agent}}$ & $\mu_{\text{hybrid}}$ & $sd$ & $d$ & $95\%$ CI & $r$ & p \\
\midrule
\rowcolor{highlight}\textbf{Systemic} & 0.86 & 0.54 & 0.39 & 0.81 & (0.63, 0.99) & 0.32 & 0.00 \\
\rowcolor{highlight}\textbf{Likelihood} & 3.79 & 3.46 & 1.10 & 0.30 & (0.16, 0.44) & 0.18 & 0.00 \\
\rowcolor{highlight}\textbf{Severity} & 3.69 & 3.26 & 1.11 & 0.39 & (0.25, 0.54) & 0.21 & 0.00 \\
\rowcolor{highlight}\textbf{Connected} & 3.92 & 3.68 & 1.00 & 0.24 & (0.09, 0.40) & 0.15 & 0.00 \\
Plausible & 3.81 & 3.81 & 1.07 & 0.00 & (-0.15, 0.16) & -0.01 & 0.89 \\
Specific & 3.02 & 3.02 & 1.27 & 0.00 & (-0.15, 0.15) & 0.00 & 0.99 \\
Novel & 2.88 & 2.75 & 1.29 & 0.10 & (-0.05, 0.25) & 0.05 & 0.24 \\
Original & 3.20 & 3.00 & 1.21 & 0.17 & (0.02, 0.32) & 0.10 & 0.02 \\
Rare & 3.02 & 2.81 & 1.25 & 0.17 & (0.02, 0.32) & 0.10 & 0.02 \\
Easy to Engage & 3.81 & 3.64 & 0.95 & 0.18 & (0.02, 0.34) & 0.08 & 0.07 \\
\rowcolor{highlight}\textbf{Need to Learn} & 3.60 & 2.90 & 1.21 & 0.58 & (0.42, 0.74) & 0.31 & 0.00 \\
Useful & 3.73 & 3.65 & 1.13 & 0.07 & (-0.09, 0.23) & 0.07 & 0.11 \\
Detailed & 3.44 & 3.33 & 1.14 & 0.10 & (-0.05, 0.26) & 0.09 & 0.04 \\
\bottomrule
\end{tabular}
\end{table*}

\begin{table*}[]
\caption{\textbf{Quantitative comparison of domain expert ratings for agent-generated versus human-identified risks (human-only condition) for the Griefbot use case.} Bolded dimensions indicate effect sizes of at least small magnitude (> 0.2).}
\Description{Table showing the results of testing for significant differences across the 13 evaluated dimensions, comparing the agent-generated and the human-identified risks in the human-only condition for the Griefbot use case. Dimensions systemic, likelihood, severity, connected, plausible, original, need to learn, useful, and detailed are printed in bold, indicating significant differences for these metrics.}
\label{tab:griefbot_human_only}
\begin{tabular}{
    p{2.2cm}
    p{1.5cm}
    p{1.7cm}
    p{1.2cm}
    p{1.2cm}
    p{2.2cm}
    p{1.0cm}
    p{1.0cm}
}
\toprule
Dimension & $\mu_{\text{agent}}$ & $\mu_{\text{human\_only}}$ & $sd$ & $d$ & $95\%$ CI & $r$ & p \\
\midrule
\rowcolor{highlight}\textbf{Systemic} & 0.78 & 0.52 & 0.43 & 0.61 & (0.46, 0.78) & 0.26 & 0.00 \\
\rowcolor{highlight}\textbf{Likelihood} & 3.71 & 3.12 & 1.16 & 0.51 & (0.35, 0.67) & 0.22 & 0.00 \\
\rowcolor{highlight}\textbf{Severity} & 3.68 & 3.40 & 1.12 & 0.25 & (0.10, 0.41) & 0.11 & 0.01 \\
\rowcolor{highlight}\textbf{Connected} & 3.76 & 3.46 & 1.17 & 0.25 & (0.10, 0.41) & 0.13 & 0.00 \\
\rowcolor{highlight}\textbf{Plausible} & 3.86 & 3.57 & 1.16 & 0.25 & (0.10, 0.40) & 0.12 & 0.00 \\
Specific & 2.82 & 2.91 & 1.26 & 0.07 & (-0.07, 0.21) & 0.04 & 0.32 \\
Novel & 2.95 & 2.78 & 1.25 & 0.14 & (0.00, 0.28) & 0.08 & 0.06 \\
\rowcolor{highlight}\textbf{Original} & 3.26 & 2.71 & 1.18 & 0.47 & (0.32, 0.62) & 0.25 & 0.00 \\
Rare & 3.06 & 2.97 & 1.26 & 0.07 & (-0.07, 0.22) & 0.04 & 0.34 \\
Easy to Engage & 3.69 & 3.52 & 0.99 & 0.17 & (0.02, 0.33) & 0.07 & 0.10 \\
\rowcolor{highlight}\textbf{Need to Learn} & 3.63 & 2.75 & 1.17 & 0.75 & (0.60, 0.92) & 0.40 & 0.00 \\
\rowcolor{highlight}\textbf{Useful} & 4.03 & 3.16 & 1.04 & 0.84 & (0.66, 1.03) & 0.42 & 0.00 \\
\rowcolor{highlight}\textbf{Detailed} & 3.73 & 3.16 & 1.05 & 0.54 & (0.38, 0.72) & 0.27 & 0.00 \\
\bottomrule
\end{tabular}
\end{table*}

\begin{table*}[]
\caption{\textbf{Quantitative comparison of domain expert ratings for agent-generated versus human-identified risks (human-plus-AI condition) for the Griefbot use case.} Bolded dimensions indicate effect sizes of at least small magnitude (> 0.2).}
\Description{Table showing the results of testing for significant differences across the 13 evaluated dimensions, comparing the agent-generated and the human-identified risks in the human-plus-AI condition for the Griefbot use case. Dimensions systemic, need to learn, useful, and detailed are printed in bold, indicating significant differences for these metrics.}
\label{tab:griefbot_hybrid}
\begin{tabular}{
    p{2.2cm}
    p{1.5cm}
    p{1.7cm}
    p{1.2cm}
    p{1.2cm}
    p{2.2cm}
    p{1.0cm}
    p{1.0cm}
}
\toprule
Dimension & $\mu_{\text{agent}}$ & $\mu_{\text{hybrid}}$ & $sd$ & $d$ & $95\%$ CI & $r$ & p \\
\midrule
\rowcolor{highlight}\textbf{Systemic} & 0.78 & 0.66 & 0.43 & 0.27 & (0.16, 0.38) & 0.12 & 0.00 \\
Likelihood & 3.71 & 3.75 & 1.07 & -0.04 & (-0.14, 0.06) & 0.00 & 0.87 \\
Severity & 3.68 & 3.45 & 1.08 & 0.21 & (0.11, 0.32) & 0.12 & 0.00 \\
Connected & 3.76 & 3.92 & 1.09 & -0.15 & (-0.26, -0.04) & -0.07 & 0.02 \\
Plausible & 3.86 & 3.90 & 1.07 & -0.03 & (-0.15, 0.08) & 0.01 & 0.81 \\
Specific & 2.82 & 3.01 & 1.24 & -0.15 & (-0.27, -0.04) & -0.09 & 0.01 \\
Novel & 2.95 & 2.89 & 1.23 & 0.05 & (-0.06, 0.17) & 0.03 & 0.38 \\
Original & 3.26 & 3.03 & 1.14 & 0.21 & (0.10, 0.32) & 0.12 & 0.00 \\
Rare & 3.06 & 2.96 & 1.23 & 0.08 & (-0.04, 0.19) & 0.04 & 0.19 \\
Easy to Engage & 3.69 & 3.77 & 0.95 & -0.08 & (-0.20, 0.03) & -0.07 & 0.03 \\
\rowcolor{highlight}\textbf{Need to Learn} & 3.63 & 3.24 & 1.17 & 0.33 & (0.22, 0.45) & 0.19 & 0.00 \\
\rowcolor{highlight}\textbf{Useful} & 4.03 & 3.80 & 1.03 & 0.22 & (0.10, 0.35) & 0.10 & 0.00 \\
\rowcolor{highlight}\textbf{Detailed} & 3.73 & 3.42 & 1.05 & 0.29 & (0.17, 0.41) & 0.13 & 0.00 \\
\bottomrule
\end{tabular}
\end{table*}

\begin{table*}[]
\caption{\textbf{Quantitative comparison of domain expert ratings for agent-generated versus human-identified risks (human-only condition) for the Death App use case.} Bolded dimensions indicate effect sizes of at least small magnitude (> 0.2).}
\Description{Table showing the results of testing for significant differences across the 13 evaluated dimensions, comparing the agent-generated and the human-identified risks in the human-only condition for the Death App use case. Dimensions systemic, severity, plausible, specific, easy to engage, and need to learn are printed in bold, indicating significant differences for these metrics.}
\label{tab:deathapp_human_only}
\begin{tabular}{
    p{2.2cm}
    p{1.5cm}
    p{1.7cm}
    p{1.2cm}
    p{1.2cm}
    p{2.2cm}
    p{1.0cm}
    p{1.0cm}
}
\toprule
Dimension & $\mu_{\text{agent}}$ & $\mu_{\text{human\_only}}$ & $sd$ & $d$ & $95\%$ CI & $r$ & p \\
\midrule
\rowcolor{highlight}\textbf{Systemic} & 0.78 & 0.63 & 0.43 & 0.37 & (0.20, 0.54) & 0.16 & 0.00 \\
Likelihood & 3.79 & 3.81 & 1.03 & -0.01 & (-0.18, 0.15) & -0.04 & 0.33 \\
\rowcolor{highlight}\textbf{Severity} & 3.78 & 3.54 & 1.13 & 0.22 & (0.03, 0.40) & 0.07 & 0.13 \\
Connected & 3.70 & 3.93 & 1.16 & -0.20 & (-0.35, -0.04) & -0.11 & 0.01 \\
\rowcolor{highlight}\textbf{Plausible} & 3.73 & 4.04 & 1.07 & -0.29 & (-0.44, -0.13) & -0.18 & 0.00 \\
\rowcolor{highlight}\textbf{Specific} & 2.87 & 3.54 & 1.31 & -0.50 & (-0.67, -0.34) & -0.27 & 0.00 \\
Novel & 2.88 & 2.98 & 1.17 & -0.08 & (-0.25, 0.09) & -0.04 & 0.42 \\
Original & 3.21 & 3.03 & 1.17 & 0.16 & (-0.02, 0.33) & 0.08 & 0.08 \\
Rare & 2.94 & 2.99 & 1.17 & -0.03 & (-0.21, 0.14) & -0.02 & 0.71 \\
\rowcolor{highlight}\textbf{Easy to Engage} & 3.50 & 3.96 & 1.03 & -0.44 & (-0.59, -0.29) & -0.25 & 0.00 \\
\rowcolor{highlight}\textbf{Need to Learn} & 3.65 & 2.95 & 1.12 & 0.63 & (0.44, 0.82) & 0.31 & 0.00 \\
Useful & 3.96 & 3.80 & 1.01 & 0.17 & (-0.01, 0.35) & 0.06 & 0.21 \\
Detailed & 3.85 & 3.73 & 1.00 & 0.12 & (-0.06, 0.30) & 0.05 & 0.33 \\
\bottomrule
\end{tabular}
\end{table*}

\begin{table*}[]
\caption{\textbf{Quantitative comparison of domain expert ratings for agent-generated versus human-identified risks (human-plus-AI condition) for the Death App use case.} Bolded dimensions indicate effect sizes of at least small magnitude (> 0.2).}
\Description{Table showing the results of testing for significant differences across the 13 evaluated dimensions, comparing the agent-generated and the human-identified risks in the human-plus-AI condition for the Death App use case. Dimensions systemic, severity, original, useful, and detailed are printed in bold, indicating significant differences for these metrics.}
\label{tab:deathapp_hybrid}
\begin{tabular}{
    p{2.2cm}
    p{1.5cm}
    p{1.7cm}
    p{1.2cm}
    p{1.2cm}
    p{2.2cm}
    p{1.0cm}
    p{1.0cm}
}
\toprule
Dimension & $\mu_{\text{agent}}$ & $\mu_{\text{hybrid}}$ & $sd$ & $d$ & $95\%$ CI & $r$ & p \\
\midrule
\rowcolor{highlight}\textbf{Systemic} & 0.78 & 0.63 & 0.45 & 0.33 & (0.24, 0.42) & 0.15 & 0.00 \\
Likelihood & 3.79 & 3.69 & 1.04 & 0.10 & (0.00, 0.19) & 0.05 & 0.06 \\
\rowcolor{highlight}\textbf{Severity} & 3.78 & 3.47 & 1.14 & 0.27 & (0.18, 0.36) & 0.13 & 0.00 \\
Connected & 3.70 & 3.55 & 1.21 & 0.13 & (0.03, 0.23) & 0.07 & 0.02 \\
Plausible & 3.73 & 3.59 & 1.17 & 0.12 & (0.02, 0.23) & 0.05 & 0.09 \\
Specific & 2.87 & 2.64 & 1.29 & 0.18 & (0.08, 0.29) & 0.10 & 0.00 \\
Novel & 2.88 & 2.84 & 1.18 & 0.04 & (-0.07, 0.14) & 0.02 & 0.40 \\
\rowcolor{highlight}\textbf{Original} & 3.21 & 2.93 & 1.15 & 0.24 & (0.14, 0.34) & 0.13 & 0.00 \\
Rare & 2.94 & 2.86 & 1.18 & 0.07 & (-0.03, 0.18) & 0.04 & 0.13 \\
Easy to Engage & 3.50 & 3.42 & 1.09 & 0.07 & (-0.03, 0.18) & 0.04 & 0.17 \\
Need to Learn & 3.65 & 3.44 & 1.12 & 0.18 & (0.08, 0.28) & 0.09 & 0.00 \\
\rowcolor{highlight}\textbf{Useful} & 3.96 & 3.69 & 1.06 & 0.25 & (0.14, 0.37) & 0.12 & 0.00 \\
\rowcolor{highlight}\textbf{Detailed} & 3.85 & 3.40 & 1.09 & 0.41 & (0.30, 0.53) & 0.21 & 0.00 \\
\bottomrule
\end{tabular}
\end{table*}

\end{document}